\documentclass[english,compress,square,super]{elsarticle}

\usepackage{lineno}
\modulolinenumbers[1]
\usepackage{geometry}
\usepackage[colorlinks=true]{hyperref}

\journal{Physical Review Fluids}

\usepackage{amsmath}
\usepackage[T1]{fontenc}
\usepackage{xcolor}
\newcommand\tb[1]{\boldsymbol{#1}}
\newcommand\sym[1]{\mathrm{sym}\left(#1\right)}
\newcommand\skw[1]{\mathrm{skw}\left(#1\right)}
\newcommand\td{\mathrm{d}}
\newcommand\pd{\partial}
\usepackage{mathrsfs}

\newcommand\ddfrac[2]{{\displaystyle\frac{\displaystyle #1}{\displaystyle #2}}}
\usepackage{stmaryrd}
\usepackage{enumerate}
\usepackage[shortlabels]{enumitem}
\usepackage{caption}
\usepackage{subcaption}
\usepackage{cleveref}
\usepackage{algorithm}  
\usepackage{algpseudocode}  
\usepackage{amssymb}
\usepackage{upgreek}
\usepackage{pgfplots}
\pgfplotsset{compat=1.16}
\usepackage{svg}
\usepackage{multirow}
\usepackage{booktabs}
\usepackage{mathtools}
\usepackage{adjustbox}
\usepackage[para,online,flushleft]{threeparttable}

\makeatletter
\usepackage{xspace}
\usepackage{accents}
\algnewcommand{\algorithmicgoto}{\textbf{go to}}%
\algnewcommand{\Goto}{\algorithmicgoto\xspace}%
\DeclareRobustCommand\bigop[1]{%
\mathop{\vphantom{\sum}\mathpalette\bigop@{#1}}\slimits@
}
\newcommand{\bigop@}[2]{%
  \vcenter{%
    \sbox\z@{$#1\sum$}%
    \hbox{\resizebox{\ifx#1\displaystyle.9\fi\dimexpr\ht\z@+\dp\z@}{!}{$\m@th#2$}}%
  }%
}

\usepackage{amsthm}
\newtheorem{remark}{Remark}

\newcommand{\revision}[1]{\textcolor{black}{#1}}

\begin{document}

\begin{frontmatter}

\title{Continuum granular flow model with restitution--derived viscoelastic damping}

\author[1]{Bodhinanda {Chandra}}\ead{bchandra@berkeley.edu}
\author[2]{Sachith {Dunatunga}}
\author[1]{Ken {Kamrin}\corref{cor1}}
\ead{kkamrin@berkeley.edu}

\address[1]{Department of Mechanical Engineering, University of California, Berkeley, CA, 94720, USA}
\address[2]{Department of Mechanical Engineering, Massachusetts Institute of Technology, 77 Massachusetts Ave, Cambridge, 02319, MA, USA}

\cortext[cor1]{Corresponding author}

\begin{abstract}
This work presents a unified viscoelastic–viscoplastic continuum framework for modeling rate-dependent granular flows across regimes. The formulation incorporates two distinct rate-dependent mechanisms, namely micro-inertia and viscoelastic dissipation, within a single continuum description. A central contribution is an explicit link between the coefficient of restitution and a continuum viscosity, derived from an analysis of wave attenuation in granular assemblies, thereby establishing a direct connection between particle-scale collision physics and macroscopic damping. This relation is introduced while retaining inertia-dependent plastic flow governed by the classical $\mu(I)$ rheology. The constitutive model is constructed by meticulously partitioning elastic and plastic responses within the model and corresponding stress-update routine, such that viscous dissipation governs wave propagation and collisional processes without altering the plastic flow rule. The framework is implemented within the material point method to simulate transient processes involving large deformations, material separation, and subsequent reconsolidation. A range of numerical examples, including steady, transient, vibrational, and impact-driven flows, demonstrates that the model captures wave propagation, diffusion, and rate-dependent granular behavior within a unified continuum setting.
\end{abstract}



\end{frontmatter}


\section{Introduction}

Dry granular flow exhibits rate-dependent behavior due to non-vanishing microscopic length and time scales. In this study, we focus on modeling grain-scale collisions and the associated energy dissipation. The presence of granular collisions gives rise to two distinct rate-dependent behaviors: (i) micro-inertia effects in dense granular flows subjected to plastic shear, and (ii) the role of restitution in bulk elastic response. Micro-inertia arises from grain-scale inertia associated with relative particle motion and velocity fluctuations, which become important when the time scale associated with macroscopic deformation is comparable to microscopic rearrangement times, a condition commonly characterized by the inertial number $I$ \cite{gdr2004dense, da2005rheophysics}. In densely flowing granular states, this mechanism modifies the effective resistance to flow and gives rise to rate-dependent apparent friction, which is commonly described by inertial rheologies such as the $\mu(I)$ model \cite{jop2006constitutive}. By contrast, restitution governs the dissipation of kinetic energy during elastic particle-to-particle collisions. Its effects manifest differently across regimes, contributing to energy loss through collisions in gas-like states and to the attenuation of stress waves and oscillations in solid-like states. It is known that the coefficient of restitution $e$ influences the $\mu(I)$ relation. This is particularly apparent in frictionless grains, but its effect diminishes for grains with friction \cite{gdr2004dense, da2005rheophysics}. In practice, this dependence can be absorbed into the calibration of $\mu(I)$ parameters. 

While micro-inertia has been widely incorporated into continuum descriptions, both within hydrodynamic models, e.g.~\cite{staron2012granular, staron2014continuum}, and elasto-viscoplastic solid formulations, e.g.~\cite{kamrin2010nonlinear, dunatunga2015continuum}, mechanisms for incorporating elastic restitution into the remaining bulk phenomenology, namely, the damping of elastic wave propagation and the physically consistent dissipation during volumetric collisions, have not yet been systematically addressed in continuum mechanics frameworks. The challenge of incorporating restitution into a continuum model in this sense is twofold: (i)~\textit{how to relate the coefficient of restitution, which is a microscopic granular property, to a well-defined continuum parameter, for example through the concept of viscosity}, and (ii)~\textit{how to introduce its dissipative effects in a manner that does not artificially add extra damping into the continuum plastic flow behavior}. Motivated by these challenges, the present study attempts to formulate a continuum mechanics model that addresses both aspects outlined above.

The empirical kinematic coefficient of restitution, $e$, was originally proposed by Newton in 1687 \cite{newton1803, cook1986newton}. As defined by Newton, the coefficient of restitution, $e$, can be computed for a particle-to-particle collision as:
\begin{eqnarray}
e = \frac{-\Tilde{v}(t_c)}{\Tilde{v}(0)}\,,
\label{eq:newton_restitution}
\end{eqnarray}
where $\Tilde{v}$ is the relative velocity in the direction of collision and $t_c$ denotes the contact time (or release time), i.e.~the time at which the contacting particles separate. For a normal collision in which no external energy is supplied to the system simultaneously, thermodynamic constraints require that $e \in [0,1]$, where $e=1$ corresponds to a fully elastic collision and $e=0$ corresponds to a fully plastic collision. The value of $e$ depends on the colliding material types as well as the contact geometry, which is influenced by particle shape and angle of contact \cite{hastie2013experimental, wang2015experimental}. Experimental observations, e.g.~\cite{bridges1984structure, kuwabara1987restitution, goldsmith2001impact}, further indicate that $e$ depends on the impact velocity, with larger values typically observed at lower $\Tilde{v}$ and decreasing values as $\Tilde{v}$ increases.

Restitution has been incorporated in the discrete element method (DEM) \cite{cundall1979discrete} through normal contact force models to introduce dissipative particle interactions. The normal contact force in DEM is often expressed as:
\begin{eqnarray}
F_n = k_n \delta_n + c_n \dot{\delta}_n\,, \qquad F_n\geq0\,,
\label{eq:dem_normal_contact_force}
\end{eqnarray}
where $k_n$ is the contact spring stiffness and $c_n$ is the normal damping factor. Here, $\delta_n$ and $\dot{\delta}_n$ denote the spring compression distance and compression rate, which in DEM are often quantified through particle interpenetration. For example, when the Kelvin–Voigt model \cite{hunt1975coefficient} is adopted and the contact duration is approximated as a half-period single-degree-of-freedom (SDOF) vibration, a closed-form relation between $c_n$ and $e$ can be obtained as:
\begin{eqnarray}
c_n = \frac{2|\ln{e}|}{\sqrt{|\ln{e}|^2+\pi^2}} \sqrt{k_n \Tilde{m}}\,,
\label{eq:normal_damping_coefficient_DEM}
\end{eqnarray}
where $\Tilde{m}$ is the effective mass, which for two colliding particles $i$ and $j$ is given by $\Tilde{m}=m_im_j/(m_i+m_j)$ \cite{schwager2008coefficient}. Different choices of spring–dashpot contact models lead to different expressions for $c_n$ \cite{butcher2000characterizing, ismail2008impact}. For example, for the Maxwell model, which is the dual (or reciprocal) formulation of the Kelvin–Voigt model, the corresponding normal damping coefficient is given by:
\begin{eqnarray}
c_n = \frac{\sqrt{|\ln{e}|^2+\pi^2}}{2|\ln{e}|} \sqrt{k_n \Tilde{m}}\,.
\label{eq:normal_damping_coefficient_DEM_maxwell}
\end{eqnarray}

The above expressions are derived by solving a damped SDOF oscillator, with the contact time $t_c$ assumed to correspond to the half-period of the displacement oscillation, i.e.~at $t=t_c$, the displacement satisfies $\delta_n(t_c)=0$. This assumption, however, is inaccurate for two contacting grains, since evaluating the relative velocity at this instant occurs after the normal contact force has already changed sign in the oscillator solution. This violates the unilateral contact condition, where contact forces can only be compressive and cannot sustain tension (cf.~Eq.~\eqref{eq:dem_normal_contact_force}$_2$). Moreover, in the case of high damping, the contact time may not admit a real solution, meaning that particles stick together after a head-on collision, leading to the \textit{dissipative capture} condition \cite{schwager2007coefficient}. As highlighted by \citet{schwager2007coefficient}, a more appropriate way to compute the relation between the damping coefficient $c_n$ and $e$ is to define $t_c$ as the instant at which the contact force vanishes as particles separate from one another. This corresponds to the half-period of the velocity vibration, occurring when the releasing velocity is maximum or, correspondingly, when the acceleration decreases to zero, i.e.~$\ddot{\delta}_n(t_c)=0$ at $t=t_c$.

Moreover, depending on the considered elasticity model, the relation between the damping factor and restitution may differ from that of the classical linear spring–dashpot model. For Hertzian contact model, $c_n$ becomes a function of $\delta_n^{1/2}$, since $k_n \sim \delta_n^{1/2}$ \cite{tsuji1992lagrangian}. In this case, the damping factor also depends on the viscous material parameters of the contacting grain, $\vartheta_g$ and $\eta_g$ (bulk and shear viscosities, with subscript $g$ denoting grain properties), analogous to the dependence of $k_n$ on the elastic moduli of the grains, e.g.~$E_g$ and $\nu_g$ \cite{brilliantov1996model}. For further derivations and details of nonlinear spring–dashpot contact models, the reader is referred to Chapter 3 in \citet{brilliantov2010kinetic}.

Beyond the normal contact response, the tangential interaction between grains plays an equally important role in governing frictional resistance and associated energy dissipation. In DEM, the tangential contact force magnitude for a frictional grain is often modeled using a regularized Coulomb friction law:
\begin{eqnarray}
F_t=\min\left(\mu_g F_n,\,k_t\delta_t+ c_t\dot{\delta}_t\right)\,,
\end{eqnarray}
where $\mu_g$ denotes the intergranular friction coefficient, while $k_t$ and $c_t$ indicate the tangential spring and damping coefficients, respectively. The tangential accumulated slip deformation and slip rate are denoted by $\delta_t$ and $\dot{\delta}_t$. While many models have been developed to quantify the normal damping coefficient, $c_n$, and its relation to $e$, far fewer studies have focused on defining the corresponding tangential or shear damping coefficient, $c_t$. Nevertheless, shear damping must also exist in granular media, since pure shear elastic waves are naturally dissipated in granular beds.

The tangential damping coefficient is often modeled using the simple relation
\begin{eqnarray}
c_t = \kappa\, c_n\,,
\end{eqnarray}
where $\kappa$ is a scaling factor, typically set to unity, yielding $c_t = c_n$ \cite{tsuji1992lagrangian, silbert2001granular}. Efforts to derive more physically motivated closure models for $c_t$ have been made in the past. For example, P\"oschel et al. \cite{becker2008coefficient, schwager2008coefficientb, glielmo2014coefficient} proposed characterizing changes in the tangential velocity component using a coefficient of tangential restitution. However, as later pointed out \cite{brilliantov2010kinetic}, the tangential coefficient of restitution is a rather complicated parameter, since tangential restitution is not independent of normal restitution due to the coupling of tangential and normal forces through Coulomb friction. Its value, therefore, depends on both the normal and tangential components of the relative contact velocity \cite{schwager2008coefficient, becker2008coefficient}. Furthermore, when particles rotate during collision, additional deformation rates arise due to rotational motion, generating viscous stresses that counteract rotation \cite{brilliantov1998rolling}. This results in a torque that decelerates particle rotation and consequently reduces relative tangential motion. Owing to these additional complexities, the assumption $c_t = c_n$ remains a commonly adopted and practical choice in current DEM implementations, implicitly treating dissipation in the normal and tangential directions in a comparable manner.

Building on the aforementioned contact-level descriptions of dissipation, one may examine how such mechanisms manifest in continuum-scale models of granular media. In the field of kinetic theory, the effect of intergranular restitution enters continuum descriptions primarily through energy dissipation during collisions, which is commonly quantified by the cooling rate, $\zeta$. For smooth inelastic particles, the cooling rate scales as $\zeta \sim (1-e^2)$, reflecting the loss of kinetic energy at each collision \cite{lun1984kinetic, brilliantov2010kinetic}. Starting from this microscopic description of collisional dissipation, the inelastic Enskog–Boltzmann equation is typically approximated to obtain macroscopic transport equations expressed in terms of hydrodynamic fields \cite{jenkins1985grad}. For weakly non-uniform granular gases, this is commonly achieved using the Chapman–Enskog expansion, originally developed for molecular gases \cite{chapman1990mathematical} and later extended to inelastic systems \cite{jenkins1985grad, brey1998hydrodynamics, garzo1999dense}. At first order in the expansion, Navier–Stokes constitutive relations can be recovered \cite{brey1998hydrodynamics}, in which the shear viscosity $\eta$ can be expressed explicitly in terms of the restitution coefficient through the cooling rate, $\zeta$, and its dependence on the granular temperature. This establishes a direct link between microscopic collisional dissipation and continuum transport properties \cite{lun1984kinetic, garzo1999dense, brilliantov2010kinetic}. While the Chapman–Enskog framework is strictly applicable to dilute and moderately dense granular gases, restitution more generally acts as a microscopic mechanism of viscous dissipation that remains relevant in dense, solid-like granular packings. In such regimes, energy is dissipated through the attenuation of transient motions, particle vibrations, and stress waves, even when there is no persistent macroscopic flow \cite{sadd2000simulation,marketos2013micromechanics, gu2020discrete}.

The objective of the current study is to incorporate the two rate-dependent mechanisms, namely micro-inertia and viscoelastic damping, seamlessly into a continuum mechanics framework. The developed framework shall permit simulations of granular behavior across different regimes, including solid-like quasi-static and dynamic responses with realistic wave transmission and attenuation, liquid-like plastic flow at large strains governed by inertia-dependent rheology, as well as gas-like states with a proper distinction between material separation and reconsolidation.

To achieve this, we begin by deriving and introducing an explicit relationship between the coefficient of restitution, $e$, and the continuum bulk and shear viscosities, $\vartheta$ and $\eta$. This is accomplished by analyzing how a granular assembly responds to a volumetric compression impulse that induces a half-period vibrational motion. To properly assess the link between restitution and viscosity, the contact time $t_c$ is taken as the moment when the contact forces between all particles — or, equivalently, the stress — within the assembly drop to zero, consistent with the observation of \citet{schwager2007coefficient}. This definition can be directly embedded in the continuum framework of \citet{dunatunga2015continuum}, where dry granular assemblies are assumed to carry no tensile stress and become stress-free once their bulk density decreases below a critical value. It is important to note that, within this regime, collisional dissipation is neglected; consequently, the model does not account for energy losses arising from particle interactions once the material density drops below the critical density threshold.

\begin{figure}[h!]
    \centering
    \includegraphics[width=0.65\linewidth]{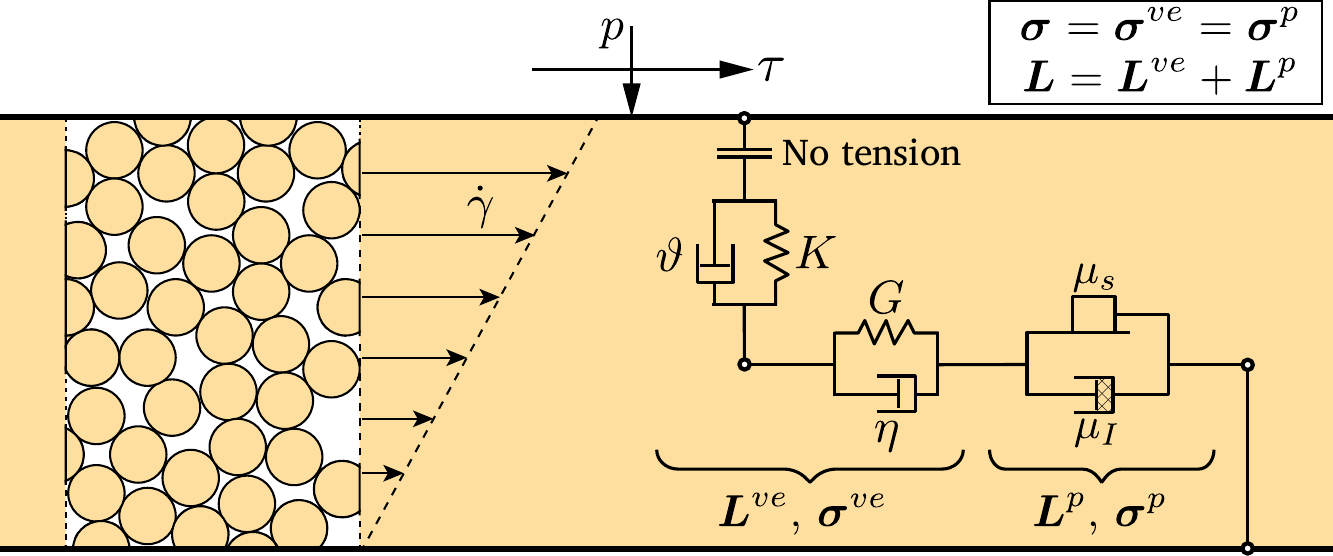}
    \caption{Schematic illustration of the considered rheological model representing a volume element of a dense granular assembly under confined shearing. The spring elements, $K$ and $G$, denote the bulk and shear moduli, whereas the linear dashpots, $\vartheta$ and $\eta$, indicate the bulk and shear viscosities. The frictional plastic elements consist of a static frictional slider and a complex dashpot, denoted $\mu_s$ and $\mu_I$, respectively. These elements correspond to the rate-independent and rate-dependent frictional resistance, respectively. Furthermore, the no-tension element separates when the granular assembly undergoes tensile loading. Since the plastic elements are arranged in series with the viscoelastic components, the stress is the same in each branch, $\tb \sigma=\tb \sigma^{ve}=\tb \sigma^p$, whereas the strain rate follows an additive decomposition, i.e.~$\tb L=\tb L^{ve} + \tb L^p$.}
    \label{fig:1_rheology_diagram}
\end{figure}

Furthermore, the viscoelastic model is designed to only damp elasticity and wave propagation within the granular medium without altering the plastic flow state, which is governed solely by the considered $\mu(I)$ rheology, following the implementation of \citet{dunatunga2015continuum}. This is achieved by carefully separating the elastic and plastic components of the kinematic fields, and applying the viscous stress contribution only to the elastic part, following the Kelvin–Voigt arrangement. The resulting analog model is illustrated in Fig.~\ref{fig:1_rheology_diagram}. Without this treatment, viscous effects from the linear dashpot would also modify plastic deformation, altering the intended $\mu(I)$ behavior and preventing the realization of energetically accurate granular flows. In addition, correct attenuation of elastic wave propagation plays a critical role in obtaining accurate dense flow predictions: undamped pressure oscillations transiently elevate and reduce the pressure, and since the shear strength is proportional to the pressure, these transient drops can trigger flow events that would not otherwise occur at the mean pressure in the absence of oscillations. By correctly damping these oscillations, the designed framework avoids spurious overactivation of plastic deformation while preserving the correct rate-dependent response.

The developed viscoelastic–viscoplastic constitutive model is implemented within the material point method (MPM) \cite{Sulsky1994}, a hybrid Eulerian–Lagrangian numerical framework that combines the strengths of both descriptions. In MPM, a continuum body is discretized by a set of Lagrangian material points that carry the material kinematics and history-dependent state variables, while a background Eulerian grid is employed to approximate spatial gradients and evaluate the balance equations. The background grid is reset at every time step, allowing MPM to avoid mesh distortion issues commonly encountered in mesh-based methods such as the finite element method (FEM). Owing to its ability to accommodate large deformations while retaining the accuracy of FEM in elasto-static regions, MPM is particularly well suited for simulating granular materials that exhibit both solid-like and flowing behavior.

Over the past two decades, numerous studies have applied MPM to model granular flow problems using a wide range of constitutive descriptions. Early work by \citet{wikeckowski2004material} employed MPM for granular flow simulations, utilizing viscoplastic regularization to prevent tensile stress states and ensure numerical stability, although the formulation remained rate-independent in its asymptotic limit. Rate-independent plasticity models based on Mohr–Coulomb and Matsuoka–Nakai criteria were adopted by \citet{andersen2009analysis} and \citet{mast2015simulating}, respectively, in simulations of granular column collapse. The introduction of the $\mu(I)$ rheology into a continuum elasto-viscoplastic framework by \citet{dunatunga2015continuum} enabled inertia-dependent plastic flow together with explicit treatment of granular separation and reconsolidation, suited to model granular media in all solid, liquid, and gas phases. \citet{fern2016role} further examined the influence of constitutive modeling choices on simulations of granular collapse, demonstrating their impact on run-out distance and failure mechanisms in loose and dense sands. Furthermore, the mechanisms of projectile impact and penetration into dry granular media were investigated by \citet{dunatunga2017continuum}. Extensions of the $\mu(I)$ rheology to account for fluid–granular interactions through the $\mu(I_v)$ framework were later proposed for modeling fluid–sediment systems and fine-particle suspensions by \citet{baumgarten2019general, baumgarten2019generalb}. More recent developments include the adoption of non-isochoric flow rules with evolving packing density by \citet{agarwal2021efficacy} within a rate-independent framework. Efforts to modify MPM spatial interpolation schemes to distinguish granular separation were introduced by \citet{seyedan2021solid}. More recently, nonlocal extensions of the $\mu(I)$ rheology based on granular fluidity concepts were introduced independently by \citet{dunatunga2022modelling} and \citet{haeri2022three}, following earlier work by \citet{kamrin2012nonlocal} and \citet{ henann2013predictive}. Approaches that incorporate collisional stresses derived from kinetic theory have also been proposed for large-deformation granular modeling within MPM \cite{marveggio2022phase, marveggio2024granular, feng2025material}. These studies typically assume a generalized Maxwell-type model, in which the collisional stress is placed in parallel with an elasto-plastic element arranged in series. In such formulations, collisional dissipation is directly coupled to plastic deformation, which may, in certain regimes, limit the representation of energetic flow features or lead to deviations from the standard $\mu(I)$ behavior for dry flowing granular materials.

The present manuscript is organized as follows. In \Cref{sec:continuum}, the theoretical formulation is elaborated, encompassing the governing equations and the proposed constitutive theory. \Cref{sec:modeling} then presents the implementation details within the MPM framework, with particular emphasis on the stress update algorithm. To keep the explanation brief, details of the MPM implementation are kept short in this section; additional information is provided in \ref{app:mpm}. Furthermore, \Cref{sec:num_ex} presents numerical examples that verify the correctness of the implementation and highlight the improved features obtained from the proposed model. Lastly, \Cref{sec:conclusion} provides a summary and the main conclusions of the study.

\section{Theory}
\label{sec:continuum}

In this section, we establish the governing equations and constitutive framework for modeling the viscoelastic–viscoplastic behavior of cohesionless granular media. Before proceeding, the mathematical operators and notations used in this study are summarized. $\dot{\square}$ and $\ddot{\square}$ denote the first and second material time derivatives; $\square\cdot\square$ and $\square:\square$ represent single and double tensor contractions; and $\square\otimes\square$ indicates the dyadic product. For a second-order tensor $\tb A$, the trace is denoted as $\mathrm{tr}(\tb A)$ and the deviatoric part as $\tb A_{\mathrm{dev}} := \tb A - (\mathrm{tr}(\tb A)/3) \tb I$ where $\tb I$ is the second order identity tensor. The symmetrization and skew-symmetrization operators are defined as $\sym{\tb A} := (\tb A + \tb A^T)/2$ and $\skw{\tb A} := (\tb A - \tb A^T)/2$, respectively. Throughout this work, bold or blackboard bold symbols denote spatial variables expressed in tensorial form.

\subsection{Governing equations and constitutive theory}
\label{subsec:gov_eq_cons}

The conservation of mass and linear momentum for a granular continuum body $\mathcal{B}$ is expressed as:
\begin{eqnarray}
    \dot{\rho} + \rho \nabla \cdot \dot{\tb u} = 0\,, \qquad
    \rho \ddot{\tb u} = \nabla \cdot \tb{\sigma} + \rho \tb{b}\,,
\label{eq:mass_momentum_balance}
\end{eqnarray}
where the bulk mass density is given by $\rho = \phi\, \rho_s$. Here, $\phi$ is the solid (packing) fraction and $\rho_s$ is the intrinsic density of the grains, which remains constant because the grains are assumed incompressible. Here, $\tb b$ is the body force per unit mass, and $\tb{\sigma}$ is the symmetric Cauchy stress tensor. The displacement field is denoted by $\tb u$, with its first and second material time derivatives corresponding to the velocity and acceleration fields, $\dot{\tb u} = \tb v$ and $\ddot{\tb u} = \dot{\tb v} = \tb a$, respectively. These balance equations must be solved together with an appropriate set of boundary conditions defining the boundary-value problem.

The spatial velocity gradient is defined as $\tb L = \nabla \tb v$, which can be decomposed into its symmetric and skew-symmetric components:
\begin{eqnarray}
    \tb D = \sym{\tb L}\,, \qquad \tb W = \skw{\tb L}\,, \qquad \tb L = \tb D + \tb W\,,
    \label{eq:kinematic_desc}
\end{eqnarray}
where $\tb D$ and $\tb W$ are referred to as the strain-rate and spin tensors, respectively. We consider the additive decomposition of $\tb L$, $\tb D$, and $\tb W$ into viscoelastic and plastic components (see Fig.~\ref{fig:1_rheology_diagram}), i.e.,
\begin{eqnarray}
    \tb L = \tb L^{ve} + \tb L^p\,, \qquad \tb D = \tb D^{ve} + \tb D^p\,, \qquad \tb W = \tb W^{ve} + \tb W^p\,.
    \label{eq:kinematic_additive_decomp}
\end{eqnarray}
Assuming the granular material is isotropic, the plastic deformation does not induce material rotation \cite{anand2005theory}; hence, the plastic spin tensor vanishes, i.e~$\tb W^p = \tb 0$.

Cohesionless granular media do not support tension. This behavior can be mathematically represented using a separation rule, as in \citet{dunatunga2015continuum}. When the bulk density falls below a critical value, $\rho_c = \phi_c\, \rho_s$, the material becomes stress-free, i.e.~$\tb{\sigma} = \tb{0}$, as it enters an open state where all grains are separated and lose contact. In this state, the material is free to expand or contract without the presence of stress, and its motion becomes purely ballistic. As the material reconsolidates and the bulk density exceeds $\rho_c$, inter-granular contacts are deemed to be reestablished, and the material transitions into a dense state where stress transmission resumes. In this regime, the stress is compressive, and the constitutive response can be described by inviscid (elastic) and viscous components. These viscoelastic components are further connected in series with a plastic element, which governs the rate-dependent inelastic flow behavior (cf.~Fig.~\ref{fig:1_rheology_diagram}; more on this in \Cref{subsec:inertia_plasticity}).

The total Cauchy stress can be defined and decomposed following the Kelvin–Voigt model (cf.~Fig.~\ref{fig:1_rheology_diagram}) as:
\begin{eqnarray}
    \tb \sigma = \tb \sigma^{ve} = \tb \sigma^{p}\,, \qquad  \tb \sigma  = \tb \sigma^{e} + \tb \sigma^v\,,
    \label{eq:stress_decomposition}
\end{eqnarray}
where $\tb \sigma^e$ and $\tb \sigma^v$ denote the elastic and viscous stresses, respectively. Considering the granular assembly to consist of stiff, non-breakable grains, a linear elastic model with constant bulk and shear moduli, $K$ and $G$, is assumed. To ensure objectivity, the Jaumann stress rate is considered:
\begin{eqnarray}
    \overset{\nabla}{\tb \sigma^e} \equiv \dot{\tb \sigma}^e - \tb{W} \tb \sigma^e + \tb \sigma^e \tb{W}=\mathbb{C}:\tb D^{ve}\,,
    \label{eq:jaumann_rate}
\end{eqnarray}
where the fourth-order elastic constitutive tensor is defined as:
\begin{eqnarray}
    \mathbb{C}= K \tb I \otimes \tb I + 2 G \left( \mathbb{I}-\frac{1}{3} \tb I \otimes \tb I \right)\,,
    \label{eq:elastic_tangent_matrix}
\end{eqnarray}
with $\tb I$ and $\mathbb{I}$ denoting the second- and fourth-order identity tensors, respectively. The numerical oscillations associated with the Jaumann rate can be minimized when elastic stretches remain small, i.e.~$\tb D^{ve} \approx \tb 0$, and hence $\tb D \approx \tb D^p$. These conditions are typically satisfied in dense granular flows composed of stiff grains.

The viscous stress is expressed as:
\begin{eqnarray}
    \tb \sigma^v= \vartheta \,\mathrm{tr}\left(\tb D^{ve}\right) \tb I+2\eta \,\tb D^{ve}_{\mathrm{dev}}\,,
    \label{eq:viscous_stress}
\end{eqnarray}
where $\vartheta$ and $\eta$ denote the bulk and shear viscosities, respectively. In conventional continuum formulations of granular media, these viscosities are often assigned empirically to damp elastic waves, with their values typically chosen in an ad hoc manner or calibrated through vibration experiments. A common approach employs Rayleigh-type stiffness-proportional damping \cite{rayleigh1896theory, caughey1960classical}, where a nondimensional coefficient ${\beta}$ is often adjusted to fit experimental data. In this study, one of our main objectives is to establish a physical basis for these parameters by relating $\vartheta$ and $\eta$ to the coefficient of restitution, $e$. This connection allows $e$, a measurable granular property, to be directly linked to the continuum viscosities. The derivation of this relationship is presented in the subsequent sections.

\subsection{Bulk viscosity and its relation with coefficient of restitution}
\label{subsec:bulk_visc}

\begin{figure}[h!]
    \centering
    \includegraphics[width=0.9\textwidth]{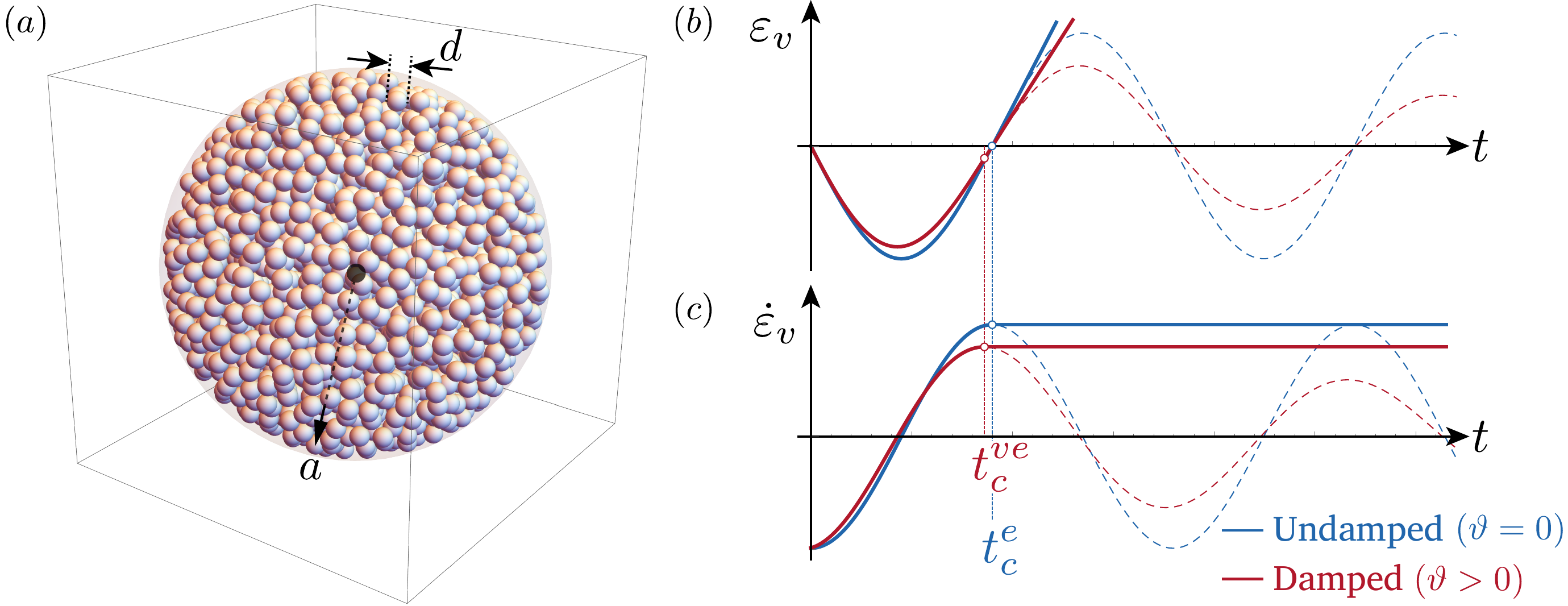}
    \caption{($a$) A schematic illustration of a spherical assembly with radius $a$ consisting of particles with grain diameter $d$. The spherical assembly deforms volumetrically under spherical compression impact, characterized by ($b$) volumetric strain, $\varepsilon_v$, and ($c$) volumetric strain rate, $\dot{\varepsilon}_v$, which evolve over time. As the granular assembly cannot carry any tension, the contact time (or release time) is considered to be the time at which the volumetric strain rate reaches its maximum value. This contact time is observed to differ for undamped and damped contact, i.e.~$t^e_c$ for $\vartheta=0$ and $t^{ve}_c$ for $\vartheta>0$, respectively. For a damped assembly, the grains may lose their contact, which leads to zero stress and further separation, even before reaching zero volumetric strain. Bold lines illustrate the evolution of strain and strain rate in the granular assembly, while dashed lines represent the typical oscillatory response of (visco)elastic solids, which permits tensile states.}
    \label{fig:2_spherical_assembly}
\end{figure}

Consider a spherical assembly composed of grains of diameter $d$. The assembly has a radius $a$ and an effective density of $\rho = \phi\,\rho_s$ (see Fig.~\ref{fig:2_spherical_assembly}$(a)$). No external surface or body forces act on the assembly. At $t = 0$~s, it is at rest with zero displacement everywhere, $\tb u(r,0) = \tb 0$, where $r$ denotes the radial coordinate in the spherical coordinate system. The assembly then undergoes a spherical compression impact, prescribed as an initial radial velocity $\tb v(r,0) = v_r(r,0)\,\tb r$, where $\tb r$ is the outward radial basis vector. This impact induces a volumetric deformation characterized by the volumetric strain $\varepsilon_v$ and its rate $\dot{\varepsilon}_v$ (see Fig.~\ref{fig:2_spherical_assembly}$(b)$ and $(c)$).

We define a continuum-level restitution measure, analogous to the definition proposed by Newton, i.e.~Eq.~\eqref{eq:newton_restitution}, denoted as $\mathcal{E}$. This quantity is defined as the ratio between the volumetric strain rate after and before a volumetric collisional event, leading to the following expression:
\begin{eqnarray}
    \mathcal{E}=\frac{-\dot{\overline{\varepsilon}}_v (t_c)}{\dot{\overline{\varepsilon}}_v (0)}\,.
    \label{eq:continuum_restitution}
\end{eqnarray}
Here, $\dot{\overline{\varepsilon}}_v$ is the averaged volumetric strain rate of the granular assembly, which can be obtained for a spherical assembly with purely radial deformation by invoking the Gauss divergence theorem as:
\begin{equation}
\begin{split}
        \dot{\overline{\varepsilon}}_v (t) =& \frac{1}{\Omega} \int_{\Omega} \nabla \cdot \tb v(t)\, \td \Omega \\
        =& \frac{1}{\Omega} \int_{\pd \Omega} \tb v(t)\cdot \tb n \, \td \Gamma \\
        =& v_r(a,t) \frac{\Gamma_\circ(t)}{\Omega_\circ(t)}\,,
\end{split}
\end{equation}
where $\Gamma_\circ$ and $\Omega_\circ$ are the circumferential area and volume of the spherical assembly. At the limit of low-speed collision \cite{stronge2018impact}, where the granular assembly is not significantly deformed by impact (i.e.~no inelastic grain deformation and particle breakage), $\Gamma_\circ/\Omega_\circ$ can be considered as constant throughout the collision. Thereby, upon simplification, we arrive at a simpler kinematic-based relation:
\begin{eqnarray}
    \mathcal{E}=\frac{-v_r (a,t_c)}{v_r (a, 0)}\,.
    \label{eq:continuum_restitution_a}
\end{eqnarray}
Note that $\mathcal{E}$ is a purely kinematic quantity rather than an intrinsic material constant, because its value can depend on the selected characteristic length scale, which in this case is chosen to be the radius $a$. In the subsequent steps, we intend to use $\mathcal{E}$ to determine an equivalent bulk viscosity, thereby relating this kinematic measure to a single continuum parameter.


The radial component of the equilibrium equation in spherical coordinates $\{r, \theta,\varphi\}$ can be expressed as:
\begin{eqnarray}
    \frac{\pd \sigma_{rr}}{\pd r} + \frac{2}{r} \left(\sigma_{rr}-\sigma_{\theta \theta} \right) = \rho \ddot{u}_r\,.
\end{eqnarray}
Upon substituting the constitutive and kinematic relations:
\begin{subequations}
    \begin{eqnarray}
    &\sigma_{rr}=\lambda \left(\varepsilon_{rr} + 2 \varepsilon_{\theta \theta}\right) + 2G \varepsilon_{rr} + \vartheta \left(\dot{\varepsilon}_{rr} + 2 \dot{\varepsilon}_{\theta \theta}\right)\,,\\
    &\sigma_{\theta \theta}=\lambda \left(\varepsilon_{rr} + 2 \varepsilon_{\theta \theta}\right) + 2G \varepsilon_{\theta \theta} + \vartheta \left(\dot{\varepsilon}_{rr} + 2 \dot{\varepsilon}_{\theta \theta}\right)\,,\\
    &\varepsilon_{rr}={\pd u_r}/{\pd r} = u_r'\,, \qquad \varepsilon_{\theta \theta} = \varepsilon_{\varphi \varphi} = {u_r}/{r}\,,
    \end{eqnarray}
\end{subequations}
the following expression can be obtained:
\begin{eqnarray}
    M \left( u''_r + \frac{2 u'_r}{r} - 2\frac{u_r}{r^2}\right) + \vartheta \left( \dot{u}''_r + \frac{2 \dot{u}'_r}{r} - 2\frac{\dot{u}_r}{r^2}\right)  = \rho \ddot{u}_r\,,
    \label{eq:spherical_equilibrium}
\end{eqnarray}
where $\lambda$ and $G$ are the two Lam\'e coefficients, while the compression modulus is denoted as $M=\lambda+2G$. 

We attempt to find a solution that satisfies Eq.~\eqref{eq:spherical_equilibrium} as well as the boundary and initial conditions, assuming the separation of variables, such that $u_r$ can be constructed as the product of separated spatial (or radial) and temporal terms, i.e.,
\begin{eqnarray}
    u_r(r,t)=R(r)\, T(t)\,.
    \label{eq:separation_variables}
\end{eqnarray}
Substituting Eq.~\eqref{eq:separation_variables} into \eqref{eq:spherical_equilibrium}, we arrive to the following expression: 
\begin{eqnarray}
    \frac{R''}{R} + \frac{2R'}{rR} - \frac{2}{r^2}=\frac{\rho \ddot{T}}{(M+\vartheta \frac{\dot{T}}{T})T}:=-k^2\,.
    \label{eq:space_time_ode}
\end{eqnarray}
Since the right and left-hand-side terms respectively depend only on $r$ (space) and $t$ (time), both sides should be equal to a constant; here we denote it as $-k^2$, where $k\in\mathcal{R}^+$.

We can first solve the \revision{left-hand-side} ODE given as:
\begin{eqnarray}
    R'' + \frac{2}{r}R'+\left(k^2 - \frac{2}{r^2} \right)R=0\,,
\end{eqnarray}
subjected to the spatial boundary conditions at the center and surface of the granular assembly, that is:
\begin{subequations}
\begin{eqnarray}
    &\mathrm{at}& \quad r=0\,, \qquad R(0)=0\,,\\
    &\mathrm{at}& \quad r=a\,, \qquad \lambda\left(R'(a)+2R(a)/a\right)+2G R'(a)=0\,.
    \label{eq:traction_free_spherical}
\end{eqnarray}
\end{subequations}
Here, Eq.~\eqref{eq:traction_free_spherical} refers to the traction-free BC at the free surface. The solution to such spatial systems can be obtained in the following form:
\begin{eqnarray}
    R(r)=A_1\, J_1\left( k r\right)\,,
\end{eqnarray}
where $A_1$ is a constant and $J_1$ denotes the spherical Bessel function of the first kind $J_\alpha$ with order $\alpha=1$.

Through the given boundary conditions, we find that the constant $k$, with dimension $[L^{-1}]$, should satisfy the following condition:
\begin{eqnarray}
    \frac{\lambda}{G} = \frac{4 J_1 (k a)}{\sin\left(k a \right)}-2\,.
\end{eqnarray}
Here, the value of $k$ is unknown while the ratio of the material moduli $\lambda/G$ and the assembly radius $a$ are known. To obtain a closed-form approximation for $k$, we consider the asymptotic limit of large $\lambda/G$, where, through Taylor expansion and perturbation analysis, we obtain:
\begin{eqnarray}
    k = \frac{\pi}{a} + O\left(\frac{1}{\lambda/G}\right)\,,
    \label{eq:zeroth_order_k}
\end{eqnarray}
such that, as $\lambda/G \rightarrow \infty$, the wavenumber approaches its leading-order value $k \rightarrow \pi/a$. This approximation is consistent with the constitutive assumptions of stiff grains outlined in \Cref{subsec:gov_eq_cons}, and will be adopted throughout the remainder of the derivation.

Next, the temporal \revision{right-hand-side} ODE (Eq.~\eqref{eq:space_time_ode}) should be solved, which can be rearranged as:
\begin{eqnarray}
    \ddot{T}+k^2 \frac{\vartheta}{\rho} \dot{T}+k^2 \frac{M}{\rho} {T}=0\,.
\end{eqnarray}
The above ODE is subjected to the previously mentioned initial condition, where the displacement field is initially zero everywhere,
\begin{eqnarray}
    \mathrm{at}& \quad t=0\,, \qquad T(0)=0\,.
\end{eqnarray}
This results in the following solution:
\begin{eqnarray}
    T(t)=A_2\, \exp\left({\frac{-\tilde{\xi}}{2}t}\right) \sin{\left(\tilde{\omega} t \right)}\,,
    \label{eq:time_component}
\end{eqnarray}
where $A_2$ is a constant, and $\tilde{\xi}$ and $\tilde{\omega}$ are parameters indicating the temporal oscillation decay rate and frequency, respectively. They read:
\begin{eqnarray}
    \tilde{\xi}=\frac{k^2 \vartheta}{\rho}\,, \qquad \tilde{\omega} = \frac{k}{2}\sqrt{4 c_p^2 - \frac{\tilde{\xi} \vartheta}{\rho}}\,.
    \label{eq:damping_constants}
\end{eqnarray}

We then arrive at the combined solution:
\begin{eqnarray}
    u_r (r,t) = A J_1\left( k r\right) \exp\left({\frac{-\tilde{\xi}}{2}t}\right) \sin{\left(\tilde{\omega} t \right)}\,,
    \label{eq:displacement_radial_solution}
\end{eqnarray}
where $A=A_1 A_2$ is a problem-specific constant denoting the spatiotemporal amplitude. In the following steps, we show that the coefficient $A$ will cancel out when deriving the continuum restitution coefficient and thus can be ignored for now.

The initial radial velocity at $r=a$ can be obtained by differentiating Eq.~\eqref{eq:displacement_radial_solution} with respect to time and setting $t=0$, i.e.,
\begin{eqnarray}
    v_r (a, 0) = A \, \tilde{\omega} \, J_1(k a)\,.
    \label{eq:velocity_radial_initial}
\end{eqnarray}
Next, we need to compute the contact time, $t_c$, which is the first instance when the volumetric stress goes to zero after a $\sim$half-cycle of oscillation. At this instant, the radial velocity $v_r$ also reaches its maximum value. Thus, to obtain $t_c$, we solve for $t$ at which the radial acceleration goes to zero: $a_r(r, t_c) = 0$. This leads to:
\begin{eqnarray}
    t_c = \frac{1}{\tilde{\omega}} \left(\pi - \tan^{-1}\left(\frac{4 \tilde{\xi} \tilde{\omega}}{4 \tilde{\omega}^2-\tilde{\xi}^2 }\right)
 \right)\,,
 \label{eq:contact_time}
\end{eqnarray}
and the corresponding radial velocity at $r=a$ is:
\begin{eqnarray}
    v_r (a, t_c) = - A \, \tilde{\omega} \, J_1(k a)\, \exp{\left(-\frac{\tilde{\xi} t_c}{2}\right)}\,.
    \label{eq:velocity_radial_tc}
\end{eqnarray}
Here, when $t > t_c$, the stress state would otherwise transition to volumetric tension, and hence, the resulting stress becomes zero, causing the spherical assembly to expand indefinitely without resistance (cf.~Fig.~\ref{fig:2_spherical_assembly}$(b)$ and $(c)$). 

Substituting Eqs.~\eqref{eq:velocity_radial_initial} and \eqref{eq:velocity_radial_tc} into \eqref{eq:continuum_restitution_a}, we arrive at the following expression of continuum restitution:
\begin{eqnarray}
    \mathcal{E} = \exp{\left( -\frac{\tilde{\xi}}{2 \tilde{\omega}} \left(\pi - \tan^{-1}\left(\frac{4 \tilde{\xi} \tilde{\omega}}{4 \tilde{\omega}^2-\tilde{\xi}^2 }\right)
 \right)\right)} \,.
 \label{eq:continuum_restitution_b}
\end{eqnarray}

Our subsequent objective is to derive a link between $\mathcal{E}$ and the bulk viscosity $\vartheta$, which has been incorporated within $\tilde{\xi}$ and $\tilde{\omega}$ through Eq.~\eqref{eq:damping_constants}. To move forward, we first introduce the non-dimensional form of viscosity defined as:
\begin{eqnarray}
    \Tilde{\vartheta} = \frac{\vartheta}{a \sqrt{M \rho}}\,.
    \label{eq:nondimensional_viscosity}
\end{eqnarray}
Then, by substituting $\vartheta= a \sqrt{M \rho} \, \Tilde{\vartheta}$ and $k$ (Eq.~\eqref{eq:zeroth_order_k}), together with Eq.~\eqref{eq:damping_constants}, into Eq.~\eqref{eq:continuum_restitution_b}, and fixing the branch cut associated with the $\tan^{-1}(\square)$ term, we obtain an expression for $\mathcal{E}$ that depends only on $\Tilde{\vartheta}$:
\begin{eqnarray}
    \mathcal{E} = \tilde{\mathcal{E}}\left(\tilde{\vartheta}\right) = \exp\left(-\frac{\Tilde{\vartheta} \pi \left(\left(1 - \, H\left(\Tilde{\vartheta} - \frac{\sqrt{2}}{\pi}\right)\right)\pi + \tan^{-1}\left(\frac{\Tilde{\vartheta} \pi \sqrt{4 - \Tilde{\vartheta}^2 \pi^2}}{\Tilde{\vartheta}^2 \pi^2 -2 }\right)\right)}{\sqrt{4 - \Tilde{\vartheta}^2 \pi^2}}\right)\,.
    \label{eq:continuum_restitution_d}
\end{eqnarray}
Here, $H(x)$ denotes the Heaviside function, defined as $H(x)=1$ for $x>0$ and $H(x)=0$ for $x \leq 0$.

\begin{figure}[h!]
    \centering
    \includegraphics[width=0.5\textwidth]{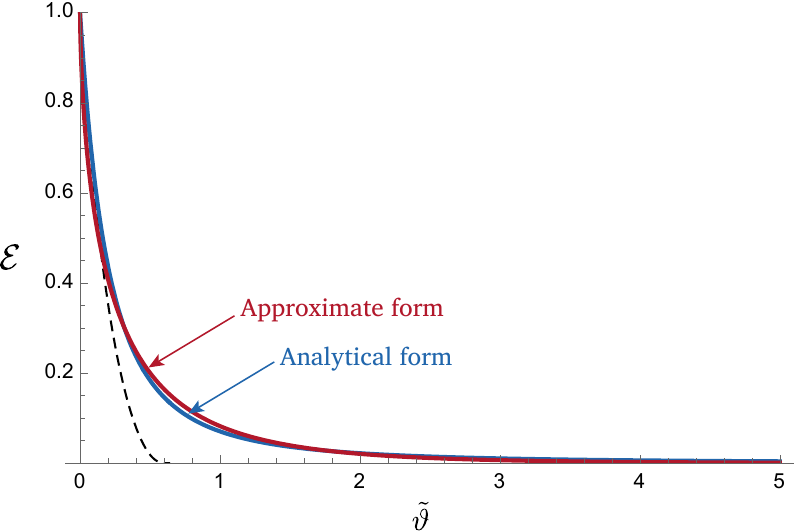}
    \caption{Comparison between the nonlinear profiles of continuum coefficient of restitution $\mathcal{E}$ depending on $\Tilde{\vartheta}$. The analytical form (Eq.~\eqref{eq:continuum_restitution_d}) is plotted in blue, while the approximate form (Eq.~\eqref{eq:continuum_restitution_e}) is plotted in red line. The dashed black line plots the form given in Eq.~\eqref{eq:continuum_restitution_f}, which was derived considering the limit of high $\lambda/G$ ratio and small viscosity $\Tilde{\vartheta}$ (cf.~\Cref{rem:limit_line}).}
    \label{fig:2_restitution_2D}
\end{figure}

The function given in Eq.~\eqref{eq:continuum_restitution_d} is relatively simple in form, but it is still highly nonlinear. This nonlinearity prevents an explicit inversion of $\Tilde{\vartheta}$ as a function of $\mathcal{E}$. Nonetheless, the exponential trend predicted by Eq.~\eqref{eq:continuum_restitution_d} (see Fig.~\ref{fig:2_restitution_2D}) motivates a simplified exponential representation of the relationship between $\mathcal{E}$ and $\Tilde{\vartheta}$. For instance, the expression
\begin{eqnarray}
    \mathcal{E} \approx \exp\left(-\frac{5}{2}\, \Tilde{\vartheta}^{\frac{2}{\pi}}\right)\,,
    \label{eq:continuum_restitution_e}
\end{eqnarray}
provides a reasonable approximation of Eq.~\eqref{eq:continuum_restitution_d}. Both expressions exhibit exponential behavior: when $\Tilde{\vartheta} = 0$, the coefficient of restitution is $\mathcal{E} = 1$, corresponding to a fully elastic collision; whereas as $\mathcal{E} \to 0$, $\Tilde{\vartheta} \to \infty$, indicating a plastic collision.

By inverting the expression in Eq.~\eqref{eq:continuum_restitution_e}, we obtain an approximate relation for the non-dimensional viscosity in terms of the continuum restitution:
\begin{eqnarray}
    \Tilde{\vartheta} = 0.237 | \ln{(\mathcal{E})}|^{\frac{\pi}{2}}\,.
    \label{eq:tilde_theta_approx}
\end{eqnarray}
The bulk viscosity $\vartheta$ can then be estimated from the non-dimensional expression above using Eq.~\eqref{eq:nondimensional_viscosity}. In the limit of particle-to-particle contact, such as in a kissing-number configuration, setting $a \rightarrow d$ gives
\begin{eqnarray}
    \vartheta = 0.237 d \sqrt{M \phi \rho_s} \,| \ln{(e)}|^{\frac{\pi}{2}}\,.
    \label{eq:bulk_viscosity}
\end{eqnarray}
This expression provides a direct means to convert the experimentally measurable coefficient of restitution, $e$, for grains of size $d$, into the corresponding continuum bulk viscosity, $\vartheta$.

\begin{remark}
\label{rem:limit_line}
    At the limit of high restitution, $\mathcal{E}\to1$, the obtained analytical and approximate solutions coincide with the following expressions:
\begin{eqnarray}
    \mathcal{E}= \exp{\left(-\frac{\pi ^2 \Tilde{\vartheta}}{\sqrt{2-\pi ^2 \Tilde{\vartheta}^2}}\right)} \qquad \Leftrightarrow \qquad
    \Tilde{\vartheta} = \frac{2 \,|\ln{(\mathcal{E})}|}{\pi\sqrt{\pi^2+\ln{(\mathcal{E})}^2}}\,,
    \label{eq:continuum_restitution_f}
\end{eqnarray}
    which are obtained by defining the contact time $t_c$ as half the period of volumetric vibration. Here, $t_c$ denotes the time at which $\varepsilon_v = 0$, i.e.~$t_c = \pi / \tilde{\omega}$, corresponding to $t^e_c$ depicted in Fig.~\ref{fig:2_spherical_assembly}$(b)$ and $(c)$. It is worth noting that the expression for the viscosity $\tilde{\vartheta}$ in Eq.~\eqref{eq:continuum_restitution_f}$_2$ is analogous to the normalized damping coefficient $c_n/\sqrt{k_n \Tilde{m}}$ in Eq.~\eqref{eq:normal_damping_coefficient_DEM}, derived for the damping term in a SDOF oscillator with a Kelvin–Voigt spring-dashpot arrangement \cite{butcher2000characterizing, schwager2008coefficient}. 
    
    While this model provides a good approximation of the relationship between $\mathcal{E}$ and $\Tilde{\vartheta}$ in the high-restitution limit, it deviates from the analytical solution at low restitution. For a given value of bulk viscosity, Eq.~\eqref{eq:continuum_restitution_f} underestimates the coefficient of restitution. Meanwhile, in the limit $\mathcal{E} \to 0$, the model yields $\Tilde{\vartheta} \to 2 / \pi \approx 0.637$, implying a finite maximum viscosity. This is inconsistent with the physical expectation that a fully plastic collision requires complete dissipation of elastic energy, corresponding to $\Tilde{\vartheta} \to \infty$.
\end{remark}

\subsection{Shear viscosity}

For granular materials subjected to small-amplitude vibrations, \citet{henann2013small} demonstrated that Rayleigh-type stiffness-proportional damping parameters, $\beta$, produce continuum simulations in excellent agreement with experimental data when predicting the displacement modes of tungsten grains across different vibration frequencies. In their study, $\beta$ was treated as a constant calibrated from a series of experiments. Now that we have a closed-form expression for $\vartheta$ through Eq.~\eqref{eq:bulk_viscosity}, we follow the same rationale and assume that, given the bulk viscosity, $\vartheta$, and the linear elastic moduli, $K$ and $G$, the corresponding shear viscosity can be evaluated as:
\begin{eqnarray}
\beta=\frac{\vartheta}{K}=\frac{\eta}{G} \qquad \Rightarrow \qquad
\eta = \frac{G}{K} \,\vartheta\,.
\label{eq:shear_viscosity_rayleigh}
\end{eqnarray}

For typical granular materials with Poisson’s ratio $0.15 \lesssim \nu \lesssim 0.3$, the elastic shear modulus, $G$, is often smaller than the bulk modulus, $K$. Consequently, when Rayleigh stiffness-proportional damping is considered, i.e.~as expressed in Eq.~\eqref{eq:shear_viscosity_rayleigh}, the resulting shear viscosity, $\eta$, is typically smaller than the bulk viscosity, $\vartheta$. Nevertheless, this relationship is not general. Experimental observations from wave-propagation tests and relatively large-amplitude vibration (often at low confinement) on granular media suggest that dissipation can be dominated by shear mechanisms, in which case $\eta$ may be comparable to or exceed $\vartheta$; e.g.~see \cite{morozov2015relation}. From a thermodynamic perspective, the only requirement imposed on $\vartheta$ and $\eta$ is that they remain non-negative to ensure positive dissipation \cite{gurtin2010mechanics}. However, no further constraint exists on their relative magnitudes.







\subsection{Inertia-based frictional viscoplasticity}
\label{subsec:inertia_plasticity}

Following the concept of $\mu(I)$ rheology, as introduced by \citet{da2005rheophysics} and \citet{jop2006constitutive}, the flow of dense granular material can be modeled as a viscoplastic solid with a rate-dependent frictional yield criterion. The yield surface follows the conical form of the Drucker–Prager model, expressed as
\begin{eqnarray}
    \tau=\mu(I)\,p\,,
    \label{eq:drucker_prager_yield}
\end{eqnarray}
where $\tau = \sqrt{\left(\tb \sigma_{\mathrm{dev}} : \tb \sigma_{\mathrm{dev}}\right)/2}$ denotes the equivalent deviatoric stress, and $p = -(1/3)\mathrm{tr}(\tb \sigma)$ represents the mean pressure, taken as positive in compression. The friction coefficient is defined according to the $\mu(I)$ rheology as
\begin{eqnarray}
    \mu(I)=\mu_s+\mu_I(I)=\mu_s + \frac{\mu_2-\mu_s}{1+I_0/I}\,,\qquad \mathrm{where} \qquad I = \dot{\gamma}^p \sqrt{\frac{d^2 \rho_s}{p}}\,,
    \label{eq:mu_i}
\end{eqnarray}
and $I_0$ is a material constant. Here, $\mu_s$ and $\mu_2$ denote the friction coefficients at zero and high shear rates, respectively, whereas $\dot{\gamma}^p$ denotes the equivalent plastic shear rate. The parameters appearing in the $\mu(I)$ relation above can be calibrated using DEM simulations or experimental measurements.

During plastic flow, the ratio of shear stress to pressure is not constant but depends on the flow rate through the inertial number. This ensures that the stress state remains on the rate-dependent yield surface defined by the $\mu(I)$ model, which may be viewed as a rate-dependent generalization of \textit{Prager’s consistency condition}. Together with the no-tension condition discussed in \Cref{subsec:gov_eq_cons}, the admissible stress states can be expressed using the following Karush–Kuhn–Tucker (KKT) conditions:
\begin{gather}
    \dot{\gamma}^p (\tau-\mu\,p)=0\,, \quad \dot{\gamma}^p\geq 0\,, \quad \tau\leq \mu\,p\,,\label{eq:yield_cond}\\
    \left(\rho-\rho_c\right) p = 0\,, \quad p\geq 0\,, \quad \rho\leq \rho_c\,. \label{eq:separation_cond}
\end{gather}
Note that, strictly speaking, the separation condition given in Eq.~\eqref{eq:separation_cond} actually corresponds to a rigid-plastic formulation, in which $p$ plays the role of a Lagrange multiplier. For the present viscoelastic model, it is more suitable to express the separation condition as
\begin{eqnarray}
    p = \begin{cases}
        0\,, & \text{if}\quad \rho\leq\rho_c\,,\\
        K \ddfrac{(\rho-\rho_c)}{\rho} + \vartheta \ddfrac{\dot{\rho}}{\rho}\,,& \text{if}\quad \rho>\rho_c\,.
    \end{cases}
    \label{eq:pressure_separation_cond}
\end{eqnarray}

The admissible stress states are obtained through a proper choice of the plastic flow rule defining $\tb D^p$. The plastic flow rule adopted in this study follows the isochoric plasticity model:
\begin{eqnarray}
    {\tb D}^p = \dot{\gamma}^p  \frac{\tb \sigma_{\mathrm{dev}}}{2\tau} \,.
    \label{eq:plastic_flow_rule}
\end{eqnarray}
where the invariants of the plastic strain rate are given by
\begin{eqnarray}
    \mathrm{tr}(\tb D^p)=0\,, \qquad \dot{\gamma}^p=\sqrt{2}||\tb D^{p}_{\mathrm{dev}}||\,.
    \label{eq:flow_rule_2}
\end{eqnarray}

In the present work, the dilative behavior of granular media under plastic flow is not considered. Such behavior can be modeled using the concept of Reynolds dilation \cite{reynolds1885lvii} or the rate-dependent $\phi(I)$ rheology, which relates the solid volume fraction to the inertial number \cite{jop2006constitutive}. It is worth emphasizing that these dilatancy effects play an important role in characterizing flow initiation during the solid-to-liquid transition, particularly due to sample preparation, where shear banding may emerge in densely packed specimens. However, as the present study focuses on characterizing material flow at the critical state, these effects are omitted for simplicity. Furthermore, the coefficient of restitution is assumed to have a negligible influence on the overall trend of the friction coefficient; any residual dependence is effectively absorbed into the calibration of the $\mu(I)$ parameters, consistent with observations for frictional grains \cite{gdr2004dense, da2005rheophysics}. A summary of the governing equations and constitutive relations derived in this section is provided in \Cref{tab:constitutive_eqs}.

\begin{table}[h!]
\centering
\caption{Summary of governing and constitutive equations derived in \Cref{sec:continuum}.}
\label{tab:constitutive_eqs}
\small
\begin{tabular}{||l c c c c||}
\hline
Rule  && Expressions && Equation \\ \hline \hline
Mass conservation   &&   $\dot{\rho} +\rho \nabla \cdot {\tb v}=0$    & & \eqref{eq:mass_momentum_balance}$_1$                    \\
Linear momentum balance && $\rho {\tb a} = \nabla \cdot \tb{\sigma} + \rho \tb{b}$ && \eqref{eq:mass_momentum_balance}$_2$                    \\
Kinematic description && $\tb L=\nabla \tb v=\tb D + \tb W\,,\quad \tb D=\sym{\tb L}\,,\quad \tb W=\skw{\tb L}$ && \eqref{eq:kinematic_desc}                    \\
Additive kinematic decomposition && $\tb L = \tb L^{ve} + \tb L^p\,, \quad \tb D = \tb D^{ve} + \tb D^p\,, \quad \tb W = \tb W^{ve}$&& \eqref{eq:kinematic_additive_decomp}                    \\
Stress equivalence and decomposition$^\dagger$ && $\tb \sigma = \tb \sigma^{ve} = \tb \sigma^p\,,\quad \tb \sigma = \tb \sigma^{e} + \tb \sigma^v$ && \eqref{eq:stress_decomposition}\\
Elastic stress rate$^\dagger$ && $\dot{\tb \sigma}^e = \mathbb{C}:\tb D^{ve} + \tb{W} \tb \sigma^e - \tb \sigma^e \tb{W}$ && \eqref{eq:jaumann_rate}\\
Elastic constitutive tensor && $\mathbb{C}= K \tb I \otimes \tb I + 2 G \left( \mathbb{I}-\frac{1}{3} \tb I \otimes \tb I \right)$ && \eqref{eq:elastic_tangent_matrix}\\
Viscous stress$^\dagger$ && $\tb \sigma^v= \vartheta \,\mathrm{tr}\left(\tb D^{ve}\right) \tb I+2\eta \,\tb D^{ve}_{\mathrm{dev}}$ && \eqref{eq:viscous_stress}\\ 
Bulk viscosity$^\dagger$ && $\vartheta = 0.237 d \sqrt{M \rho} \,| \ln{(e)}|^{\frac{\pi}{2}}$ && \eqref{eq:bulk_viscosity}\\ 
Shear viscosity$^\dagger$ && $\eta=(G/K)\vartheta$ && \eqref{eq:shear_viscosity_rayleigh}$_2$ \\ 
Frictional yield condition && $ \dot{\gamma}^p (\tau-\mu\,p)=0\,, \quad \dot{\gamma}^p\geq 0\,, \quad \tau\leq \mu\,p $ && \eqref{eq:yield_cond} \\  
Internal friction coefficient && $ \mu = \mu_s + {(\mu_2-\mu_s)}/{(1+I_0/I)} $ && \eqref{eq:mu_i}$_1$\\  
Inertia number && $ I = \dot{\gamma}^p \sqrt{{(d^2 \rho_s)}/{p}}$ && \eqref{eq:mu_i}$_2$\\ 
Plastic flow rule && $\tb D^p= \dot{\gamma}^p \tb \sigma_{\text{dev}}/(2\tau)$ && \eqref{eq:plastic_flow_rule}\\ 
Separation rule$^\dagger$ && $    p = \begin{cases}
        0\,, & \text{if}\quad \rho\leq\rho_c\\
        K \ddfrac{(\rho-\rho_c)}{\rho} + \vartheta \ddfrac{\dot{\rho}}{\rho}\,,& \text{if}\quad \rho>\rho_c
    \end{cases}$ && \eqref{eq:pressure_separation_cond} \\ \hline

\end{tabular}
\begin{tablenotes}
\footnotesize
\raggedright
\item[$\dagger$] \revision{Newly added or modified from  \citet{dunatunga2015continuum}.}
\end{tablenotes}
\end{table}

\section{Implementation}
\label{sec:modeling}

One of the objectives of this work is to propose a stress-update algorithm that can be incorporated into a large-deformation continuum-based solver. The developed algorithm should account for viscoelastic damping and should not modify the plastic flow at dense, fluid-like state, which is determined solely by inertia-based plasticity following the $\mu(I)$ rheology. This section elaborates on how this algorithm is formulated and implemented.

\subsection{Material point discretization}

\begin{figure}[h!]
    \centering
    \includegraphics[width=0.92\linewidth]{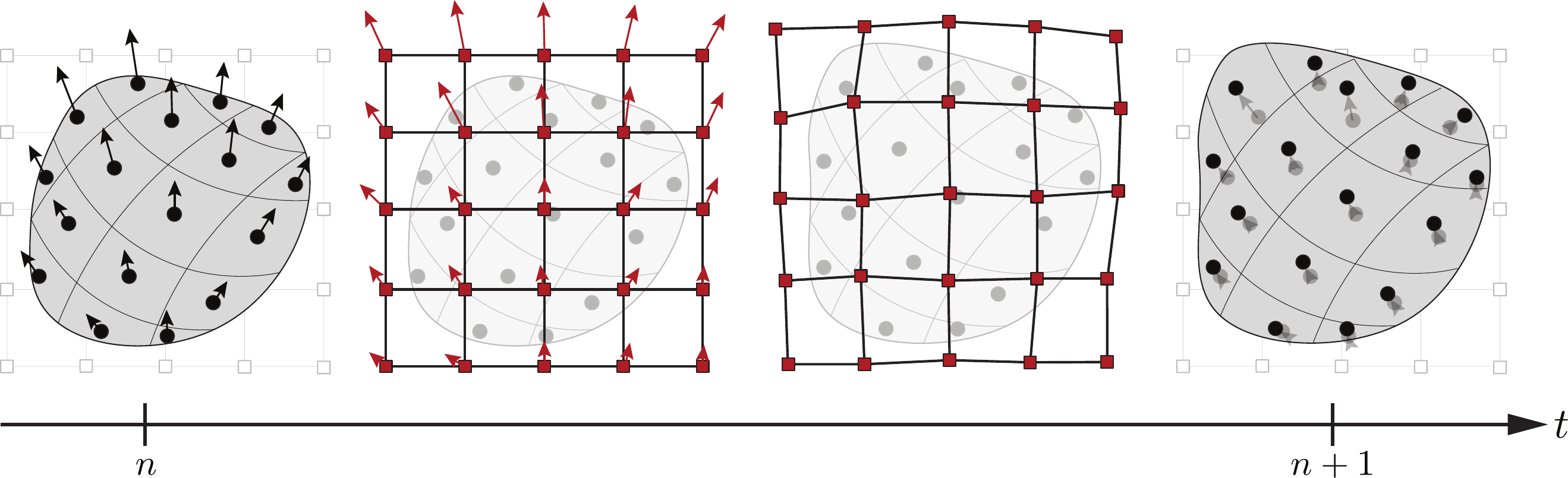}
    \caption{Schematic illustration of the MPM computational cycle. The continuum body is discretized into Lagrangian material points that store all history-dependent variables. At the beginning of each time step, kinematic quantities are mapped from the material points to the background grid, where the mass and momentum equations are solved at the grid nodes. The updated deformation and kinematic fields are then transferred back to the material points, whose positions are subsequently updated. Because the grid carries no history information, it can be reset after each step, enabling the method to accommodate large deformations without mesh distortion issues.}
    \label{fig:3_mpm}
\end{figure}

In this study, the material point method (MPM) \citep{Sulsky1994} is employed as the numerical framework. MPM is a continuum-based approach originating from the Fluid–Implicit–Particle (FLIP) method \citep{brackbill1986flip}, developed to simulate the behavior of materials undergoing large deformations. In this method, the continuum body $\mathcal{B}$ is represented by a collection of Lagrangian material points that carry all history-dependent state variables, while a background Eulerian grid is utilized to compute the equations of motion and evaluate spatial gradients. Since all physical quantities are stored at the material points, the background grid can be reset at each time step, enabling the method to capture large deformations without suffering from mesh distortion errors as experienced by typical mesh-based methods, like the finite element method (FEM). A schematic illustration of the MPM procedure is presented in Fig.~\ref{fig:3_mpm}, where the implementation details of the MPM used in this study can be found in \ref{app:mpm}. 

\subsection{Stress-update algorithm}

The stress-update algorithm adopted in this study follows the approach of \citet{dunatunga2015continuum}, with several modifications to account for the viscoelastic constitutive formulation described in \Cref{subsec:gov_eq_cons}. The stress integration algorithm can be summarized into the following expression:
\begin{equation}
\begin{split}
    \tb \sigma^{n+1}= \tb \sigma^{e,n}  + \Delta t \Biggl(\mathbb{C}:\left(\tb D^{n+1} - \tb D^{p,n+1} \right)  + \tb W^{n+1} \tb \sigma^{e,n} - \tb \sigma^{e,n} \tb W^{n+1}\Biggr) \\
    + \vartheta\, \mathrm{tr}\left(\tb D^{n+1}- \tb D^{p,n+1}\right) \tb I + 2 \eta \left( \tb D^{n+1}_{\mathrm{dev}} - \tb D^{p,n+1}_{\mathrm{dev}} \right)\,,
\end{split}
\end{equation}
where variables with superscript $(n+1)$ denote the unknown quantities to be evaluated at each time step.

The stress update follows a predictor–corrector scheme commonly used in computational inelasticity. First, the trial stress $\tb{\sigma}^{tr}$ is computed by assuming the total strain rate $\tb D^{n+1}$ is fully elastic. We adopt the rotationally neutralized objective stress-update scheme proposed by \citet{simo2006computational}. Here, the rotational rate is assumed to influence only the elastic stress update, which implies that the viscous component remains invariant under superposed rigid-body motion. The trial stress is therefore written as:
\begin{eqnarray}
    \tb \sigma^{tr} = \hat{\Sigma}\left(\tb \sigma^{e,n}, \tb F^n,\Delta \tb F^{n+1}, \mathbb{C} \right) + \vartheta\, \mathrm{tr}\left(\tb D^{n+1}\right) \tb I + 2 \eta \tb D^{n+1}_{\mathrm{dev}}\,,
\end{eqnarray}
where $\hat{\Sigma}$ denotes the rotationally neutralized elastic trial operator. For brevity, the detailed formulation of this operator is not rewritten here; interested readers may refer to \citet{simo2006computational}. The operator takes as input the previous elastic stress $\tb \sigma^{e,n}$, the deformation gradient from the previous step $\tb F^n$, the incremental deformation gradient $\Delta \tb F^{n+1}$, and the elastic constitutive tensor $\mathbb{C}$. The deformation gradient, strain-rate tensor, and spin tensor appearing in the above expressions are evaluated once $\tb L^{n+1}$ is obtained from the kinematic relations (in MPM, this is computed through Eq.~\eqref{eq:velocity_gradient}):
\begin{eqnarray}
    \Delta \tb F^{n+1} = \tb I + \Delta t \tb L^{n+1}\,, \qquad \tb F^{n+1}=\Delta \tb F^{n+1} \tb F^n\,,\qquad
    \tb D^{n+1}= \sym{\tb L^{n+1}}\,,\qquad
    \tb W^{n+1}= \skw{\tb L^{n+1}}\,.
\end{eqnarray}

From the trial stress, the corresponding stress invariants $p^{tr}$ and $\tau^{tr}$ can be computed. If the trial pressure is negative, the density has dropped below $\rho_c$, and the updated stresses should be set to zero, $\tb \sigma^{n+1} = \tb \sigma^{e,n+1} = \tb 0$, since the material is about to separate. If $p^{tr} > 0$, the material may be either in an elastic state or in a plastically flowing state. To distinguish these cases, we check whether the stress lies within the static yield surface, which corresponds to $\dot{\gamma}^{p,n+1} = 0$. Following the notation of \citet{dunatunga2015continuum}, we define
\begin{eqnarray}
    S_0 = \mu_s p^{tr}\,,
\end{eqnarray}
where, if $\tau^{tr} < S_0$, the material is considered to be in the elastic state, and the trial stress is accepted as an admissible stress, i.e.~$\tb \sigma^{n+1} = \tb \sigma^{tr}$.

If $\tau^{tr} > S_0$, the granular material is in its flowing state, corresponding to $\dot{\gamma}^{p,n+1} > 0$. In this case, the stress state must be mapped back onto the yield surface according to the flow rule given in Eq.~\eqref{eq:plastic_flow_rule}. At the same time, since $\dot{\gamma}^p > 0$ and consequently $I > 0$, the yield surface hardens following the $\mu(I)$ relation defined in Eq.~\eqref{eq:mu_i}$_1$. The main objective of the return-mapping algorithm is to determine the plastic shear flow rate $\dot{\gamma}^p$ that is consistent with the value of $\mu(I)$ defining the yield surface. At the end of the return-mapping process, the updated stress can be written as
\begin{eqnarray}
    \tb \sigma^{n+1} = \tb \sigma^{tr} - \Delta t \,\mathbb{C}:\tb D^{p, n+1}-\vartheta \, \mathrm{tr}\left( \tb D^{p, n+1} \right)-2\eta \tb D^{p, n+1}_{\mathrm{dev}}\,.
\label{eq:return_mapping_algebraic}
\end{eqnarray}
Because the plastic flow rule is isochoric, the third term on the right-hand side vanishes (see Eq.~\eqref{eq:flow_rule_2}$_1$). Furthermore, the deviatoric stress and the plastic strain rate are co-directional, i.e.,
\begin{eqnarray}
    \tb N =\frac{\tb \sigma^{tr}_{\mathrm{dev}}}{||\tb \sigma^{tr}_{\mathrm{dev}}||}=\frac{\tb \sigma^{n+1}_{\mathrm{dev}}}{||\tb \sigma^{n+1}_{\mathrm{dev}}||}=\frac{\tb D^{p}_{\mathrm{dev}}}{||\tb D^{p}_{\mathrm{dev}}||}\,.
    \label{eq:deviatoric_director}
\end{eqnarray}
This allows the return mapping to be performed directly in the stress-invariant space. The stress update reduces to updating the volumetric and deviatoric stress invariants:
\begin{eqnarray}
    p^{n+1} = p^{tr}\,, \qquad \tau^{n+1} = \tau^{tr} - \left(G \Delta t + \eta \right) \dot{\gamma}^{p,n+1}\,.
\end{eqnarray}

Following the implicit radial return algorithm proposed by \citet{dunatunga2015continuum}, the updated deviatoric stress can be computed as
\begin{eqnarray}
    \tau^{n+1} = \frac{2H}{B+\sqrt{B^2-4H}}\,,
\end{eqnarray}
where the relevant variables are defined as:
\begin{equation}
\begin{gathered}
    B = S_2 + \tau^{tr} + \alpha\,, \qquad 
    H = S_2 \tau^{tr} + S_0 \alpha\,,  \\
    S_2 = \mu_2 p^{tr}\,, \qquad 
    \alpha = \xi \widetilde{G} \Delta t \sqrt{p^{tr}}\,, \qquad  \xi=I_0/\sqrt{d^2 \rho_s}\,.
    \label{eq:mu_i_params}
\end{gathered}
\end{equation}
Here, the only modified parameter is $\widetilde{G} = G + \eta/\Delta t$, which incorporates the contribution of the shear viscosity, $\eta$. The equivalent plastic shear strain rate and the corresponding plastic strain rate tensor can be updated subsequently as:
\begin{eqnarray}
    \dot{\gamma}^{p,n+1}=\frac{\tau^{tr}-\tau^{n+1}}{\widetilde{G}\Delta t}\,, \qquad \tb D^{p,n+1}=\tb D^{p,n+1}_{\mathrm{dev}}=\frac{\dot{\gamma}^{p,n+1}}{\sqrt{2}} \tb N\,.
\end{eqnarray}

Knowing the updated stress invariants, the stress tensor at the end of the step can be assembled as:
\begin{eqnarray}
    \tb \sigma^{n+1} = \frac{\tau^{n+1}}{\tau^{tr}} \tb \sigma_{\mathrm{dev}}^{tr} - p^{n+1} \tb I\,.
\end{eqnarray}
Finally, the elastic stress must be updated and stored for use in the next stress update:
\begin{eqnarray}
    \tb \sigma^{e,n+1} = \tb \sigma^{n+1} - \vartheta \, \mathrm{tr}\left(\tb D^{n+1}\right) \tb I - 2 \eta \left(\tb D^{n+1}_{\mathrm{dev}} -  \frac{\dot{\gamma}^{p,n+1}}{\sqrt{2}} \tb N\right)\,.
    \label{eq:stress_update}
\end{eqnarray}

Following the proposed algorithm, we can correctly resolve the three-state constitutive relation, which transitions between the viscoelastic solid-like state, the liquid-like flowing state, and the separated gas-like state. The framework also ensures that the rate-dependent behavior is treated consistently within each regime. Viscous damping is applied only to the elastic component of the stress and does not dampen or modify the plastic flow. This is physically reasonable because the bulk and shear viscosities are related to the coefficient of restitution, $e$, which governs the energy loss during intergranular elastic collisions. In contrast, the micro-inertia influences only the flowing state, where it modifies the apparent friction through the inertial rheology $\mu(I)$, allowing the model to capture the rate-dependent hardening observed in dense granular flows.


\section{Numerical examples}
\label{sec:num_ex}

Five numerical examples are presented to highlight the capability and assess the accuracy of the proposed formulation. \revision{The purpose of each example, together with the specific limitations of previous models that it addresses, is summarized in \Cref{tab:test_cases_summary}.} The computations employ structured quadrilateral (in 2D) or hexahedral (in 3D) background grids. The basis functions used for spatial interpolation are problem dependent, and we use either linear basis functions or a quadratic B-spline basis with weighted-least-square kernel correction near the domain boundary as suggested by \citet{nakamura2023taylor}. Because computed material-point stress fields in MPM often contain high-frequency noise, we adopt the post-processing strategy similar to \citet{dunatunga2015continuum}, in which stresses are first mapped from material points to grid nodes and then projected back to the material points for visualization purposes. This mapped stress field is used solely for post-processing and does not replace the actual stress values used in the computation. The procedure is applied only to stress measures, such as pressure and the deviatoric stress, and not to kinematic quantities such as velocity and strain rate.

\begin{table}[h!]
\centering
\caption{\revision{Summary of the conducted numerical examples in \Cref{sec:num_ex}.}}
\label{tab:test_cases_summary}
\small
\begin{tabular}{||p{0.08\textwidth} p{0.42\textwidth} p{0.42\textwidth}||}
\hline
\revision{\textbf{Section}} & \revision{\textbf{Purpose}} & \revision{\textbf{Limitation of previous models}} \\
\hline
\hline
\ref{subsec:spherical_compaction} &
\revision{Verify the analytical solution for a spherical granular assembly undergoing spherical compaction for different values of the bulk viscosity $\vartheta$.} &
\revision{The previous model recovers only the purely elastic solution, corresponding to $\vartheta = 0$.} \\
\hline

\ref{subsec:bagnold} &
\revision{Verify that the Bagnold solution for granular chute flow can be reproduced across various values of restitution.} &
\revision{Improper viscoelastic damping often damps the plastic flow response, preventing recovery of the Bagnold velocity profile.} \\
\hline

\ref{subsec:silo} &
\revision{Investigate how variations in restitution influence the flow characteristics in a flat-bottom silo and the corresponding angle of repose.} &
\revision{Previous models often use ad-hoc damping to stabilize repose response, but the resulting numerical diffusion can hinder accurate resolution of energetic granular dynamics.} \\
\hline

\ref{subsec:impactor} &
\revision{Examine how restitution attenuates elastic waves and impact-induced oscillations, while keeping cratering and splashing behavior unchanged.} &
\revision{Previous models fail to adequately attenuate stress oscillations, leading to unrealistic persistent stress fluctuations. When damping is introduced, it often suppresses splashing behavior, reduces cratering depth, and decreases the amount of material ejected.} \\
\hline

\ref{subsec:pattern} &
\revision{Validate the nonlinear interaction between friction and restitution, which is required to reproduce diamond or square pattern formations in vibrating granular systems.} &
\revision{In the absence of friction, previous continuum models can only generate stripe patterns, whereas without restitution, no pattern forms at all.} \\
\hline
\end{tabular}
\end{table}

\subsection{Compaction of a spherical granular assembly}
\label{subsec:spherical_compaction}

Numerical simulations of the spherical compaction test described in \Cref{subsec:bulk_visc} are performed using MPM to verify the derived analytical solution. Here, a spherical granular assembly is constructed with radius $a=0.5$ m. The grains are considered to be stiff, with Lam\'e's modulus ratio set to a considerably large value, i.e.~$\lambda/G=10^7$, to model the limit considered in the derivation. The solid density and initial packing fraction are set to be $\rho_s=2500$ kg/m$^3$ and $\phi_0=\pi/6$, respectively. The critical density is, therefore, can be set as $\rho_c = \phi_0 \rho_s =1309$ kg/m$^3$, such that upon spherical expansion ($\rho<\rho_c$), the stress-free condition is immediately reached as the assembly enters the separation state\footnote{Since the \textit{update-stress-first} (USF) scheme \cite{zhang2016material} is adopted in the current MPM formulation, we set $\rho_c$ to be slightly lower than the initial density so that compressive elastic stress can be developed immediately at the first simulation step.}. To quantify the relationship between the bulk viscosity and the measured continuum restitution, $\vartheta \geq 0$ is varied. Meanwhile, the shear viscosity is set to zero, $\eta=0$.

\begin{figure}[h!]
    \centering
    \includegraphics[width=0.9\textwidth]{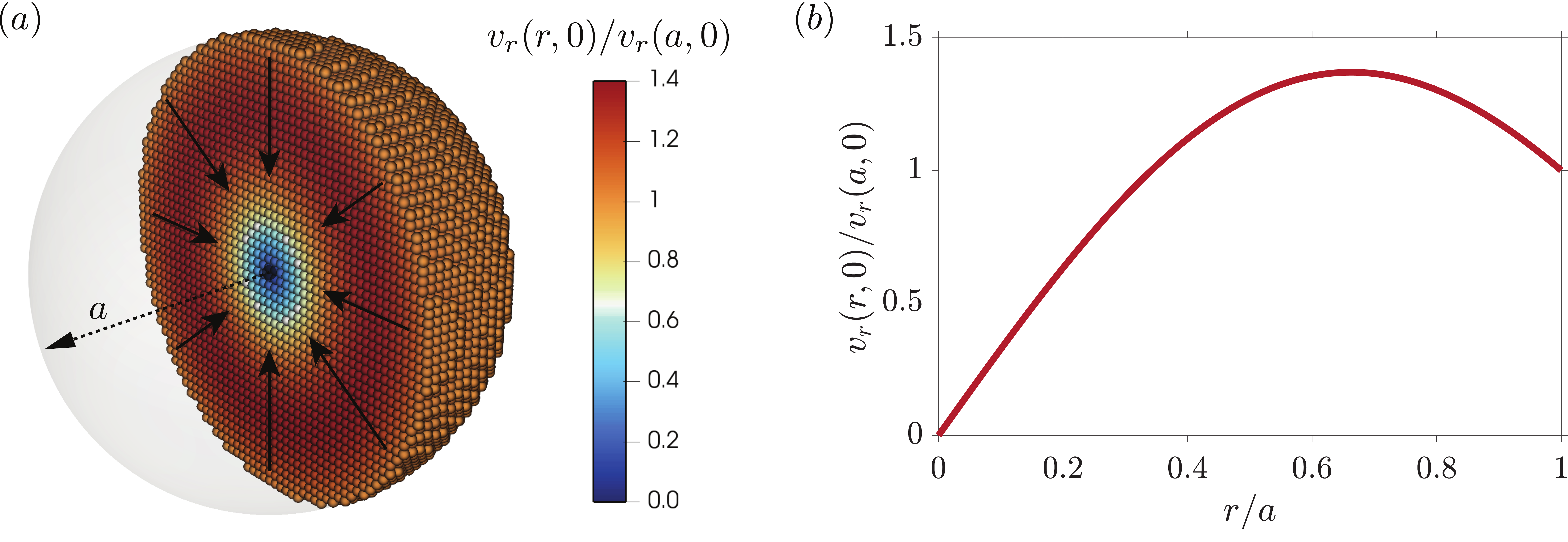}
    \caption{Compaction of a spherical assembly: $(a)$ cross-section view of the model geometry and the initial radial velocity field, normalized by the velocity at $r=a$. Here, the plotted points represent continuum material points of the MPM solver, each representing many granular particles with packing fraction $\phi$. $(b)$ Normalized initial radial velocity as a function of normalized radial distance, $r/a$.}
    \label{fig:4_1_model_spherical}
\end{figure}

The initial radial velocity for each material point located at a radial distance $r$ is set as
\begin{eqnarray}
    \tb v(0)=-v_r (r, 0) \tb r = - A_r \, J_1(k r)\, \tb r\,,
\end{eqnarray}
where $\tb r$ is a unit vector pointed radially outward from the centroid of the spherical assembly, and $A_r$ is the prescribed velocity amplitude; here is set as $A_r = 0.001~\mathrm{m/s}$. Meanwhile, $J_1(kr)$ denotes the spherical Bessel function of the first kind and $k$ is the constant defined in Eq.~\eqref{eq:zeroth_order_k}. The model geometry and the initially prescribed velocity field are shown in Fig.~\ref{fig:4_1_model_spherical}.

The continuum restitution $\mathcal{E}$ is obtained by measuring the velocity of the material point located nearest to the radial distance $a$ at $t = 0$ and $100~\mu\mathrm{s}$ (due to the high ratio of $\lambda/G$, the contact time is relatively short, $t_c \lesssim 60~\mu\mathrm{s}$). To perform this test, we use a relatively fine material-point discretization to better conserve energy, i.e.~more than 65,000 material points are employed, arranged with 125 material points per cell, and the element size $h$ is chosen such that $2a/h = 10$, where quadratic B-spline basis functions are adopted. We observed that higher-order basis functions yield substantially more accurate results for resolving the nonlinear spatial variability of the kinematic fields. The time increment for this problem is set to $\Delta t= 0.1~\mu\mathrm{s}$.

\begin{figure}[h!]
    \centering
    \includegraphics[width=0.9\textwidth]{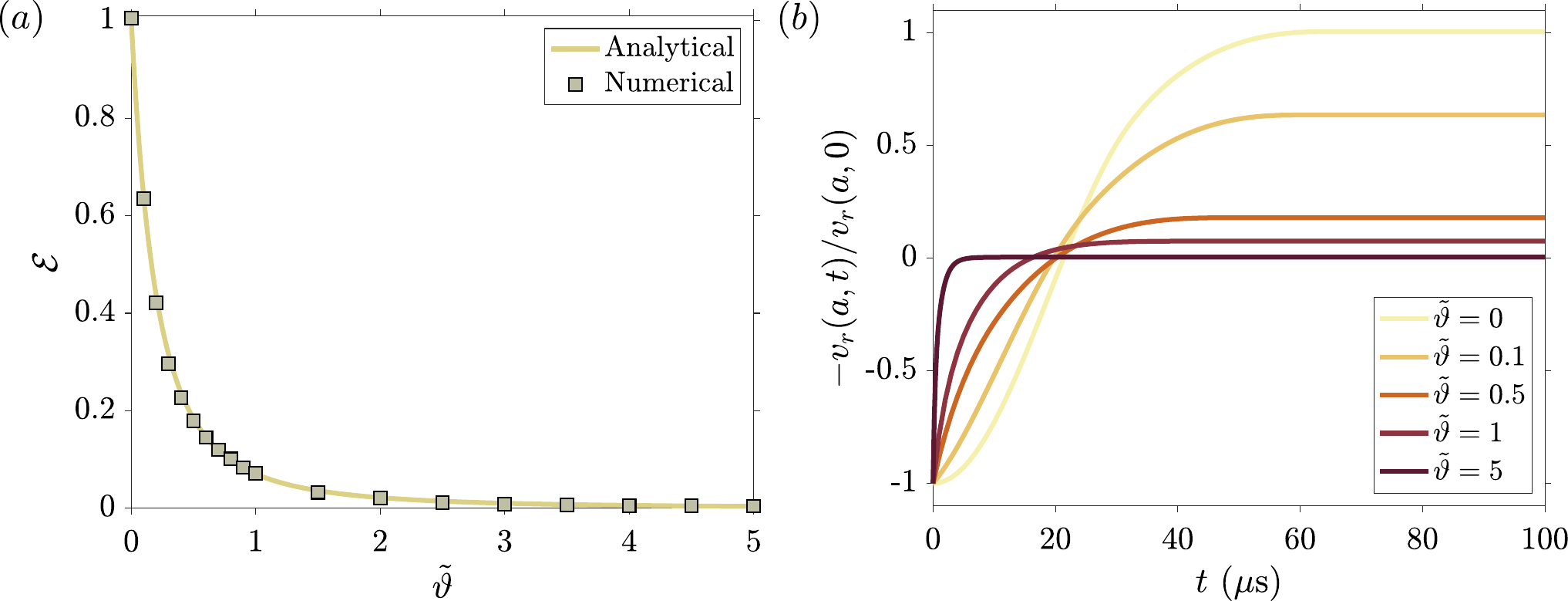}
    \caption{Compaction of a spherical assembly: $(a)$ comparison between the derived analytical solution (Eq.~\eqref{eq:continuum_restitution_d}) and the measured continuum restitution from numerical tests conducted using MPM for different values of bulk viscosity. The continuum restitution is measured by taking the velocity ratio at the surface of the spherical assembly as given by Eq.~\eqref{eq:continuum_restitution}. $(b)$ Time evolution of the velocity ratio at $r=a$ for several values of bulk viscosity, i.e.~$\Tilde{\vartheta} = \{0, 0.1, 0.5, 1, 5\}$.}
    \label{fig:4_1_anal_num_comparison}
\end{figure}

The measured restitution and the time evolution of the velocity ratio at $r = a$ for different values of normalized bulk viscosity are plotted in Fig.~\ref{fig:4_1_anal_num_comparison}$(a)$ and compared with the analytical solution given by Eq.~\eqref{eq:continuum_restitution_d}, showing very good agreement. The time evolution of the ratio of the radial velocity at $r = a$ to its initial value is presented in Fig.~\ref{fig:4_1_anal_num_comparison}$(b)$, further illustrating the damping behavior for selected values of $\Tilde{\vartheta} = \{0, 0.1, 0.5, 1, 5\}$. These results verify the continuum relationship between continuum restitution and bulk viscosity and show that the formulation reproduces the expected viscoelastic damping and the transition to separation.

\subsection{Flow on an inclined plane}
\label{subsec:bagnold}

The next verification test investigates whether variations in viscoelastic damping, or equivalently in the coefficient of restitution, affect steady dense granular flow behavior. In the dense flowing regime, the material response is governed by the modeled frictional $\mu(I)$ rheology \cite{gdr2004dense, da2005rheophysics}, implying that the viscoelastic framework derived solely from elastic restitution should not change the flow behavior in this regime, thereby avoiding any double counting of dissipation mechanisms. To examine this, we consider the inclined-plane flow configuration, which provides a well-established setting for isolating dense-flow behavior.  

\begin{figure}[h!]
    \centering
    \includegraphics[width=1\linewidth]{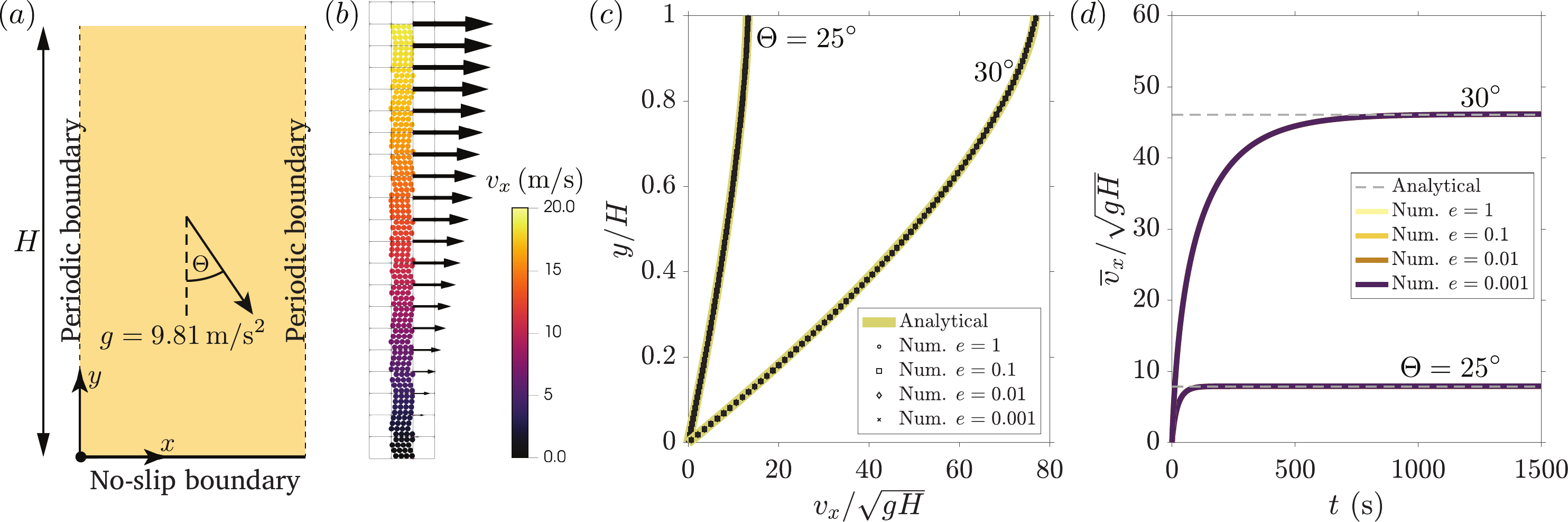}
    \caption{Flow on an inclined plane: $(a)$ model geometry, tilted gravity, and boundary conditions. $(b)$ Simulated steady-state velocity profile using MPM with $\Theta=25^\circ$, $e=1$. $(c)$ Steady-state horizontal velocity profile at $t=1500$ s and $(d)$ time evolution of the height-average velocity, both shown for two tilt angles $\Theta=25^\circ$ and $30^\circ$ and for different restitution coefficient values.}
    \label{fig:4_2_bagnold}
\end{figure}

The inclined-plane flow problem is idealized in a quasi-one-dimensional setting, where in the MPM an infinitely long granular slab is modeled as a single-element column with periodic boundary conditions on the lateral sides and a no-slip boundary at the base; see Fig.~\ref{fig:4_2_bagnold}$(a)$ and $(b)$. The inclination of the plane is introduced by rotating the gravitational acceleration vector by an angle $\Theta$. The column height is $H = 20$ cm, and it is initially stress-free. To avoid abrupt transitions to equilibrium, which may induce numerical oscillations, the gravitational acceleration ($g = 9.81$ m/s$^2$) is smoothly ramped up during the first tenth of a second. The flow then evolves toward a distinct steady-state condition, yielding the steady-state Bagnold velocity profile:
\begin{eqnarray}
    \frac{{v}_x(y)}{\sqrt{gH}}=\frac{2}{3} \xi \frac{\tan\Theta-\mu_s}{\mu_2-\tan\Theta}\sqrt{{\phi} \rho_s \cos\Theta} \left( \frac{H^{3/2}-(H-y)^{3/2}}{H^{1/2}} \right)\,,
    \label{eq:bagnold}
\end{eqnarray}
where $\xi$ is the constant defined in Eq.~\eqref{eq:mu_i_params}$_5$. Integrating the horizontal velocity over the entire slab height and dividing by $H$ gives the depth-averaged horizontal velocity, i.e.,
\begin{eqnarray}
    \frac{\overline{v}_x}{\sqrt{gH}}=\frac{2}{5} \xi H \frac{\tan\Theta-\mu_s}{\mu_2-\tan\Theta}\sqrt{\phi \rho_s \cos\Theta}\,.
    \label{eq:bagnold_avg}
\end{eqnarray}
The analytical solutions given by Eqs.~\eqref{eq:bagnold} and \eqref{eq:bagnold_avg} are derived under the assumption of rigid-plastic flow, where $\phi=\phi_0$ and $H=H_0$ are taken to be constant. Although volumetric deformation is allowed in our framework, it remains very small because the bulk modulus, $K$, is chosen to be sufficiently large relative to the maximum compressive stress, and the plastic flow rule is considered to be isochoric. Therefore, the initial values of $\phi$ and $H$ are used when evaluating the velocity solution.

The material and numerical parameters are set as follows. The Young's modulus and Poisson's ratio are set to $E = 20$ MPa and $\nu = 0.3$, respectively, which correspond to $K = 16.7$ MPa and $G = 7.7$ MPa. These values are modeled considering solid grains with a Young's modulus on the order of $E \sim 1$ GPa \cite{kamrin2008stochastic, dunatunga2017continuum}. The solid density is set to $\rho_s = 2500$ kg/m$^3$, the grain size is set to $d = 1$ mm, and the initial packing fraction is set to $\phi_0 = 0.59$. The frictional flow parameters for the $\mu(I)$-rheology are chosen to represent typical monodisperse glass beads \cite{jop2006constitutive, pouliquen2006flow}, namely $\mu_s = \tan 21^\circ \approx 0.3839$, $\mu_2 = \tan 33^\circ \approx 0.6494$, and $I_0 = 0.3$. In this study, viscoelastic damping is varied by changing the coefficient of restitution, $e = \{1, 0.1, 0.01, 0.001\}$, which correspondingly modifies the bulk and shear viscosities via Eqs.~\eqref{eq:bulk_viscosity} and \eqref{eq:shear_viscosity_rayleigh}.

The MPM simulation uses an element resolution of $H/h = 20$, with 16 material points per cell, totaling 320 particles. Both linear and quadratic B-spline basis functions can be employed for this problem; we choose the former to slightly reduce computational cost as the simulation must be run for a long duration to reach a steady state. The time step is set to $\Delta t = 10^{-5}$ s following the critical time-step estimate described in \ref{app:critical_time_step}, and the simulation is run until $T = 1500$ s. Two tilt angles, $\Theta = 25^\circ$ and $30^\circ$, are considered, both within the range bounded by $\mu_s$ and $\mu_2$ to yield a steady-state solution.

Fig.~\ref{fig:4_2_bagnold}$(c)$ plots the steady-state velocity profiles at the two tilt angles, $\Theta=25^\circ$ and $30^\circ$, over the slab height, compared with the analytical solution obtained from Eq.~\eqref{eq:bagnold}. It can be seen that varying the coefficient of restitution $e$, which influences the magnitude of the viscoelastic bulk and shear viscosities, does not affect the steady-state flow velocity; the velocity profiles collapse to the same curve, which is governed only by variation in $\Theta$. This is further elaborated by Fig.~\ref{fig:4_2_bagnold}$(d)$, where not only do the steady-state velocity profiles coincide, but the evolution of the height-averaged velocity over time also remains independent of coefficients of restitution. Since the deformation is mainly governed by plastic flow, the influence of viscoelastic damping does not manifest in the transient evolution of the velocity field. Upon reaching steady state, the averaged velocity approaches the prediction as given by Eq.~\eqref{eq:bagnold_avg}. Consistent with this expectation, discrete-element simulations with non-zero grain friction for similar configurations have also reported no measurable dependence on the restitution coefficient \cite{silbert2001granular, da2005rheophysics}.

\subsection{Flat bottom silo flow}
\label{subsec:silo}

In the next numerical example, we aim to demonstrate the capability of the proposed framework to simulate granular media that flow plastically, undergo phase change into a separated gas-like state, reconsolidate, and subsequently support stress as an elasto-plastic solid. In particular, we investigate whether resolving collisional dissipation through the coefficient of restitution influences the flow dynamics and the resulting repose behavior. To examine this, we consider a two-dimensional silo flow simulation, similar to the configuration studied by \citet{dunatunga2015continuum}, but with additional length so that the flow behavior is not affected by the side boundary conditions.

\begin{figure}[h!]
    \centering
    \includegraphics[width=0.95\linewidth]{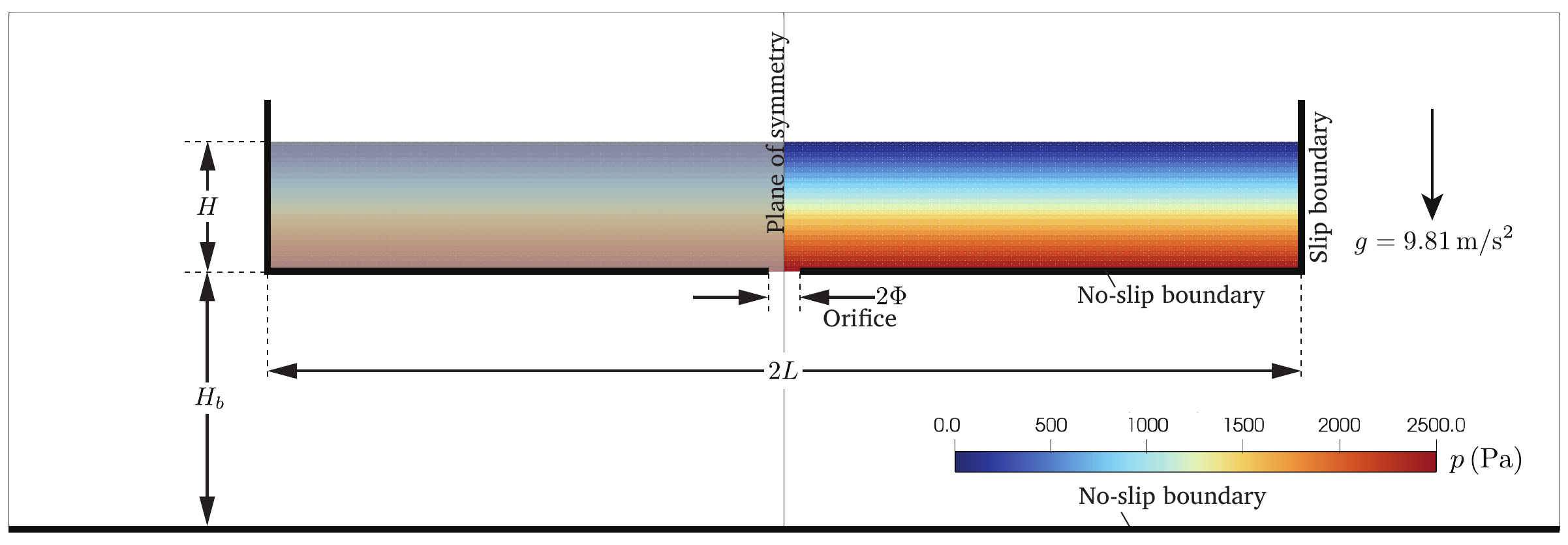}
    \caption{Flat bottom silo flow: model geometry, boundary conditions, and lithostatic mean stress profile.}
    \label{fig:4_3_model_silo}
\end{figure}

The silo is elevated \revision{$H_b = 0.5$ m} above the base and has an initial length $2L = 2$ m and a fill height $H = 0.25$ m. At the center of the silo floor, an orifice of width $2\Phi = 0.06$ m allows the granular material to discharge and form a sink flow. The silo base and side walls are modeled as rough (no-slip) and smooth (slip), respectively, so that the initial stress and packing conditions follow the lithostatic state (see Fig.~\ref{fig:4_3_model_silo}). Once the initial stress equilibrium is reached, the orifice is opened instantaneously to initiate drainage. As the grains exit through the orifice, they enter the separation state and therefore become stress-free. Upon impacting the base, the material points re-enter a compressive state and begin to develop stress. The base is modeled as a fully rough (no-slip) boundary. Owing to the flow symmetry, we model and simulate only half of the silo, with the symmetry boundary imposed as a roller (or slip) boundary.

The simulation domain is discretized into structured quadrilateral elements with size $h = 0.01$ m and initially four material points per cell, yielding a total of 10,000 material points. The slip and no-slip boundary conditions on the base and side walls are imposed by prescribing the appropriate combinations of normal and tangential nodal kinematics. The orifice is modeled by deactivating the nodal boundary constraints at the center of the silo base. Since a half model is used, the orifice of size $\Phi = 0.03$ m is represented by removing the velocity constraints at three consecutive nodes counted from the left-most node on the symmetry plane. Due to the symmetry boundary condition, the horizontal constraint at this left-most node remains enforced. It is worth noting that, since higher-order basis functions enlarge the support influence of the nodal constraints around the orifice, we would require a finer mesh to resolve the opening accurately. For this reason, we decided to employ bi-linear basis functions in this problem.

\begin{figure}[h!]
    \centering
    \includegraphics[width=0.95\linewidth]{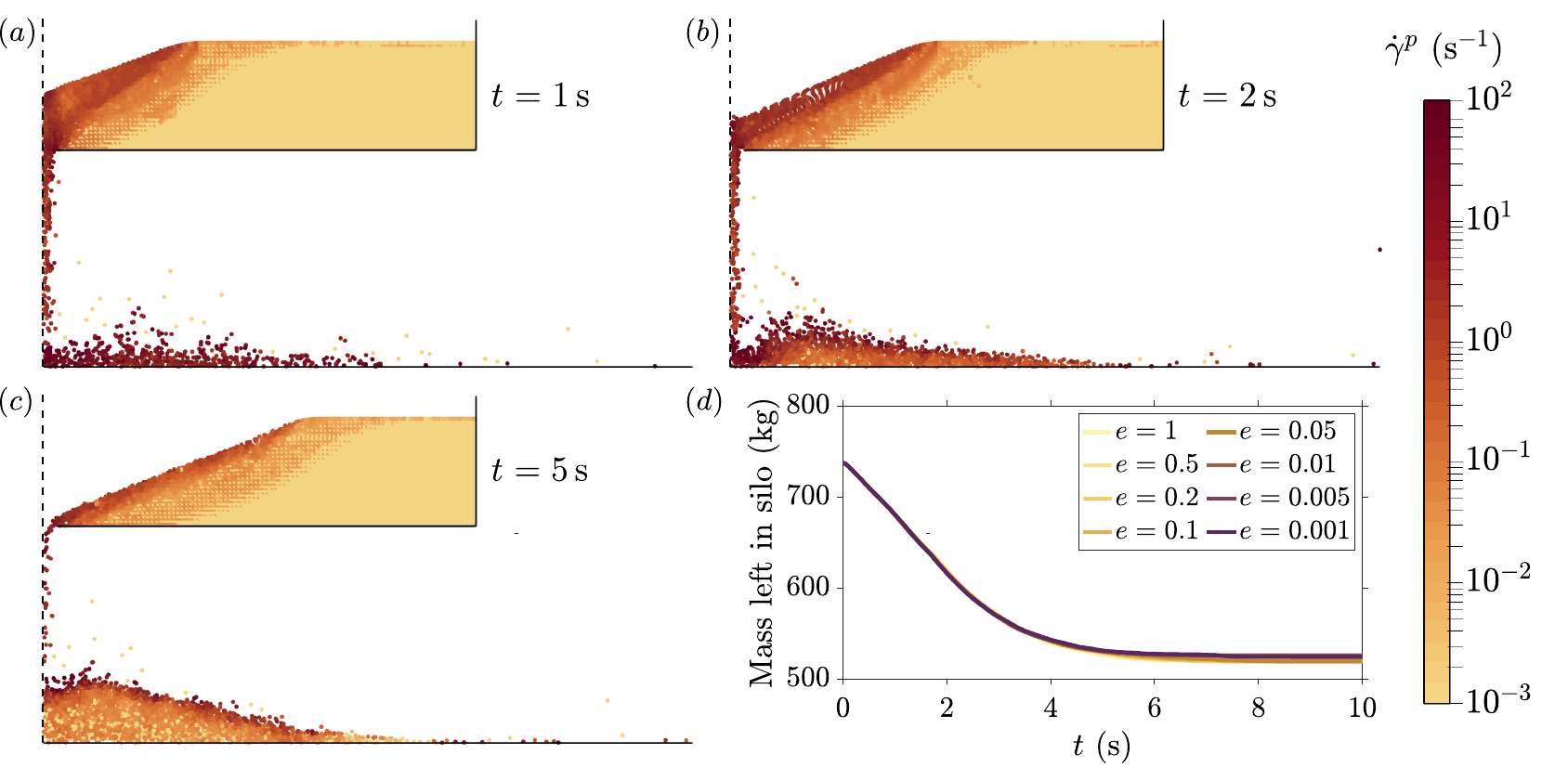}
    \caption{Flat-bottom silo flow: $(a)$--$(c)$  MPM simulation snapshots showing the evolution of the plastic shear strain rate, $\dot{\gamma}^p$, during silo discharge and subsequent reconsolidation for $e = 1$. $(d)$ The mass remaining in the silo for different coefficient of restitution values. The total mass measurements are adjusted to the full model with a silo dimension of $2L\times H$.}
    \label{fig:4_3_snapshot_mass_time}
\end{figure}

The material parameters considered in this study are the same as those used in \Cref{subsec:bagnold}, which are suitable for monodisperse glass beads with $d = 1$ mm. Eight different cases of the coefficient of restitution are investigated, i.e.~$e = \{1, 0.5, 0.2, 0.1, 0.05, 0.01, 0.005, 0.001\}$. The time step is set to $\Delta t = 2 \times 10^{-6}$ s and the simulation is run until $t = 10$ s. Fig.~\ref{fig:4_3_snapshot_mass_time}$(a)$–$(c)$ shows the flow evolution at $t = 1$, 2, and 5 s for the case with $e = 1$, corresponding to no viscoelastic damping. Meanwhile, Fig.~\ref{fig:4_3_snapshot_mass_time}$(d)$ presents the time evolution of retained mass inside the silo, which shows a relatively steady decrease during the first four seconds before gradually decelerating around $t = 5$ s. Since the flow inside the silo is generally a dense plastic flow, similar to the inclined-plane problem discussed in \Cref{subsec:bagnold}, the discharge rate is insensitive to the coefficient of restitution. This is evident in Fig.~\ref{fig:4_3_snapshot_mass_time}$(d)$, where all curves collapse onto a similar trend regardless of the value of $e$ (see correlation with DEM simulations \cite{coetzee2016calibration}).

\begin{figure}[h!]
    \centering
    \includegraphics[width=0.9\linewidth]{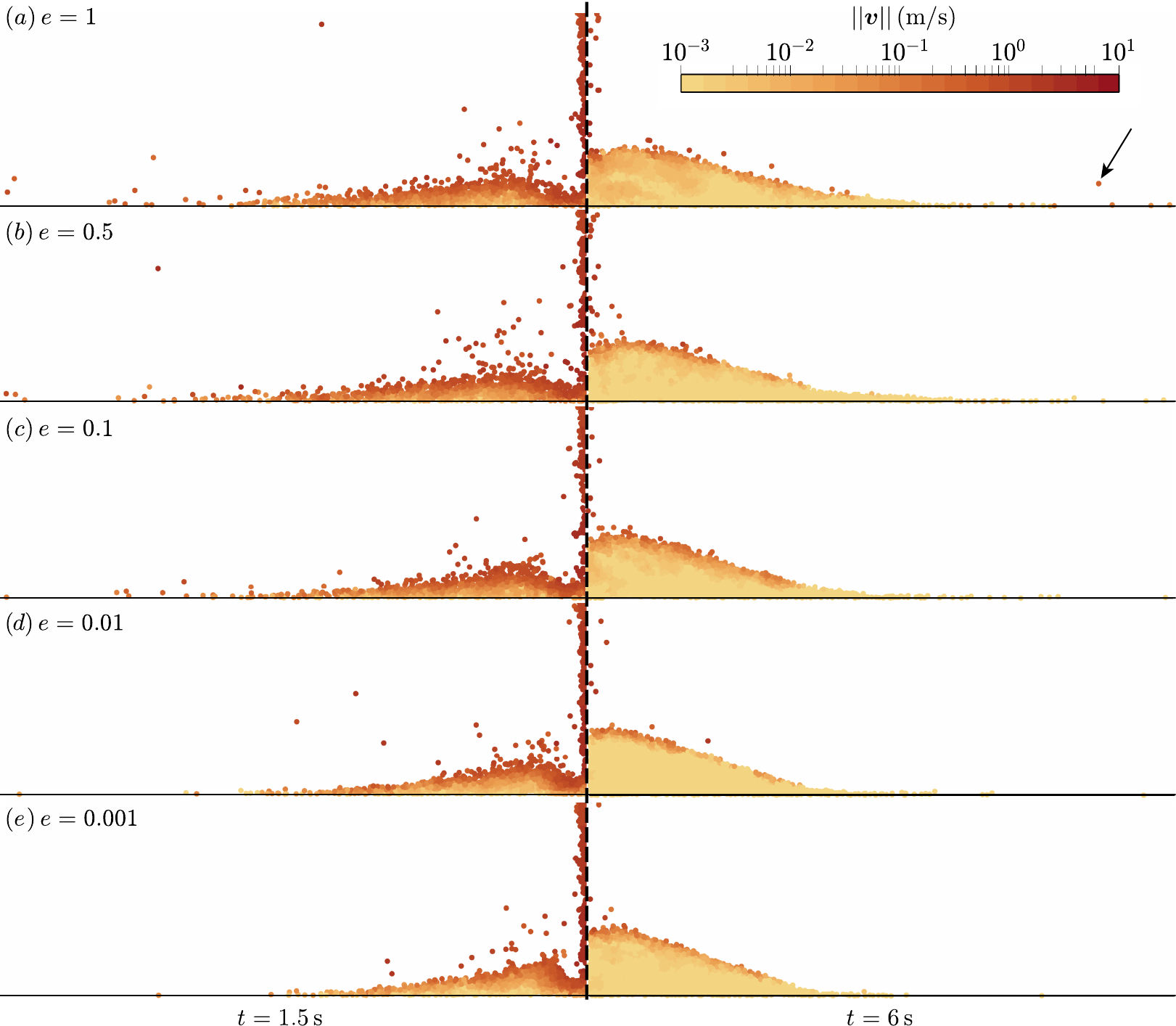}
    \caption{Flat-bottom silo flow: snapshots of silo-flow reconsolidation obtained from MPM simulations, showing the formation of a heap at the lower base. Five different coefficients of restitution are plotted for comparison: $(a)$ $e = 1$, $(b)$ $e = 0.5$, $(c)$ $e = 0.1$, $(d)$ $e = 0.01$, and $(e)$ $e = 0.001$. Two time instances are shown, (left) $t = 1.5$ s and (right) $t = 6$ s, to illustrate the development of the heap. Black arrows highlight unphysical oscillations of material points due to undamped elastic contact. The visualized velocity field is not the direct particle velocity, but is interpolated from the grid velocity.}
    \label{fig:4_3_splash_slope}
\end{figure}

Fig.~\ref{fig:4_3_splash_slope} highlights how viscoelastic continuum damping, controlled by the coefficient of restitution $e$ (Eq.~\eqref{eq:bulk_viscosity}), influences the overall reconsolidation behavior. Higher restitution increases the flow energetics, as seen in the left column of the figure at $t = 1.5$ s, where larger $e$ leads to more grains flying upward due to reduced energy dissipation upon impact with the lower base. This behavior was not captured in our earlier formulation \cite{dunatunga2015continuum}, since a more dissipative scheme, the \textit{updated-stress-last} (USL) \cite{zhang2016material}, was used. Although the USL scheme enhances numerical stability, its uncontrolled numerical diffusion constrains its capability to capture highly energetic granular dynamics when compared with DEM \cite{yue2018hybrid}. 

The coefficient of restitution also affects the stability of isolated material points near the runout front. As indicated by the arrow in the right plot of Fig.~\ref{fig:4_3_splash_slope}$(a)$, material points at the leading edge often exhibit unphysical bouncing (see the supplemental materials for the animation \cite{supplemental_material}). This occurs because a constant $\Delta t$ cannot exactly resolve the collision time, that is, the instant when a separated material point (with $\rho < \rho_c$) re-enters the dense state ($\rho \geq \rho_c$). As a result, in a scheme that does not introduce artificial numerical damping, elastic collisions may inject spurious energy into the system, which may cause oscillatory behavior or, if $\Delta t$ is large enough, even lead to divergence of the numerical solution. When $\Delta t$ is chosen sufficiently small, the amount of numerical dissipation is typically reduced, and consequently, less energy is removed by numerical dissipation during elastic contact. Introducing viscoelasticity mitigates this issue in a physical way, by dissipating energy at collisions, thereby reducing oscillations that could otherwise drive the numerical solution toward unstable and unphysical material behavior.

\begin{figure}[h!]
    \centering
    \includegraphics[width=0.9\linewidth]{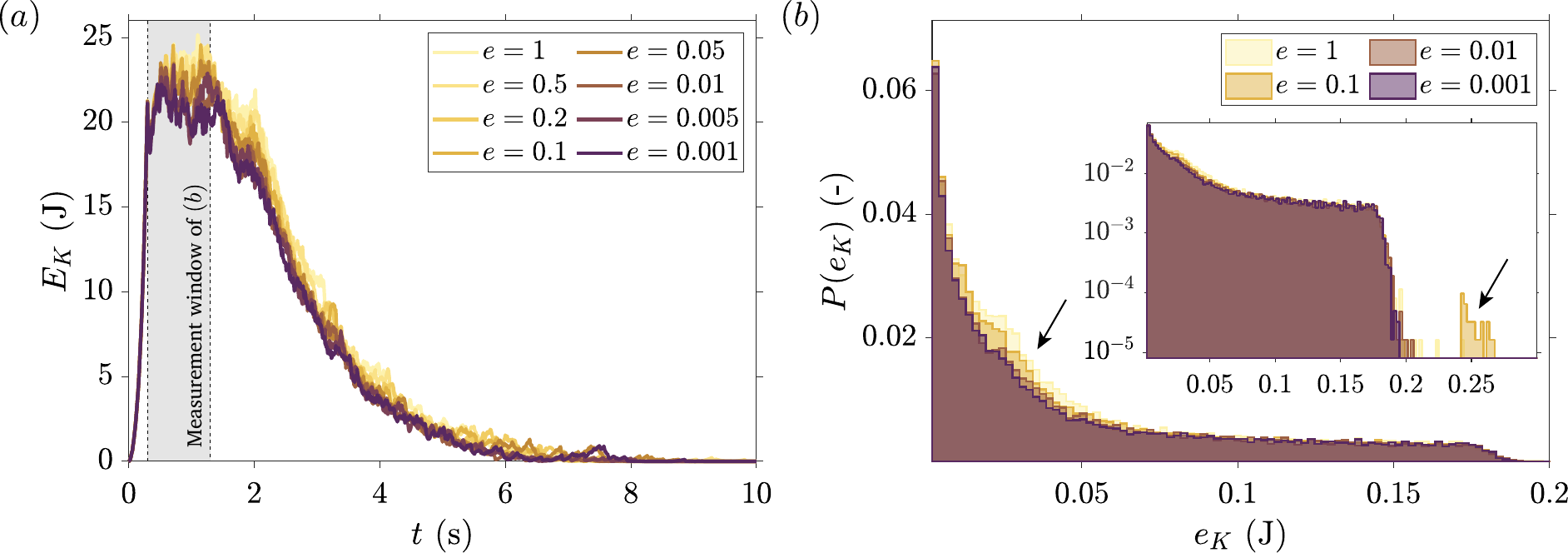}
    \caption{Flat-bottom silo flow: $(a)$ evolution of total kinetic energy over time for different coefficients of restitution, and $(b)$ normalized kinetic energy counts for all material points around the first impact, within the time window $t \in (0.3,1.3)\,\mathrm{s}$ (highlighted in $(a)$). The histogram counts are shown only for $e_K \in (0.002,0.2)\,\mathrm{J}$ to highlight the major differences among the four restitution values $e = \{1, 0.1, 0.001, 0.0001\}$. The inset shows the same distribution on a logarithmic scale to magnify the rare-event counts at high kinetic energy. Black arrows mark differences in the kinetic energy landscape around $e_K \approx 0.03\text{–}0.05$ J and in the high-energy counts at $e_{K} > 0.2$ J. Here, $e_K=\frac{1}{2}m_p (\tb v_p\cdot \tb v_p)$ and $E_K=\sum_p e_{K}$, where the subscript $p$ denotes material point quantities.}
    \label{fig:4_3_kinetic_energy}
\end{figure}

Fig.~\ref{fig:4_3_kinetic_energy} further quantifies these effects. The total kinetic energy, $E_K$, over time (Fig.~\ref{fig:4_3_kinetic_energy}$(a)$) decreases systematically as $e$ decreases. We further analyze the distribution of kinetic energy over the system, $e_K$, within a one-second window between $t \in (0.3, 1.3)$ s. This interval is chosen because the material points initially fall freely from the silo to the base in approximately $t \sim \left(2H_b/g\right)^{1/2} \approx 0.32$ s, where $H_b = 0.5$ m is the silo height. The resulting kinetic energy distribution is plotted in Fig.~\ref{fig:4_3_kinetic_energy}$(b)$, showing two dominant regions of dissipation: an energy band around $e_K \approx 0.03$–$0.06$ J and rare high-energy events above $0.2$ J. We also find that most material point kinetic energy lies below $0.18$ J, consistent with the potential energy of a material point free-falling from height $H_b$, i.e.~$e_\Pi=m_p g H_b$.

\begin{figure}[h!]
    \centering
    \includegraphics[width=0.95\linewidth]{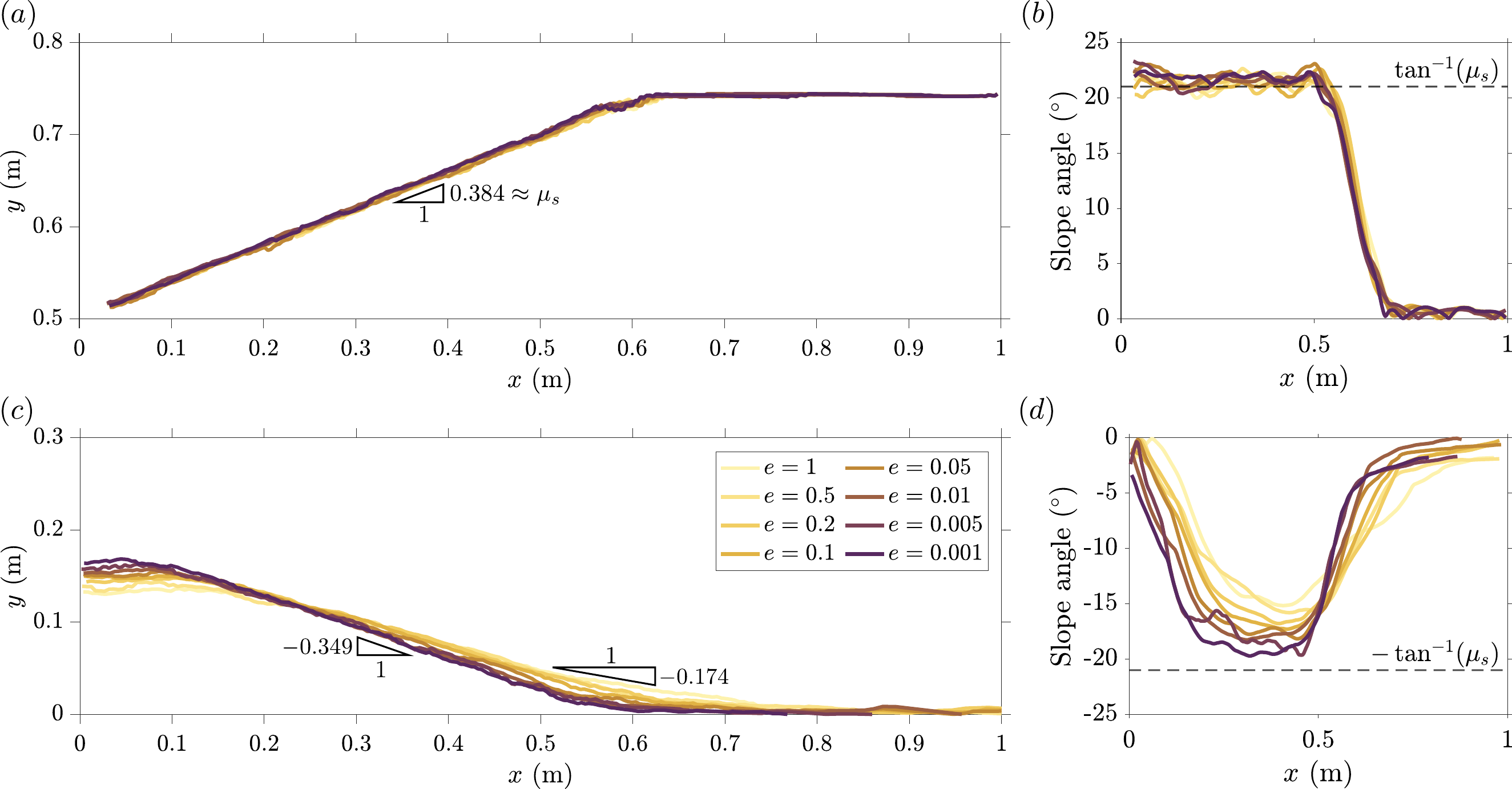}
    \caption{Flat-bottom silo flow: repose surface configuration and slope angle of $(a)$--$(b)$ the upper heap and $(c)$--$(d)$ the lower heap. The upper repose angle follows the prescribed static coefficient of friction, $\mu_s = 0.3839 = \tan(21^\circ)$. In contrast, the lower heap exhibits a smaller and spatially varying repose angle due to dynamic and collisional effects. The free surface is extracted from the material point coordinates using the principal component analysis routine in MATLAB \cite{MATLAB2024}. The local repose angle is computed from the smoothed surface using a moving-window linear regression to estimate $\partial y/\partial x$, followed by $\Theta = \tan^{-1}\!\left(\mathrm{d}y/\mathrm{d}x\right)$.}
    \label{fig:4_3_static_dynamic_repose}
\end{figure}

Once the flow comes to rest, the restitution coefficient is observed not to influence the final repose configuration of the upper heap, which attains the \textit{static repose angle} given by $\mu_s$, as expected. This is illustrated in Figs.~\ref{fig:4_3_static_dynamic_repose}$(a)$ and $(b)$, where the upper-heap slopes for all $e$ approach the angle governed by the static friction coefficient, $\tan^{-1}(\mu_s)=21^\circ$. In contrast, Fig.~\ref{fig:4_3_splash_slope}(right) and Figs.~\ref{fig:4_3_static_dynamic_repose}$(c)$ and $(d)$ show clear differences in the \textit{dynamic repose angle} at the lower heap for different coefficients of restitution, with all values measured below $21^\circ$. Near the center of the heap ($x < 0.1$ m), the slope flattens due to collisions induced by the discharge through the orifice, which push the existing material points laterally and drive flow down the slope. At $x > 0.1$ m, lower restitution yields a steeper slope and shorter runout distances. This overall trend is consistent with many DEM studies of granular flow problems (e.g.~\cite{staron2007spreading, santos2016investigation, wei2019numerical, xiao2025sensitivity}).

\subsection{Granular bed subjected to an impactor}
\label{subsec:impactor}

\begin{figure}[h!]
    \centering
    \includegraphics[width=0.9\linewidth]{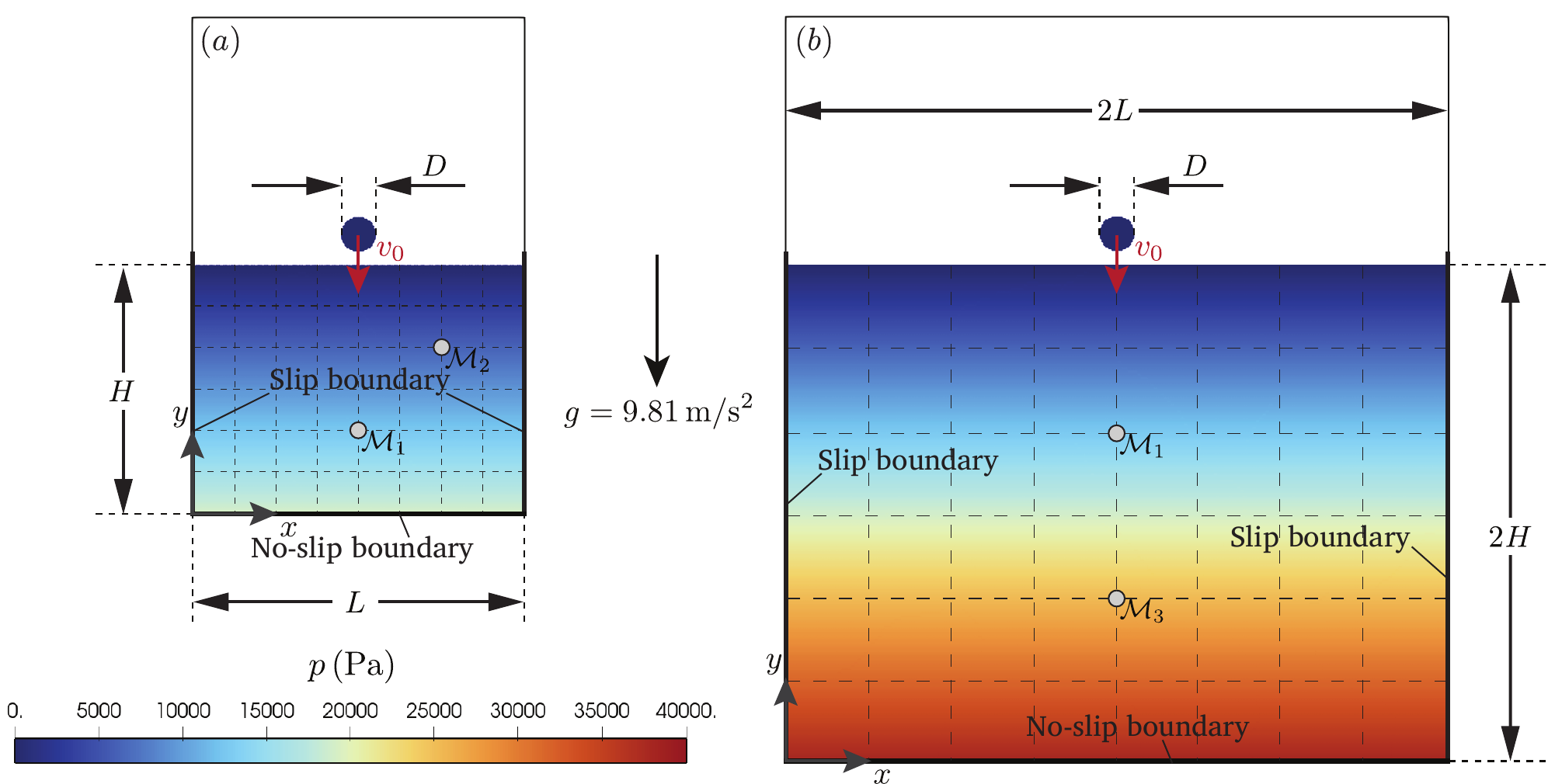}
    \caption{Granular bed with an impactor: model geometry, boundary and initial conditions, and lithostatic mean stress profile. Two models with different dimensions are considered: $(a)$ model 1 with size $L\times H$ and $(b)$ model 2 with size $2\,(L\times H)$. The gray circles highlight three locations of the main observation points. The dashed-grid vertices are used as additional measurement points to evaluate the characteristic oscillatory parameters of the granular material.}
    \label{fig:4_4_impactor_model}
\end{figure}

In the following study, we investigate the effect of the proposed viscoelastic formulation on damping stress oscillations in a granular bed subjected to impact. As shown in Fig.~\ref{fig:4_4_impactor_model}, two simulation models with different sizes are considered: (i) model 1 with dimensions $L \times H$, and (ii) model 2 with dimensions twice those of model 1, i.e.~$2\,(L \times H)$. Here, we take $L = 2$~m and $H = 1.5$~m. These two models are selected because of their differing characteristic length scales (the bed depth in this case), which influence the longest wavelengths that can be supported in the system. The granular beds are impacted by a circular object of diameter $D = 0.2$~m, initially positioned 0.1~m above the free surface and moving vertically downward with speed $v_0 = 5$~m/s.

The granular beds are assumed to be composed of dense glass beads of size $d = 5$~mm, with an initial packing fraction of $\phi_0 = 0.832$. The solid grain density, elastic moduli, and frictional parameters are selected to be identical to those in \Cref{subsec:bagnold}. The impactor, on the other hand, has a density twice that of the granular bed ($\rho_s=5000$~kg/m$^3$), while all other material parameters are set to be equal. Here, we investigate the damping characteristics of the problem by varying the coefficient of restitution of the granular media, $e = \{1,\,0.1,\,0.01,\,0.001\}$.

Both simulation domains are discretized using structured quadrilateral meshes with size $h = 0.02$ m, with four material points initialized per cell. For the circular impactor, structured material points with the same size and volume as those of the granular bed are employed. Here, quadratic B-spline basis functions are used to obtain more accurate measurements. The boundary conditions are prescribed as no-slip at the bottom boundary and slip on the two side boundaries of the granular bed. The side boundary condition is applied only up to five nodes above the initial bed surface, so that debris ejected above this height is not constrained from leaving the simulation domain. A lithostatic analysis is first performed to obtain the initial stress distribution in the granular bed, excluding the impactor. The dynamic simulation is then carried out until $T = 3$ s, with a time increment of $\Delta t = 10^{-5}$ s.

\begin{figure}[h!]
    \centering
    \includegraphics[width=0.95\linewidth]{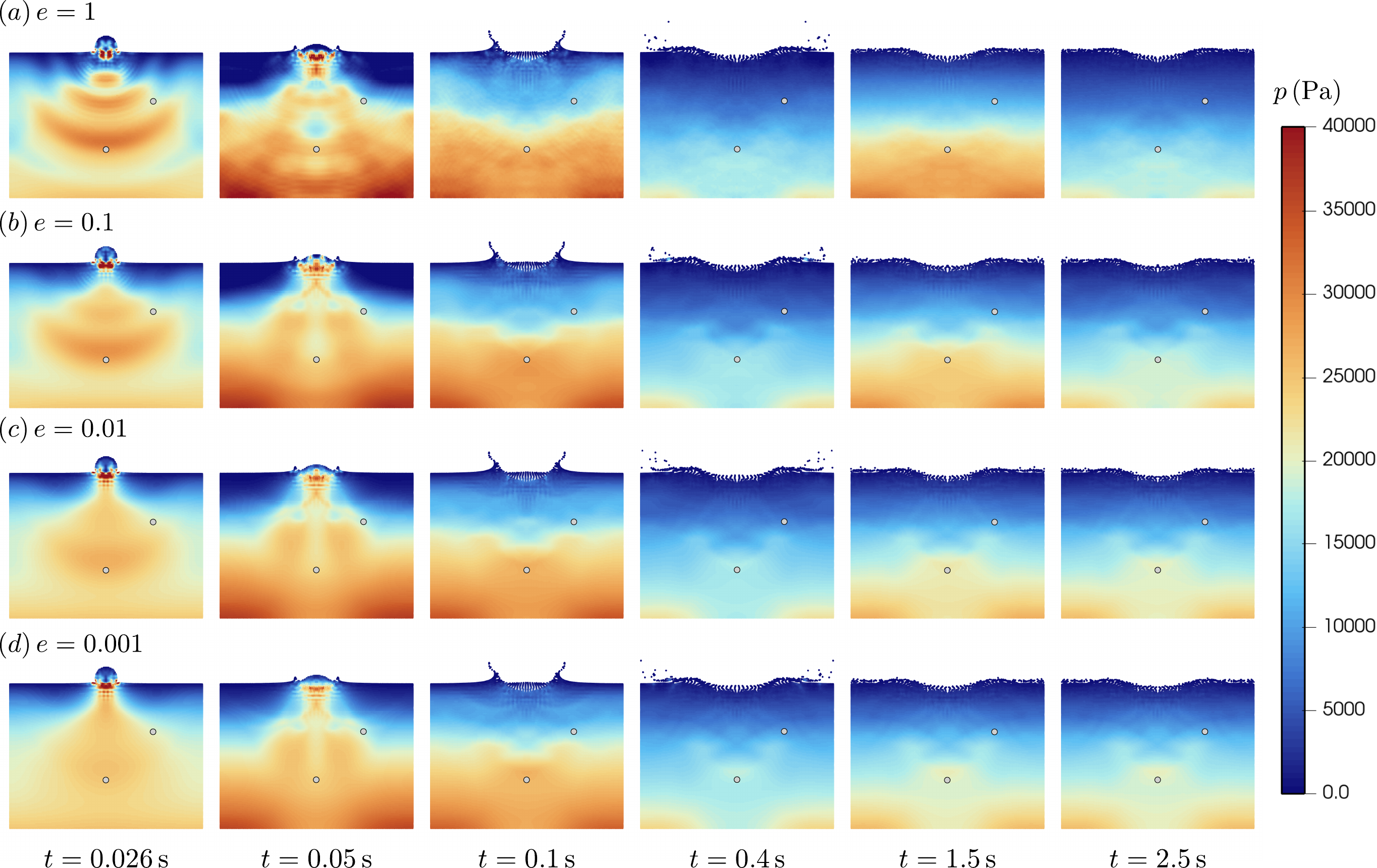}
    \caption{Granular bed with an impactor: pressure profile snapshots of model 1 at different times: (left to right) $t=$0.026 s, 0.05 s, 0.1 s, 0.4 s, 1.5 s, and 2.5 s. Four coefficient of restitution are plotted for comparison: $(a)$ $e = 1$, $(b)$ $e = 0.1$, $(c)$ $e = 0.01$, and $(d)$ $e = 0.001$. Two gray circles in each snapshot indicate the measurement points $\mathcal{M}_1$ and $\mathcal{M}_2$ (cf.~Fig.~\ref{fig:4_4_impactor_model}).}
    \label{fig:4_4_pressure_small}
\end{figure}

Several measurement points are defined within the domain; cf.~Fig.~\ref{fig:4_4_impactor_model}. First, each model contains two primary measurement points: in model 1, $\mathcal{M}_1$ and $\mathcal{M}_2$ are located at coordinates $(1, 0.5)$ and $(1.5, 1)$, respectively, and in model 2, $\mathcal{M}_1$ and $\mathcal{M}_3$ are located at $(2, 2)$ and $(2, 1)$, respectively. All coordinates are given in meters and measured relative to the origin at the bottom-left corner of the computational domain. Here, the points $\mathcal{M}_1$ are different in coordinates for the two models, but essentially located at the same location relative to the point of impact. Second, to further characterize the oscillatory response in the entire domain, we measure dynamic pressure oscillations at additional spatial locations. For this purpose, we construct measurement grids with cell sizes of 0.25~m for model 1 and 0.5~m for model 2. The grid vertices are used as additional measurement points.

\begin{figure}[h!]
    \centering
    \includegraphics[width=0.95\linewidth]{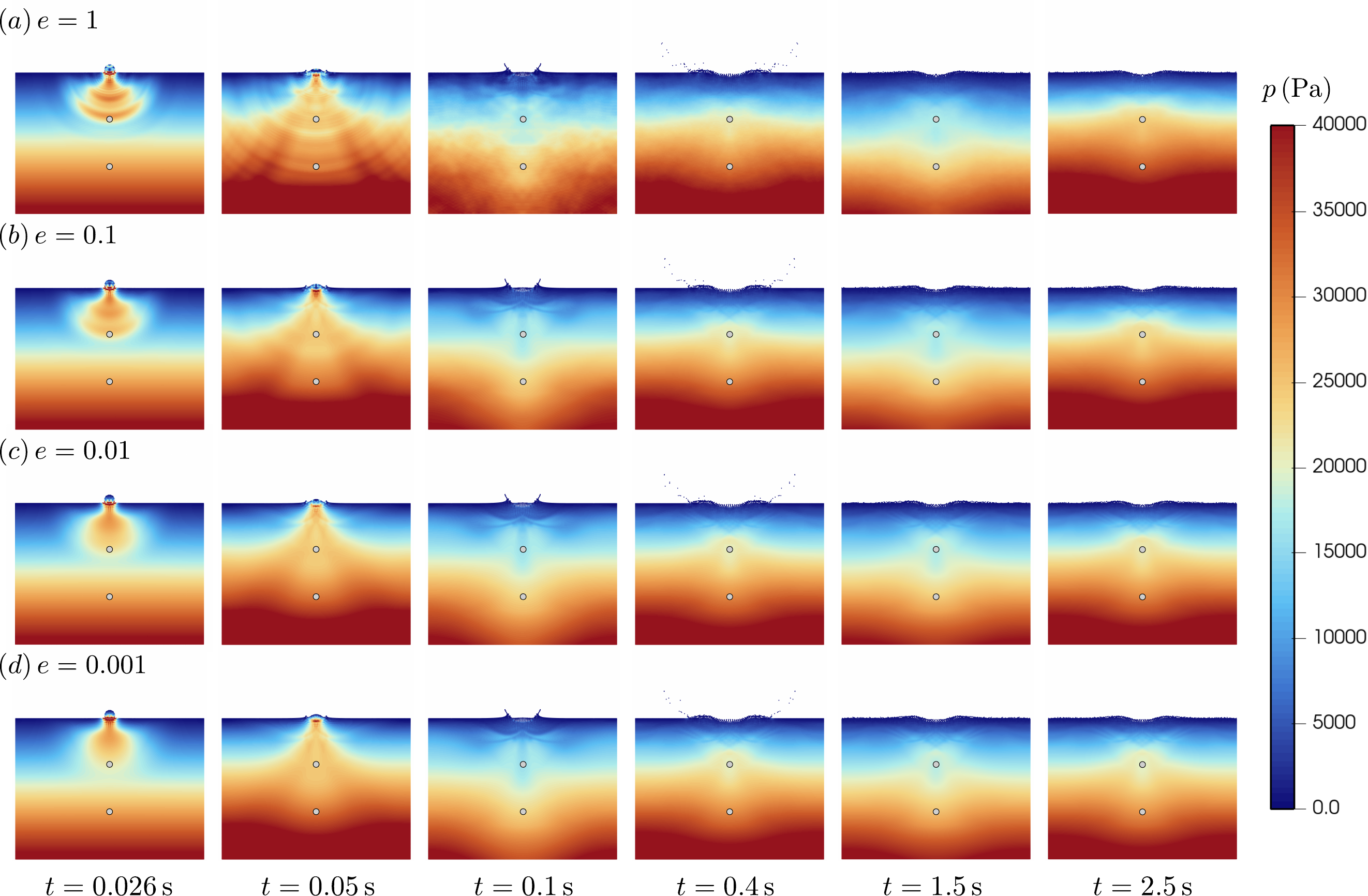}
    \caption{Granular bed with an impactor: pressure profile snapshots of model 2 at different times: (left to right) $t=$0.026 s, 0.05 s, 0.1 s, 0.4 s, 1.5 s, and 2.5 s. Four coefficient of restitution are plotted for comparison: $(a)$ $e = 1$, $(b)$ $e = 0.1$, $(c)$ $e = 0.01$, and $(d)$ $e = 0.001$. Two gray circles in each snapshot indicate the measurement points $\mathcal{M}_1$ and $\mathcal{M}_3$ (cf.~Fig.~\ref{fig:4_4_impactor_model}).}
    \label{fig:4_4_pressure_large}
\end{figure}

\begin{figure}[h!]
    \centering
    \includegraphics[width=0.85\linewidth]{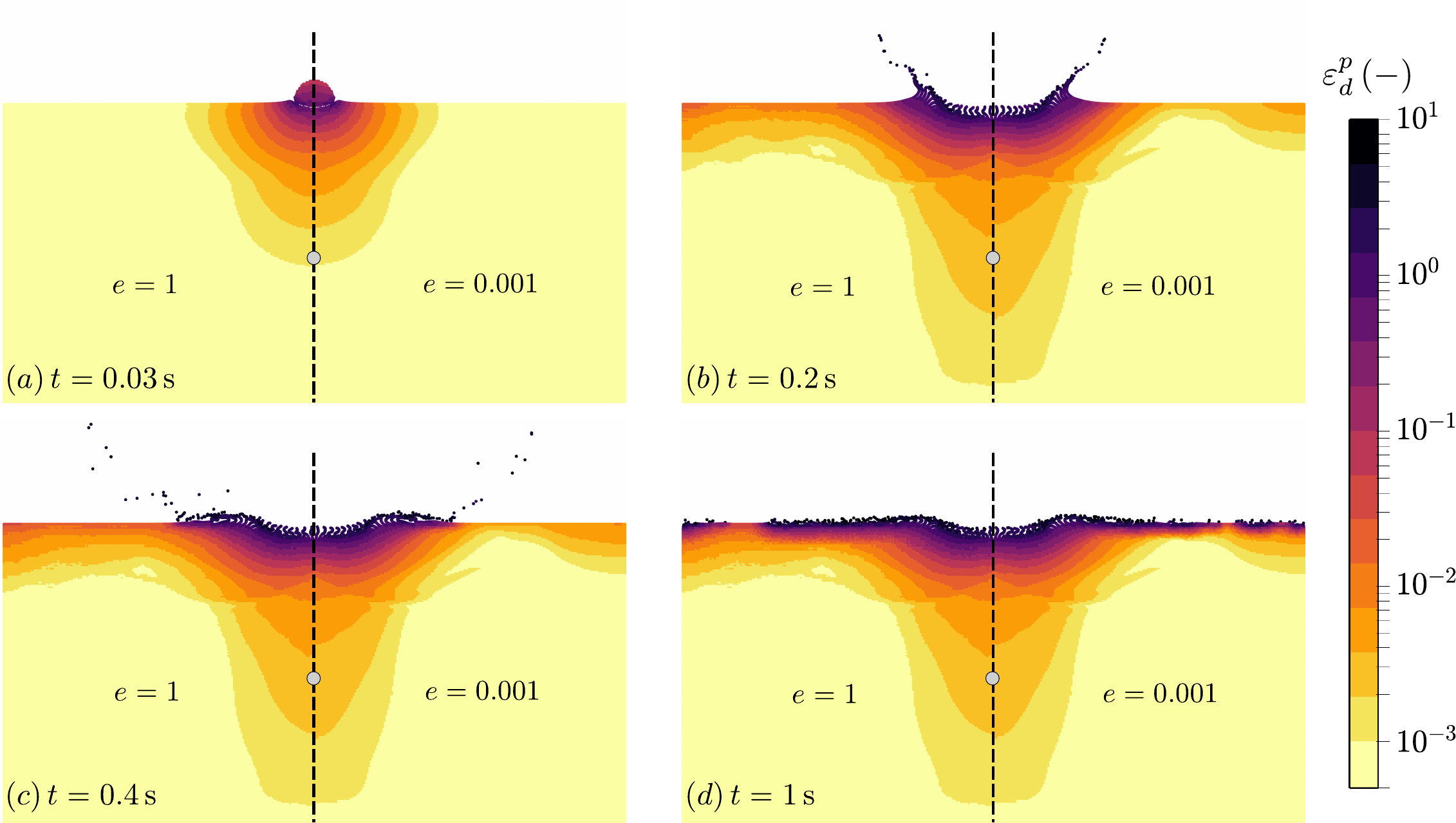}
    \caption{Granular bed with an impactor: accumulated plastic deviatoric strain near the surface at four different time snapshots, $(a)\,t=0.03$ s, $(b)\,0.2$ s, $(c)\,0.4$ s, and $(d)\,1$ s for model 2. The contours for two different restitution values, (left) $e=1$ and (right) 0.001, are compared side by side. The accumulated plastic deviatoric strain is computed as $\varepsilon_d^p = \int_0^t(\dot{\gamma}^p(t')/\sqrt{3})\, \td t'$. The location of the measurement point $\mathcal{M}_1$ is indicated by a gray circle.}
    \label{fig:4_4_pdstrain_compare_full}
\end{figure}

Snapshots of the pressure contours at different times are shown in Figs.~\ref{fig:4_4_pressure_small} and \ref{fig:4_4_pressure_large} for models 1 and 2, respectively, comparing pressure-wave propagation upon impact across the four different coefficients of restitution. Several observations can be made from these plots. At $t = 0.026$~s, we observe that as the coefficient of restitution decreases, the peaks and valleys of the travelling wave upon impact become smoother spatially. Upon reflection of the wave from the bottom and side boundaries, at $t = 0.05$~s, we further see that decreasing restitution effectively damps the higher-order frequency components that remain prominent in the undamped case ($e = 1$). At $t = 0.1$ s, the splash or ejecta patterns are largely the same in all scenarios, suggesting that viscoelastic damping has little influence on the plastic flow behavior. This observation is further supported by Fig.~\ref{fig:4_4_pdstrain_compare_full}, which shows side-by-side contours of the accumulated plastic deviatoric strain, defined as $\varepsilon^p_d=\int_0^t(\dot{\gamma}^p(t')/\sqrt{3})\, \td t'$, in model 2 for the undamped, $e=1$, and the highly damped cases, $e = 0.001$. However, since the viscoelastic damping can damp the stress vibration within the elastic state, this may occasionally prevent a material point from entering the separated regime. This is visible in Fig.~\ref{fig:4_4_pdstrain_compare_full}$(d)$, where the plastic zone near the bed surface and close to the side walls is slightly smaller in the highly damped case compared to the undamped case. Finally, for times $t > 1$~s, the undamped case continues to exhibit noticeable oscillations, whereas the damped cases still oscillate but with substantially smaller variations in the pressure contours. See the supplementary material for the corresponding animation for models 1 and 2 \cite{supplemental_material}.

\begin{figure}[h!]
    \centering
    \includegraphics[width=\linewidth]{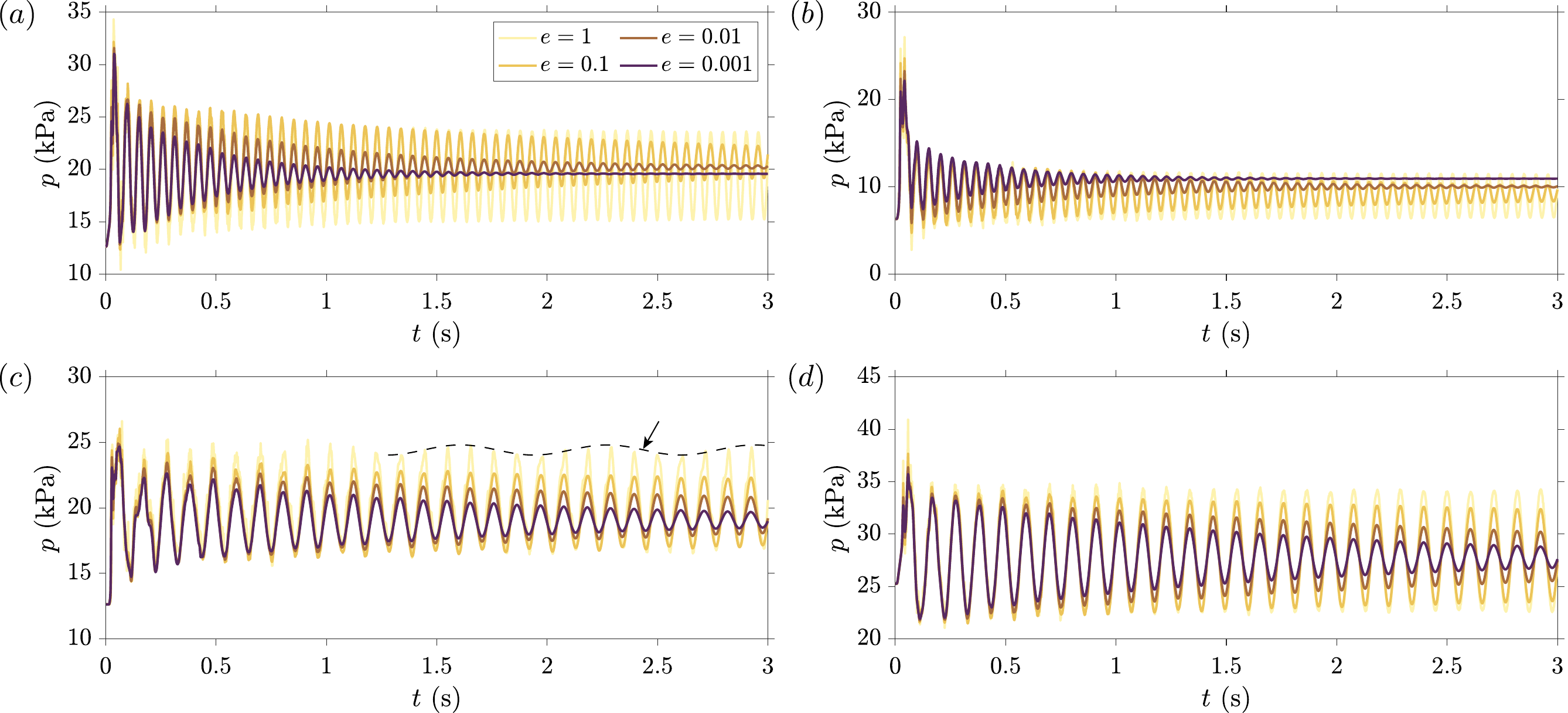}
    \caption{Granular bed with an impactor: pressure evolution over time at the three observation points for two different models: $(a)$ model 1 -- $\mathcal{M}_1$, $(b)$ model 1 -- $\mathcal{M}_2$, $(c)$ model 2 -- $\mathcal{M}_1$, and $(d)$ model 2 -- $\mathcal{M}_3$. The black arrow and dashed line highlight the beat frequency observed in undamped elastic vibration.}
    \label{fig:4_4_pressure}
\end{figure}

Next, the measured oscillations at the measurement points are examined. Fig.~\ref{fig:4_4_pressure} shows the pressure time histories at $\mathcal{M}_1$, $\mathcal{M}_2$, and $\mathcal{M}_3$ for the two models. Model 1 (Figs.~\ref{fig:4_4_pressure}$(a)$ and $(b)$) exhibits noticeably higher oscillation frequencies than model 2 (Figs.~\ref{fig:4_4_pressure}$(c)$ and $(d)$), consistent with its shorter characteristic length scale, which limits the spatial extent over which stress waves can develop. The difference in domain size also governs the damping time scale of the oscillations. As previously shown in \Cref{subsec:bulk_visc}, the characteristic decay time can be written as: $\tilde{\tau} \sim 2/\tilde{\xi} = 2\rho/(k^2\vartheta)$. Here, $k$ is the constant introduced through the separation of variables of the momentum balance equation. Its admissible value is subsequently fixed by the boundary conditions, which introduces an explicit dependency on the characteristic domain size $a$, with $k \approx \pi/a$ to leading order\footnote{In the spherical compaction problem discussed in \Cref{subsec:bulk_visc}, $a$ denotes the radius of the spherical assembly. However, in the context of the granular impactor example, $a$ is more closely associated with the granular bed depth, as the side boundaries oscillate predominantly in the vertical direction, making the depth the relevant characteristic length scale.}. As a result, increasing the depth reduces $k$ and leads to a quadratic increase in the decay time, explaining the more persistent oscillations observed in model 2. 

In the damped cases, measured pressure values near the surface at $\mathcal{M}_1$ and $\mathcal{M}_2$ show a slight shift in their steady-state magnitudes compared to the undamped case (cf.~Figs.~\ref{fig:4_4_pressure}$(a)$--$(c)$). This is because these measurement points are located within or near regions that undergo plastic deformation. At steady state, multiple stress states may satisfy the static equilibrium condition, particularly when residual shear stresses do not fully recover. In contrast, no such shift is observed at the deeper measurement point $\mathcal{M}_3$ (Fig.~\ref{fig:4_4_pressure}$(d)$), where the material response remains elastic throughout the simulation. 

Another notable observation is the presence of beat frequencies in the undamped case ($e = 1$), which is especially pronounced in the pressure measurement of model 2, as shown in Fig.~\ref{fig:4_4_pressure}$(c)$. These beat patterns arise from the superposition of multiple oscillatory modes with closely spaced frequencies. In the damped cases, higher-frequency modes decay more rapidly, suppressing the interference effects responsible for beat phenomena. As a result, no clear beat frequency is observed when viscoelastic damping is present.

\begin{figure}[h!]    
    \centering
    \includegraphics[width=0.9\linewidth]{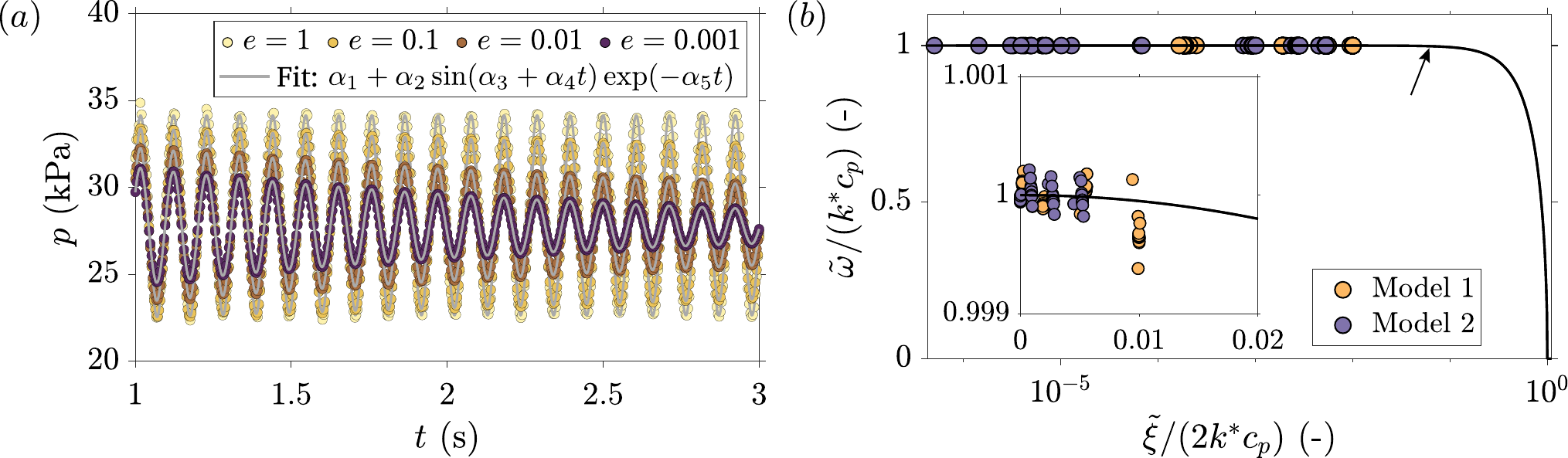}
    \caption{Granular bed with an impactor: $(a)$ fitting the measured pressure--time response for model 2 at point $\mathcal{M}_3$ using a damped sinusoidal function with fitting parameters $\alpha_i$ $(i=1\sim5)$. Each curve displayed in the figure corresponds to a specific fitted function associated with a particular choice of $e$. $(b)$ The oscillation frequency $\tilde{\omega}$ and the damping rate $\tilde{\xi}$ (each scaled by the primary wave velocity $c_p$ and the calibrated length-scale parameter $k^*$), measured at different locations in the domains of model 1 and model 2. The black arrow highlights the relation between $\tilde{\omega}$ and $\tilde{\xi}$ previously described in Eq.~\eqref{eq:damping_constants}, which corresponds to a semicircle relation. The inset plot enlarges the slight decreasing trend of oscillation frequency with increasing damping (or decay rate).}
    \label{fig:4_4_fit_analysis}
\end{figure}

Lastly, we examine whether the proposed formulation can be approximated as an SDOF vibration, given that the response is predominantly vertical. To check this, pressure oscillations are recorded at all interior grid vertices shown in Fig.~\ref{fig:4_4_impactor_model}, excluding vertices which are located at the material boundaries, with pressures obtained by interpolation from surrounding material points. To avoid the initial irregular waveforms associated with impact and near-surface plastic deformation, only the last two seconds of the simulation ($t = 1$--$3$ s) are used for analysis. The extracted pressure data are then fitted to a decaying sinusoidal function, $\alpha_1 + \alpha_2 \sin(\alpha_3 + \alpha_4 t)\exp(-\alpha_5 t)$, where $\alpha_1$ denotes the steady-state pressure, $\alpha_2$ the oscillation amplitude, $\alpha_3$ a phase shift, $\alpha_4$ the oscillation frequency, and $\alpha_5$ the exponential decay rate (see Fig.~\ref{fig:4_4_fit_analysis}$(a)$). Consistent with the notation in Eq.~\eqref{eq:time_component}, $\alpha_4$ directly corresponds to the oscillation frequency $\tilde{\omega}$, while $\alpha_5$ relates to the attenuation rate $\tilde{\xi}/2$. The inferred $\tilde{\omega}$ and $\tilde{\xi}$ values are then normalized by the primary wave speed $c_p$ and a calibrated length-scale parameter $k^*$, which is calibrated separately for models 1 and 2 based on their characteristic domain sizes. As plotted in Fig.~\ref{fig:4_4_fit_analysis}$(b)$, the resulting data show good agreement with the frequency–damping relationship given in Eq.~\eqref{eq:damping_constants}, corresponding to a semicircular relation. As the damping rate increases, a slight reduction in oscillation frequency is observed. However, since the system remains in the underdamped regime, the frequency shift is minimal since $\tilde{\xi} \ll 2 k^* c_p$.

\subsection{Patterns in vibrated granular media}
\label{subsec:pattern}

In the final numerical example, we investigate a particularly interesting yet challenging problem in granular vibration. Experiments have shown that a thin layer of granular material, subjected to vertical vibrations, can develop distinct spatial patterns for certain combinations of driving frequency and amplitude \cite{melo1994transition, umbanhowar1996localized}. These patterns, along with the associated parameter ranges in frequency and amplitude, have been well characterized experimentally by \citet{bizon1998patterns}. Meanwhile, a number of discrete-particle simulations using both contact-dynamics (CD) approaches \cite{bizon1998patterns, smith2012reflections} and DEM \cite{moon2004role, watson20253d} have been performed to validate the experimental findings. Although discrete methods can relatively straightforwardly reproduce the emerging patterns, only limited studies have been attempted using continuum-based methods, most of which rely on simplified analytical treatments \cite{eggers1999continuum, venkataramani2001pattern} or on kinetic-theory-based hydrodynamic models that typically assume frictionless particles \cite{bougie2005onset, bougie2011continuum}. As \citet{bougie2005onset} pointed out, neglecting interparticle friction confines the analysis to only stripe patterns. Simulations without friction have not been able to reproduce the square and hexagonal patterns seen in experiments with frictional grains \cite{bizon1998patterns, moon2004role}. 

\begin{figure}[h!]
    \centering
    \includegraphics[width=0.6\linewidth]{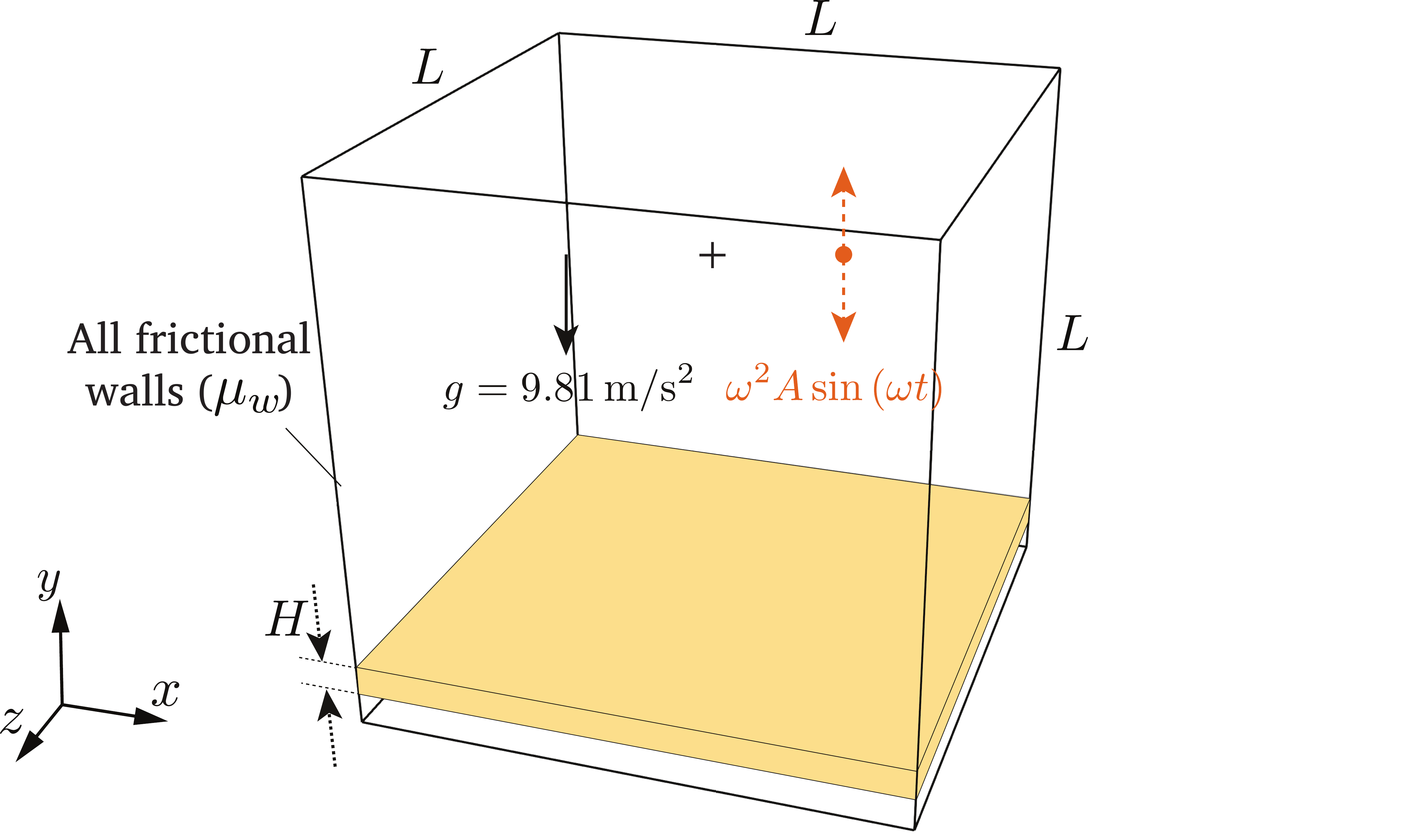}
    \caption{Patterns in vibrated granular media: initial geometry, loading, and boundary conditions.}
    \label{fig:4_5_model_pattern}
\end{figure}

Using the proposed continuum framework, which accounts for both the granular restitution and inertia-dependent friction, we seek to reproduce the granular vibration experiment of \citet{bizon1998patterns}. We initialize a thin granular continuum layer of thickness $H$ placed inside a box with side length $L$ subjected to a gravitational field $g=9.81$ m/s$^2$; see Fig.~\ref{fig:4_5_model_pattern}. The granular layer is modeled as a continuous medium that represents granular assembly with an average diameter of $d = 0.55$~mm. Both the box size and the layer thickness are prescribed in terms of $d$, namely $L=100d$ and $H=5.42d$, whereas the initial packing fraction is set as $\phi_0=0.58$. The box is then subjected to vertical sinusoidal oscillations with frequency $f$ and amplitude $A$. In our simulations, we model the vibration by applying a body acceleration instead of directly prescribing the motion of the bottom plate boundary, under the assumption that the reference frame moves with the plate. The resulting additional body acceleration is given by $\omega^2 A \sin{(\omega t)}$, where $\omega=2\pi f$ denotes the angular frequency. In this work, we focus on parameter values that generate square patterns. Following \citet{bizon1998patterns}, this is achieved by setting the dimensionless vibration frequency $f^*=f \sqrt{H/g}$ and amplitude $\Gamma=4 \pi^2 f^2 A /g$ to $f^*=0.27$ and $\Gamma=3$, respectively.

The granular material is modeled to match the experimental setup, which used lead beads. The solid density is set to $\rho_s = 11{,}000$~$\text{kg/m}^3$, and the bulk Young's modulus and Poisson's ratio are set to $E = 200$ MPa and $\nu = 0.3$. Several parameters of the viscoplastic model are kept identical to those used for the glass bead cases, specifically $\mu_2 = 0.6494$ and $I_0 = 0.3$, except the static friction coefficient, which is taken as $\mu_s = 0.4$. This value is slightly higher than that employed for glass beads and is determined from calibration. Such an increase in $\mu_s$ is reasonable given the greater interparticle friction of lead beads. The wall friction between the granular material and the container boundaries is also set equal to $\mu_s$, i.e., $\mu_w = 0.4$, for all six walls. The calibration study shows that increasing $\mu_s$ and $\mu_w$ decreases the maximum rebound height and delays the onset of pattern formation. The coefficient of restitution is chosen as $e = 0.3$, following the measurements of \citet{goldsmith2001impact} for spherical lead beads, for an impact speed of roughly 0.4 ft/s, consistent with the mean velocity magnitude measured in the simulations. We additionally examine a perfectly elastic scenario with $e = 1$ to demonstrate that pattern formation occurs only when viscoelastic dissipation is present.

The MPM discretization employs a structured grid composed of trilinear hexahedral elements with size $h = L/50$. We previously attempted to use a quadratic B-spline basis for this problem. However, the relatively larger kernel support of the B-spline function seems to impose artificial resistance on the material points during reconsolidation, which increases the average bed height and inhibits the development of spatial patterns. To initiate the emergence of the pattern, approximately 53,000 material points are seeded in an irregular arrangement. Specifically, a Poisson disk sampling algorithm following the approach proposed by \citet{bridson2007fast} is employed, along with a Voronoi tessellation scheme to determine their respective initial volumes. The variation in volume is constrained to within $\pm10\%$. The chosen time step is $\Delta t = 10^{-6}$ s, and the simulation is carried out up to $T = 10$ s.

\begin{figure}[h!]
    \centering
    \includegraphics[width=0.8\linewidth]{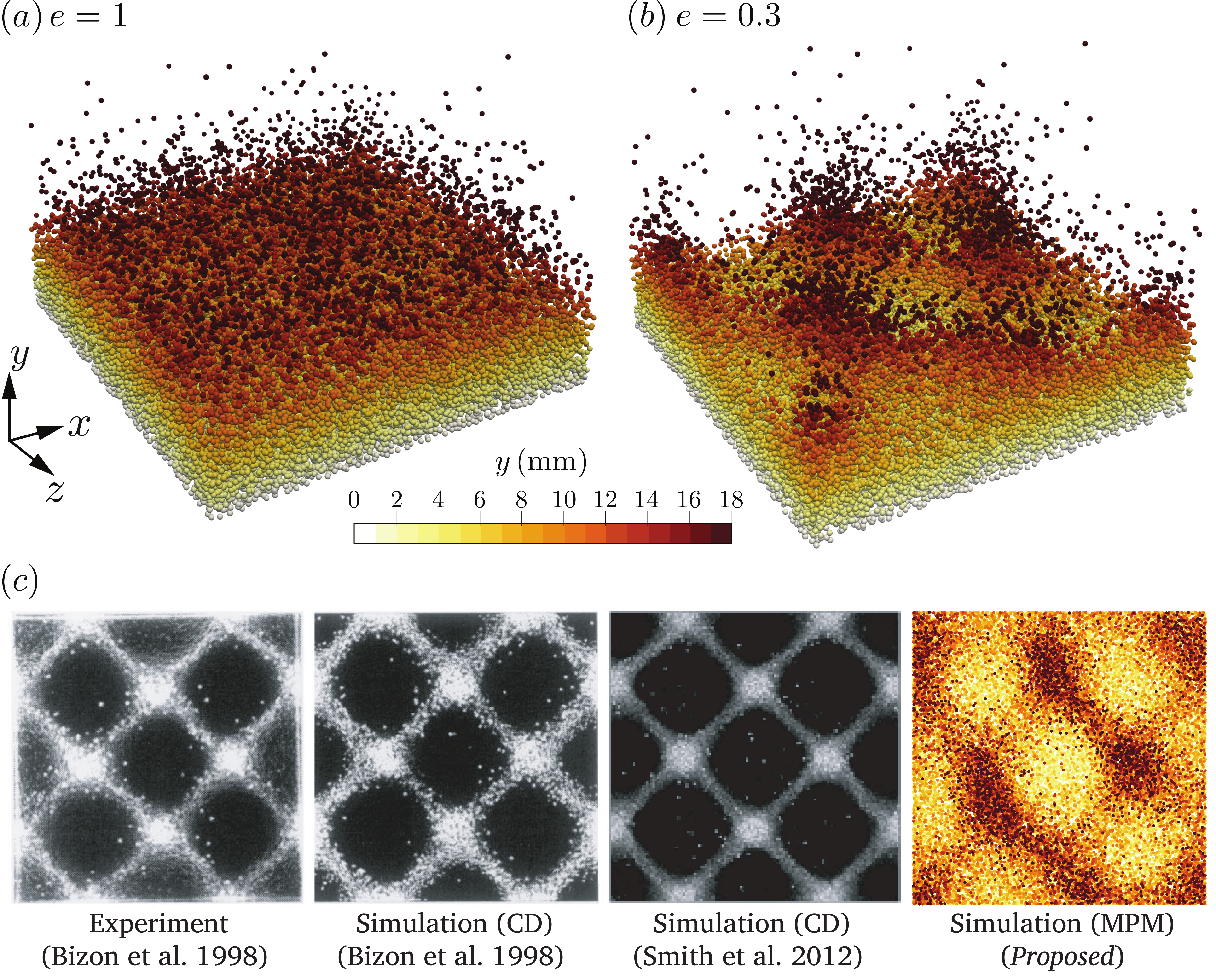}
    \caption{Patterns in vibrated granular media: Obtained continuum simulation results are shown for $(a)$ $e=1$ and $(b)$ $e=0.3$, corresponding respectively to cases without and with pattern formation under vibration. $(c)$ A top-view comparison is presented between the derived continuum-based solution (right-most) and both the experiment reported by \citet{bizon1998patterns} and the contact-dynamic-based (CD) simulations of \citet{bizon1998patterns} and \citet{smith2012reflections}.}
    \label{fig:4_5_pattern_compare}
\end{figure}

The steady-state numerical results for both $e=1$ and $e=0.3$ are shown in Fig.~\ref{fig:4_5_pattern_compare} for comparison. Under fully elastic conditions ($e=1$), which effectively replicate the method previously proposed by \citet{dunatunga2015continuum}, the simulations do not reproduce any pattern formation. In contrast, when viscoelastic damping is introduced with a suitable restitution coefficient, a square pattern develops after a period of sustained vibration. This pattern forms a stable square or diamond-shaped configuration with distinct cusps and troughs. A top-view comparison is presented in Fig.~\ref{fig:4_5_pattern_compare}$(c)$, together with corresponding experimental results and discrete simulations \cite{bizon1998patterns, smith2012reflections}. The resulting numerical simulation successfully reproduces the diamond-shaped pattern, potentially marking the first instance in which a continuum-based method has captured a stable diamond configuration in a granular vibration simulation. Animations illustrating the cases presented here are available in \cite{supplemental_material}.

While the obtained patterns agree well with experimental observations and discrete simulations, two points merit further investigation. First, as shown in Fig.~\ref{fig:4_5_pattern_compare}$(a)$ and $(b)$, the material points are observed to bounce higher than in simulations by \citet{smith2012reflections}. This behavior may be attributed to the constitutive assumption that the material becomes stress-free in the separated state, i.e.~$\rho < \rho_c$. In reality, even in the loose gaseous regime, granular materials continue to dissipate energy through dilute-state collisions in mid-air, leading to a finite cooling rate and the development of kinetic stresses, particularly during recontraction. In the present model, such kinetic stresses associated with granular temperature and inelastic collisions are neglected for both theoretical and numerical convenience \cite{dunatunga2015continuum}. However, improving the predictions may require introducing additional dissipation, for example, by adding viscous stresses in the gaseous state \cite{baumgarten2019general}.

Secondly, as shown in Fig.~\ref{fig:4_5_pattern_compare}$(c)$, the predicted pattern wavelength is slightly larger than that observed experimentally. This difference may be attributed, at least in part, to the treatment of the gaseous regime. In particular, the elevated bouncing height influences the oscillation period, which in turn stretches the wavelength in the lateral ($x$–$z$) directions. From a numerical standpoint, the observed discrepancy may also reflect the influence of spatial resolution. The present discretization employs on the order of $\sim$60,000 material points, comparable to discrete particle simulations, and thus already represents a computationally demanding configuration. Nevertheless, this resolution appears to be close to the minimum required to robustly capture the onset of the pattern, since coarser discretizations tend to suppress pattern formation by not sufficiently resolving the characteristic wavelength.  Notwithstanding these points, the MPM results show that the current continuum-based model still captures the key characteristics of the pattern at a usable resolution, with the square pattern appearing only when both the restitution coefficient and friction are accurately represented.

\section{Conclusions}
\label{sec:conclusion}

In this work, we developed a unified viscoelastic–viscoplastic continuum framework for granular materials that incorporates two sources of rate dependency: micro-inertia and viscoelastic damping. A closed-form relationship was first derived to link the granular coefficient of restitution to continuum viscosities, particularly with the bulk viscosity, which is further related to the shear viscosity through Rayleigh-stiffness-proportional damping. The derivation explicitly accounts for granular separation conditions, recognizing that dry granular media do not support tensile stress, and integrates naturally with the separation treatment proposed in our earlier work \cite{dunatunga2015continuum}. In addition, an implementation strategy was proposed to consistently integrate viscoelastic and viscoplastic responses through a unified constitutive stress update algorithm, ensuring that viscoelastic damping affects elastic wave propagation without altering plastic deformation, which remains governed solely by the inertial $\mu(I)$ rheology.

The proposed model was implemented within the MPM to address large-deformation continuum granular problems, and its performance was demonstrated through five numerical examples. A spherical compression impact test first verified the derived relationship between continuum restitution and bulk viscosity, forming the basis of the restitution–viscosity connection. An inclined-plane flow problem then confirmed that the inclusion of viscoelastic damping does not alter the analytical Bagnold velocity profile associated with the $\mu(I)$ rheology, consistent with previous observations from discrete element simulations. A flat-bottom silo flow example further highlighted the ability of the framework to introduce controlled dissipation during reconsolidation, leading to increased dynamic repose angles without affecting the discharge flow rate. Furthermore, simulations of a granular bed subjected to an impactor demonstrated effective attenuation of stress vibrations while preserving surface plastic deformation and ejecta formation around the impact crater. Finally, simulations of pattern formation induced by vibrating granular media are carried out, demonstrating that viscoelastic damping together with friction is essential to reproduce the stable square and diamond patterns, in agreement with experimental observations and simulations based on contact dynamics.


\section*{Acknowledgments}
The authors acknowledge the support from the U.S. National Science Foundation (NSF) under Award No.~2532010 and the Berkeley Department of Mechanical Engineering. This research used the Savio computational cluster resource provided by the Berkeley Research Computing program at the University of California, Berkeley (supported by the UC Berkeley Chancellor, Vice Chancellor for Research, and Chief Information Officer). 

\section*{Declaration of competing interest}
The authors declare that they have no conflict of interest.

\section*{Data availability statement}
The data that support the findings of this study are available from the first author upon reasonable request.

\appendix

\section{Material Point Method}
\label{app:mpm}

This appendix section discusses relevant discretization methods considered in single-phase MPM simulations. Two types of subscripts are introduced: $\square_p$ for variables associated with material point $p$, and $\square_I$ or $\square_J$ for variables allocated in nodes $I$ or $J$, respectively. The present MPM framework, along with its implementation, is accessible as open-source software on GitHub\footnote{https://github.com/geomechanics/mpm} and has been extensively developed and rigorously validated for a wide range of applications in both solid and fluid mechanics \citep{kularathna2021semi, liang2022shear, chandra2024stabilized, chandra2024mixed, kurima2025absorbing}.

\subsection{Discretization assumptions}
\label{subsec:mp_discretization}

In MPM, the geometry of body $\mathcal{B}$ is discretized into $N_p$ material points, or particles, of volume $V_p$ as (cf.~Fig.~\ref{fig:3_mpm}):
\begin{equation}
\Omega_\mathcal{B} = \int_\Omega\, \td \Omega \approx \sum_{p=1}^{N_p}V_p \,.
\end{equation}
These material points carry all the information, including masses, velocities, stresses, and history-dependent properties, throughout the entire simulation. The physical coordinate of these material points in $\mathcal{R}^3$ at time $t=n$ is defined as $\tb x^n_p$. Each of the material points carries a constant mass $m_p$, and thus, the overall mass of the system is conserved automatically. The mass density field of the continuum can be correspondingly defined by using the Dirac delta function $\delta$ as \cite{Sulsky1994}:
\begin{equation}
    \rho := \rho(\tb{x},t) = \sum_{p=1}^{N_p} m_p \delta (\tb{x}-\tb{x}_p)\,.
    \label{eq:2_density_field}
\end{equation}

Similar to FEM, MPM employs the variational formulation, i.e.~by applying the Galerkin approximation, the weak form of the momentum balance equation, Eq.~\eqref{eq:mass_momentum_balance}$_2$, is first derived as:
\begin{eqnarray}
R(\tb{u},\delta \tb{u}) \equiv
\int_\Omega  \tb{\sigma} : \nabla \delta \tb{u}\, \td \Omega - 
\int_\Omega  \rho \left( \tb{b} - {\ddot{\tb u}} \right) \cdot \delta \tb{u}\, \td \Omega - 
\int_{\Gamma_N} \bar{\tb{t}} \cdot \delta \tb{u}\, \td \Gamma = 0\,, \qquad \forall \delta \tb{u} \in \mathcal{V}_0\,.
\label{eq:momentum_balance_weak_residual}
\end{eqnarray}
Here, $\delta \tb{u}$ is an arbitrary \textit{test} or \textit{trial function}, where
$\delta \tb{u} = \left\lbrace \delta \tb{u} \in \mathcal{V}_0 | \delta \tb{u}=0 \, \mathrm{on} \, \Gamma_D \right\rbrace$, where $\mathcal{V}_0\in \mathcal{V}$ is the \textit{space of virtual displacements} and $\mathcal{V} \subset \mathcal{R}^{3}$ is the \textit{manifold of admissible configurations} defined as:
\begin{equation}
  \label{eq:manifold-admissible-configurations-model}
  \mathcal{V}\ \coloneqq\ \{ \varphi^{t} : \mathcal{B}_0 \rightarrow
  \mathcal{R}^{3} \mid J > 0\ \text{and} \ \tb{u}\mid_{\Gamma_D} = \bar{\tb{u}} \}\,.
\end{equation}
In this context, the motion function $\varphi$ defines the mapping $\varphi^t : \mathcal{B}_0 \rightarrow \mathcal{B}\in \mathcal{R}^3$, while $J=\det{(\tb F)}$ is the determinant of the local deformation gradient, which linearly transforms the reference configuration to the current configuration following:
\begin{equation}
\tb{F} \left( \tb{X}, t \right) := \mathrm{Grad} \left({\varphi}(\tb{X}, t)\right) =\frac{\pd {\varphi}(\tb{X}, t)}{\pd \tb{X}}\,.
\label{eq:def_grad}
\end{equation}
Here, $\tb X$ denotes the position vector at the reference configuration.

Following finite element discretization, we can approximate the particles' displacement field and its trial function from the corresponding nodal quantities, i.e.,
\begin{eqnarray}
\tb u_{p}:=\tb u (\tb x_p)=\sum_{I=1}^{n_n}S_{Ip} \tb u_{I}\,, \qquad 
\delta \tb u_{p}:=\delta \tb u (\tb x_p)=\sum_{I=1}^{n_n}S_{Ip} \delta  \tb u_{I}\,,
\end{eqnarray}
where $S_{Ip}\equiv S_I(\tb{x}_p)$ is a set of basis functions connecting material point $p$ and background node $I$.

The standard MPM considers an explicit time integration scheme, where the nodal acceleration $\tb{a}_J^{n+1}$ can be computed at each time step, invoking the arbitrariness of the test function $\delta \tb u_{I}$ as:
\begin{equation}
    \tb m_{IJ}^n \tb{a}_J^{n+1} = \tb{f}_I^{\mathrm{int},n} + \tb{f}_I^{\mathrm{ext},n}\,.
    \label{eq:2_explicit_time_integration}
\end{equation}
Here, the discretized consistent mass matrix $\tb m_{IJ}^n$, the internal force $\tb{f}_I^{\mathrm{int},n}$, and the external force $\tb{f}_{I}^{\mathrm{ext},n}$ are:
\begin{gather}
    \tb m_{IJ}^n = \sum_{p=1}^{n_p} S_{Ip} S_{Jp} m_p\,,
    \label{eq:mpm-mass-matrix}\\
    \tb{f}_I^{\mathrm{int},n} = - \sum_{p=1}^{n_p} \tb B^T_{Ip} {\tb \sigma}^{n+1}_p V_p^{n+1}\,,
    \label{eq:mpm-internal-force}\\
    \tb{f}_{I}^{\mathrm{ext},n} = \sum_{p=1}^{n_p} S_{Ip} \tb{b}_p m_p + \int_{ \Gamma_N} S_{I}(\tb x, n) \cdot \overline{\tb{t}}(\tb{x}, n)\, \td \Gamma\,.
    \label{eq:mpm-external-force}
\end{gather}
The variable $n_p$ denotes the number of particles associated with a selected node $I$ and/or $J$. For computational efficiency and numerical stability, the matrix $\tb m_{IJ}$ is often lumped in MPM, but at the expense of energy conservation. In the above, $\tb B_{Ip}$ is the \textit{deformation matrix} which essentially contains the shape function gradient arranged in the order of Voigt notation. Notice that the traction surface integral in Eq.~\eqref{eq:mpm-external-force} is left in its weak form as its discretized form may vary according to the considered discretization \cite{chandra2021nonconforming, liang2023imposition, liang2026virtual}. The material point stress $\tb \sigma$ can be updated by considering the constitutive relation explained previously in \Cref{sec:continuum} and \ref{sec:modeling}. In MPM literature, there are a few different schemes to update the material point stress \cite{bardenhagen2002energy}. In this study, the \textit{update-stress-first} (USF) scheme method is adopted, as it is energetically more accurate and exhibits less numerical diffusion \cite{zhang2016material}.

For constitutive purposes, the velocity gradient tensor $\tb L^{n+1}_p$ should be computed at each particle as: 
\begin{eqnarray}
    \tb L_p^{n+1} = \sum_{I=1}^{n_n} \nabla S_{Ip} \otimes \tb v_I^{n}\,.
    \label{eq:velocity_gradient}
\end{eqnarray}
The computed velocity gradient is also used to update the local volume of each material point. Assuming a constant value of $\tb L$ over each time step and considering that the material point mass remains constant, the analytical solution of Eq.~\eqref{eq:mass_momentum_balance}$_1$ allows us to update the material point volume as follows, which in turn also updates the effective mass density:
\begin{eqnarray}
    V_p^{n+1} = V_p^{n} \exp{(\Delta t \,\mathrm{tr}(\tb L_p^{n+1}))} \qquad \Rightarrow \qquad \rho^{n+1}_p = \phi^{n+1}_p \rho_s = \frac{m_p}{V_p^{n+1}}\,.
    \label{eq:initial_velocity}
\end{eqnarray}

In MPM, a fixed background grid is used to solve the aforementioned governing equations. Since no variables are stored in the nodes, information transfer from particles to background nodes is performed at the beginning of the time step. For example, nodal velocity and mass can be mapped as:
\begin{equation}
    \tb{v}_I^{n} = \frac{1}{m_I^n}\sum_{p=1}^{n_p} S_{Ip} m_p \tb v^n_p\,,\qquad m_I^n = \sum_{p=1}^{n_p} S_{Ip} m_p\,.
    \label{eq:initial_nodal_velocity}
\end{equation}
Furthermore, at the end of each step, when the solution of the balance equations is obtained at the nodes, they are also to be transferred back to the material points. The velocity and the position of material points at time $t = n+1$ can be updated following the FLIP (fluid-implicit-particle) method \cite{brackbill1986flip}:
\begin{eqnarray}
    \tb{v}_p^{n+1} = \tb{v}_p^{n}+ \Delta t \sum_{I=1}^{n_n} N_{Ip} \tb{a}_I^{n+1} \,, \qquad
     \tb{x}_p^{n+1} = \tb{x}_p^{n} + \Delta t \sum_{I=1}^{n_n} N_{Ip} \tb{v}_I^{n+1}\,,
     \label{eq:mpm-position-update}
\end{eqnarray}
where the updated nodal velocity is $\tb{v}_I^{n+1} = \tb{v}_I^{n} + \Delta t \tb{a}_I^{n+1}$. 

\subsection{Critical time increment}
\label{app:critical_time_step}

Incorporating rate-dependency into the continuum formulation modifies the stress wave propagation speed within the medium. Consequently, it becomes essential to re-establish the critical time increment through Von Neumann stability analysis. For the forward Euler explicit time integration scheme, we can derive the limiting criterion to be:
\begin{eqnarray}
    \Delta t_{cr} = h \sqrt{\frac{\rho }{M}} \left(\sqrt{{\widehat{\vartheta}}^2+1}-\widehat{\vartheta}\right)\,,
\end{eqnarray} 
where $h$ is the background mesh size and the non-dimensional viscosity $\widehat{\vartheta}$ is given as:
\begin{eqnarray}
    \widehat{\vartheta}=\frac{\vartheta}{h \sqrt{M \rho}}\,.
\end{eqnarray} 
By setting $\vartheta \rightarrow 0$, the classical Courant–Friedrichs–Lewy (CFL) condition for a purely elastic problem can be recovered. Conversely, as $\vartheta$ increases, the critical time step decreases, consistent with observations by \citet{belytschko1983overview}. Meanwhile, as $M \rightarrow 0$, $\Delta t_{cr}$ approaches the limiting criterion for pure viscous problems, i.e,
\begin{eqnarray}
    \lim_{M\rightarrow0} \Delta t_{cr}=\frac{{h}^2 \rho }{2 \vartheta }\,.
\end{eqnarray}
We can then pick a stable time step $\Delta t$ for the explicit scheme to be at least lower than $\Delta t_{cr}$, i.e.~$\Delta t\leq \Delta t_{cr}$. 

For computational efficiency and improved numerical stability, implicit versions of MPM are often employed in large-deformation solid and fluid simulations (e.g.~\cite{guilkey2003implicit, coombs2020lagrangian, chandra2024stabilized, chandra2024mixed}), where $\Delta t$ can be chosen to exceed $\Delta t_{cr}$, since the stability of implicit methods is no longer governed by the CFL condition. However, in the present formulation, excessively large $\Delta t$ values can reduce the convexity of the nonlinear problem, leading to slower convergence even when a discretization-consistent tangent matrix is used. Moreover, although large $\Delta t$ values remain numerically stable, they may fail to accurately capture the underlying physics in problems involving material separation and collision, where the elastic modulus abruptly transitions from a finite value to zero once $\rho < \rho_c$ (cf.~Eq.~\eqref{eq:pressure_separation_cond}). Therefore, a sufficiently small $\Delta t$ is still required to ensure physical accuracy in resolving such collision events. Considering these aspects, the explicit time integration scheme remains the most cost-effective and robust choice, and is therefore adopted in the present study.

\bibliography{mybibfile}

@article{Sulsky1994,
  title   = "A particle method for history-dependent materials",
  journal = "Computer Methods in Applied Mechanics and Engineering",
  volume  = "118",
  pages   = "179--196",
  year    = "1994",
  author  = "Sulsky, D and Chenb, Z and Schreyer, H L"
}

@phdthesis{kamrin2008stochastic,
  title={Stochastic and deterministic models for dense granular flow},
  author={Kamrin, Kenneth Norman},
  year={2008},
  school={Massachusetts Institute of technology}
}

@article{brey1998hydrodynamics,
  title={Hydrodynamics for granular flow at low density},
  author={Brey, J Javier and Dufty, James W and Kim, Chang Sub and Santos, Andr{\'e}s},
  journal={Physical Review E},
  volume={58},
  number={4},
  pages={4638},
  year={1998},
  publisher={American Physical Society}
}

@article{garzo1999dense,
  title={Dense fluid transport for inelastic hard spheres},
  author={Garz{\'o}, V and Dufty, JW},
  journal={Physical Review E},
  volume={59},
  number={5},
  pages={5895},
  year={1999},
  publisher={APS}
}

@article{lun1984kinetic,
  title={Kinetic theories for granular flow: inelastic particles in Couette flow and slightly inelastic particles in a general flowfield},
  author={Lun, Cli KK and Savage, Stuart B and Jeffrey, DJ and Chepurniy, Nicholas},
  journal={Journal of fluid mechanics},
  volume={140},
  pages={223--256},
  year={1984},
  publisher={Cambridge University Press}
}

@incollection{jenkins1985grad,
  title={Grad’s 13-moment system for a dense gas of inelastic spheres},
  author={Jenkins, JT and Richman, MW},
  booktitle={The Breadth and Depth of Continuum Mechanics: A Collection of Papers Dedicated to JL Ericksen on His Sixtieth Birthday},
  pages={647--669},
  year={1985},
  publisher={Springer}
}

@book{chapman1990mathematical,
  title={The mathematical theory of non-uniform gases: an account of the kinetic theory of viscosity, thermal conduction and diffusion in gases},
  author={Chapman, Sydney and Cowling, Thomas George},
  year={1990},
  publisher={Cambridge university press}
}

@book{brilliantov2010kinetic,
  title={Kinetic theory of granular gases},
  author={Brilliantov, Nikolai V and P{\"o}schel, Thorsten},
  year={2010},
  publisher={Oxford University Press}
}

@article{kuwabara1987restitution,
  title={Restitution coefficient in a collision between two spheres},
  author={Kuwabara, Goro and Kono, Kimitoshi},
  journal={Japanese journal of applied physics},
  volume={26},
  number={8R},
  pages={1230},
  year={1987},
  publisher={IOP Publishing}
}

@article{brilliantov1996model,
  title={Model for collisions in granular gases},
  author={Brilliantov, Nikolai V and Spahn, Frank and Hertzsch, Jan-Martin and P{\"o}schel, Thorsten},
  journal={Physical review E},
  volume={53},
  number={5},
  pages={5382},
  year={1996},
  publisher={APS}
}

@article{schwager2007coefficient,
  title={Coefficient of restitution and linear--dashpot model revisited},
  author={Schwager, Thomas and P{\"o}schel, Thorsten},
  journal={Granular Matter},
  volume={9},
  number={6},
  pages={465--469},
  year={2007},
  publisher={Springer}
}

@article{dunatunga2017continuum,
  title={Continuum modeling of projectile impact and penetration in dry granular media},
  author={Dunatunga, Sachith and Kamrin, Ken},
  journal={Journal of the Mechanics and Physics of Solids},
  volume={100},
  pages={45--60},
  year={2017},
  publisher={Elsevier}
}

@article{kurima2025absorbing,
  title={Absorbing boundary conditions in material point method adopting perfectly matched layer theory},
  author={Kurima, Jun and Chandra, Bodhinanda and Soga, Kenichi},
  journal={Soil Dynamics and Earthquake Engineering},
  volume={191},
  pages={109219},
  year={2025},
  publisher={Elsevier}
}

@article{jop2006constitutive,
  title={A constitutive law for dense granular flows},
  author={Jop, Pierre and Forterre, Yo{\"e}l and Pouliquen, Olivier},
  journal={Nature},
  volume={441},
  number={7094},
  pages={727--730},
  year={2006},
  publisher={Nature Publishing Group UK London}
}

@article{tsuji1992lagrangian,
  title={Lagrangian numerical simulation of plug flow of cohesionless particles in a horizontal pipe},
  author={Tsuji, Yutaka and Tanaka, Toshitsugu and Ishida, T},
  journal={Powder technology},
  volume={71},
  number={3},
  pages={239--250},
  year={1992},
  publisher={Elsevier}
}

@article{bridges1984structure,
  title={Structure, stability and evolution of Saturn's rings},
  author={Bridges, Frank G and Hatzes, A and Lin, DNC},
  journal={Nature},
  volume={309},
  number={5966},
  pages={333--335},
  year={1984},
  publisher={Nature Publishing Group UK London}
}

@article{schwager2008coefficient,
  title={Coefficient of restitution for viscoelastic spheres: The effect of delayed recovery},
  author={Schwager, Thomas and P{\"o}schel, Thorsten},
  journal={Physical Review E—Statistical, Nonlinear, and Soft Matter Physics},
  volume={78},
  number={5},
  pages={051304},
  year={2008},
  publisher={APS}
}

@book{rayleigh1896theory,
  title={The theory of sound},
  author={Rayleigh, John William Strutt Baron},
  volume={2},
  year={1896},
  publisher={Macmillan}
}

@article{caughey1960classical,
  title={Classical normal modes in damped linear dynamic systems},
  author={Caughey, TK},
  journal={Journal of applied mechanics},
  volume={27},
  number={2},
  pages={269--271},
  year={1960},
  publisher={ASME International}
}

@article{silbert2001granular,
  title={Granular flow down an inclined plane: Bagnold scaling and rheology},
  author={Silbert, Leonardo E and Erta{\c{s}}, Deniz and Grest, Gary S and Halsey, Thomas C and Levine, Dov and Plimpton, Steven J},
  journal={Physical Review E},
  volume={64},
  number={5},
  pages={051302},
  year={2001},
  publisher={APS}
}

@article{henann2013small,
  title={Small-amplitude acoustics in bulk granular media},
  author={Henann, David L and Valenza, John J and Johnson, David L and Kamrin, Ken},
  journal={Physical Review E—Statistical, Nonlinear, and Soft Matter Physics},
  volume={88},
  number={4},
  pages={042205},
  year={2013},
  publisher={APS}
}

@article{belytschko1983overview,
  title={An overview of semidiscretization and time integration procedures},
  author={Belytschko, Ted},
  journal={Computational methods for transient analysis(A 84-29160 12-64). Amsterdam, North-Holland, 1983,},
  pages={1--65},
  year={1983}
}

@article{hunt1975coefficient,
    author = {Hunt, K. H. and Crossley, F. R. E.},
    title = {Coefficient of Restitution Interpreted as Damping in Vibroimpact},
    journal = {Journal of Applied Mechanics},
    volume = {42},
    number = {2},
    pages = {440-445},
    year = {1975},
    month = {06},
    issn = {0021-8936},
    doi = {10.1115/1.3423596},
    url = {https://doi.org/10.1115/1.3423596},
    eprint = {https://asmedigitalcollection.asme.org/appliedmechanics/article-pdf/42/2/440/5454660/440\_1.pdf},
}

@article{butcher2000characterizing,
  title={Characterizing damping and restitution in compliant impacts via modified KV and higher-order linear viscoelastic models},
  author={Butcher, EA and Segalman, DJ},
  journal={J. Appl. Mech.},
  volume={67},
  number={4},
  pages={831--834},
  year={2000}
}

@book{stronge2018impact,
  title={Impact mechanics},
  author={Stronge, William James},
  year={2018},
  publisher={Cambridge university press}
}

@article{cook1986newton,
  title={Newton’s ‘experimental’law of impacts},
  author={Cook, Ian},
  journal={The Mathematical Gazette},
  volume={70},
  number={452},
  pages={107--114},
  year={1986},
  publisher={Cambridge University Press}
}

@book{newton1803,
  author    = {Isaac Newton},
  title     = {The Mathematical Principles of Natural Philosophy},
  translator = {Andrew Motte},
  year      = {1803},
  publisher = {London},
  note      = {Translated from the Latin},
}

@article{ismail2008impact,
    author = {Ismail, K. A. and Stronge, W. J.},
    title = {Impact of Viscoplastic Bodies: Dissipation and Restitution},
    journal = {Journal of Applied Mechanics},
    volume = {75},
    number = {6},
    pages = {061011},
    year = {2008},
    month = {08},
    issn = {0021-8936},
    doi = {10.1115/1.2965371},
    url = {https://doi.org/10.1115/1.2965371},
    eprint = {https://asmedigitalcollection.asme.org/appliedmechanics/article-pdf/75/6/061011/5476870/061011\_1.pdf},
}

@article{da2005rheophysics,
  title={Rheophysics of dense granular materials: Discrete simulation of plane shear flows},
  author={Da Cruz, Fr{\'e}d{\'e}ric and Emam, Sacha and Prochnow, Micha{\"e}l and Roux, Jean-No{\"e}l and Chevoir, Fran{\c{c}}ois},
  journal={Physical Review E—Statistical, Nonlinear, and Soft Matter Physics},
  volume={72},
  number={2},
  pages={021309},
  year={2005},
  publisher={APS}
}

@book{simo2006computational,
  title={Computational inelasticity},
  author={Simo, Juan C and Hughes, Thomas JR},
  volume={7},
  year={2006},
  publisher={Springer Science \& Business Media}
}

@article{agarwal2021efficacy,
  title={Efficacy of simple continuum models for diverse granular intrusions},
  author={Agarwal, Shashank and Karsai, Andras and Goldman, Daniel I and Kamrin, Ken},
  journal={Soft Matter},
  volume={17},
  number={30},
  pages={7196--7209},
  year={2021},
  publisher={Royal Society of Chemistry}
}

@article{becker2008coefficient,
  title={Coefficient of tangential restitution for the linear dashpot model},
  author={Becker, Volker and Schwager, Thomas and P{\"o}schel, Thorsten},
  journal={Physical Review E—Statistical, Nonlinear, and Soft Matter Physics},
  volume={77},
  number={1},
  pages={011304},
  year={2008},
  publisher={APS}
}

@article{brilliantov1998rolling,
  title={Rolling friction of a viscous sphere on a hard plane},
  author={Brilliantov, Nikolai V and P{\"o}schel, Thorsten},
  journal={Europhysics Letters},
  volume={42},
  number={5},
  pages={511},
  year={1998},
  publisher={IOP Publishing}
}

@article{glielmo2014coefficient,
  title={Coefficient of restitution of aspherical particles},
  author={Glielmo, Aldo and Gunkelmann, Nina and P{\"o}schel, Thorsten},
  journal={Physical Review E},
  volume={90},
  number={5},
  pages={052204},
  year={2014},
  publisher={American Physical Society}
}

@article{schwager2008coefficientb,
  title={Coefficient of tangential restitution for viscoelastic spheres},
  author={Schwager, Thomas and Becker, Volker and P{\"o}schel, Thorsten},
  journal={The European Physical Journal E},
  volume={27},
  number={1},
  pages={107--114},
  year={2008},
  publisher={Springer}
}

@article{feng2025material,
  title={Material Point Method Modeling of Granular Flow Considering Phase Transition From Solid-Like to Fluid-Like States},
  author={Feng, Hang and Liang, Weijian and Yin, Zhen-Yu and Hu, Liming},
  journal={International Journal for Numerical and Analytical Methods in Geomechanics},
  volume={49},
  number={6},
  pages={1642--1664},
  year={2025},
  publisher={Wiley Online Library}
}

@article{reynolds1885lvii,
  title={LVII. On the dilatancy of media composed of rigid particles in contact. With experimental illustrations},
  author={Reynolds, Osborne},
  journal={The London, Edinburgh, and Dublin Philosophical Magazine and Journal of Science},
  volume={20},
  number={127},
  pages={469--481},
  year={1885},
  publisher={Taylor \& Francis}
}

@article{pouliquen2006flow,
  title={Flow of dense granular material: towards simple constitutive laws},
  author={Pouliquen, Olivier and Cassar, Cyril and Jop, Pierre and Forterre, Yoel and Nicolas, Maxime},
  journal={Journal of Statistical Mechanics: Theory and Experiment},
  volume={2006},
  number={07},
  pages={P07020},
  year={2006},
  publisher={IOP Publishing}
}

@article{wang2015experimental,
  title={Experimental determination of parameter effects on the coefficient of restitution of differently shaped maize in three-dimensions},
  author={Wang, Lijun and Zhou, Wenxiu and Ding, Zhenjun and Li, Xingxing and Zhang, Chuangen},
  journal={Powder Technology},
  volume={284},
  pages={187--194},
  year={2015},
  publisher={Elsevier}
}

@article{hastie2013experimental,
  title={Experimental measurement of the coefficient of restitution of irregular shaped particles impacting on horizontal surfaces},
  author={Hastie, DB},
  journal={Chemical Engineering Science},
  volume={101},
  pages={828--836},
  year={2013},
  publisher={Elsevier}
}

@misc{MATLAB2024,
  author       = {{The MathWorks, Inc.}},
  title        = {MATLAB Statistics and Machine Learning Toolbox},
  howpublished = {\url{https://www.mathworks.com/products/statistics.html}},
  year         = {2024},
  address      = {Natick, Massachusetts},
  note         = {Version R2024b}
}

@article{gdr2004dense,
  title={On dense granular flows},
  author={GDR MiDi gdrmidi@ polytech. univ-mrs. fr http://www. lmgc. univ-montp2. fr/MIDI/},
  journal={The European Physical Journal E},
  volume={14},
  pages={341--365},
  year={2004},
  publisher={Springer}
}

@article{anand2005theory,
  title={A theory for amorphous viscoplastic materials undergoing finite deformations, with application to metallic glasses},
  author={Anand, L and Su, C},
  journal={Journal of the Mechanics and Physics of Solids},
  volume={53},
  number={6},
  pages={1362--1396},
  year={2005},
  publisher={Elsevier}
}

@book{gurtin2010mechanics,
  title={The mechanics and thermodynamics of continua},
  author={Gurtin, Morton E and Fried, Eliot and Anand, Lallit},
  year={2010},
  publisher={Cambridge university press}
}

@article{liang2022shear,
  title={Shear band evolution and post-failure simulation by the extended material point method (XMPM) with localization detection and frictional self-contact},
  author={Liang, Yong and Chandra, Bodhinanda and Soga, Kenichi},
  journal={Computer Methods in Applied Mechanics and Engineering},
  volume={390},
  pages={114530},
  year={2022},
  publisher={Elsevier}
}

@book{zhang2016material,
  title={The material point method: a continuum-based particle method for extreme loading cases},
  author={Zhang, Xiong and Chen, Zhen and Liu, Yan},
  year={2016},
  publisher={Academic Press}
}

@article{brackbill1986flip,
  title={FLIP: A method for adaptively zoned, particle-in-cell calculations of fluid flows in two dimensions},
  author={Brackbill, Jeremiah U and Ruppel, Hans M},
  journal={Journal of Computational physics},
  volume={65},
  number={2},
  pages={314--343},
  year={1986},
  publisher={Elsevier}
}

@article{chandra2021nonconforming,
  title={Nonconforming Dirichlet boundary conditions in implicit material point method by means of penalty augmentation},
  author={Chandra, Bodhinanda and Singer, Veronika and Teschemacher, Tobias and Wuechner, Roland and Larese, Antonia},
  journal={Acta Geotechnica},
  volume={16},
  number={8},
  pages={2315--2335},
  year={2021},
  publisher={Springer}
}

@article{guilkey2003implicit,
  title={Implicit time integration for the material point method: Quantitative and algorithmic comparisons with the finite element method},
  author={Guilkey, James Edward and Weiss, Jeffrey A},
  journal={International Journal for Numerical Methods in Engineering},
  volume={57},
  number={9},
  pages={1323--1338},
  year={2003},
  publisher={Wiley Online Library}
}

@article{nakamura2023taylor,
  title={Taylor particle-in-cell transfer and kernel correction for material point method},
  author={Nakamura, Keita and Matsumura, Satoshi and Mizutani, Takaaki},
  journal={Computer Methods in Applied Mechanics and Engineering},
  volume={403},
  pages={115720},
  year={2023},
  publisher={Elsevier}
}

@article{bardenhagen2002energy,
  title={Energy conservation error in the material point method for solid mechanics},
  author={Bardenhagen, SG},
  journal={Journal of Computational Physics},
  volume={180},
  number={1},
  pages={383--403},
  year={2002},
  publisher={Elsevier}
}

@article{morozov2015relation,
  title={On the relation between bulk and shear seismic dissipation},
  author={Morozov, Igor B},
  journal={Bulletin of the Seismological Society of America},
  volume={105},
  number={6},
  pages={3180--3188},
  year={2015},
  publisher={Seismological Society of America}
}

@article{marveggio2022phase,
  title={Phase transition in monodisperse granular materials: how to model it by using a strain hardening visco-elastic-plastic constitutive relationship},
  author={Marveggio, Pietro and Redaelli, Irene and di Prisco, Claudio},
  journal={International Journal for Numerical and Analytical Methods in Geomechanics},
  volume={46},
  number={13},
  pages={2415--2445},
  year={2022},
  publisher={Wiley Online Library}
}

@article{haeri2022three,
  title={Three-dimensionsal granular flow continuum modeling via material point method with hyperelastic nonlocal granular fluidity},
  author={Haeri, Amin and Skonieczny, Krzysztof},
  journal={Computer Methods in Applied Mechanics and Engineering},
  volume={394},
  pages={114904},
  year={2022},
  publisher={Elsevier}
}

@article{dunatunga2022modelling,
  title={Modelling silo clogging with non-local granular rheology},
  author={Dunatunga, Sachith and Kamrin, Ken},
  journal={Journal of Fluid Mechanics},
  volume={940},
  pages={A14},
  year={2022},
  publisher={Cambridge University Press}
}

@article{seyedan2021solid,
  title={From solid to disconnected state and back: Continuum modelling of granular flows using material point method},
  author={Seyedan, Seyedmohammadjavad and So{\l}owski, Wojciech T},
  journal={Computers \& Structures},
  volume={251},
  pages={106545},
  year={2021},
  publisher={Elsevier}
}

@article{marveggio2024granular,
  title={Granular material regime transitions during high energy impacts of dry flowing masses: MPM simulations with a multi-regime constitutive model},
  author={Marveggio, Pietro and Zerbi, Matteo and Redaelli, Irene and di Prisco, Claudio},
  journal={International Journal for Numerical and Analytical Methods in Geomechanics},
  volume={48},
  number={15},
  pages={3699--3724},
  year={2024},
  publisher={Wiley Online Library}
}

@article{sadd2000simulation,
  title={DEM simulation of wave propagation in granular materials},
  author={Sadd, Martin H and Adhikari, Gautam and Cardoso, Francisco},
  journal={Powder Technology},
  volume={109},
  number={1-3},
  pages={222--233},
  year={2000},
  publisher={Elsevier}
}

@misc{supplemental_material,
  title        = {},
  year         = {},
  howpublished = {},
  note  = {See the supplemental material at {[}URL{]} for rendered animations of the flat-bottom silo flow, the granular bed subjected to an impactor, and the patterns formed in vibrated granular media (Sections 5.3--5.5).}
}

@article{bridson2007fast,
  title={Fast Poisson disk sampling in arbitrary dimensions.},
  author={Bridson, Robert},
  journal={SIGGRAPH sketches},
  volume={10},
  number={1},
  pages={1},
  year={2007}
}

@article{gu2020discrete,
  title={Discrete element modeling of shear wave propagation using bender elements in confined granular materials of different grain sizes},
  author={Gu, Xiaoqiang and Liang, Xiaomin and Shan, Yao and Huang, Xin and Tessari, Anthony},
  journal={Computers and Geotechnics},
  volume={125},
  pages={103672},
  year={2020},
  publisher={Elsevier}
}

@article{marketos2013micromechanics,
  title={A micromechanics-based analytical method for wave propagation through a granular material},
  author={Marketos, G and O’Sullivan, C},
  journal={Soil Dynamics and Earthquake Engineering},
  volume={45},
  pages={25--34},
  year={2013},
  publisher={Elsevier}
}

@article{liang2026virtual,
  title={The Virtual Stress Boundary Method to Impose Nonconforming Neumann Boundary Conditions in the High-Order Material Point Method},
  author={Liang, Yong and Given, Joel and Chandra, Bodhinanda and Zeng, Zhixin and Zhang, Xiong and Soga, Kenichi},
  journal={International Journal for Numerical Methods in Engineering},
  volume={127},
  number={5},
  pages={e70291},
  year={2026},
  publisher={Wiley Online Library}
}

@article{henann2013predictive,
  title={A predictive, size-dependent continuum model for dense granular flows},
  author={Henann, David L and Kamrin, Ken},
  journal={Proceedings of the National Academy of Sciences},
  volume={110},
  number={17},
  pages={6730--6735},
  year={2013},
  publisher={National Academy of Sciences}
}

@article{kamrin2012nonlocal,
  title={Nonlocal constitutive relation for steady granular flow},
  author={Kamrin, Ken and Koval, Georg},
  journal={Physical review letters},
  volume={108},
  number={17},
  pages={178301},
  year={2012},
  publisher={APS}
}

@article{santos2016investigation,
  title={Investigation of particle dynamics in a rotary drum by means of experiments and numerical simulations using DEM},
  author={Santos, Dyrney A and Barrozo, Marcos AS and Duarte, Claudio R and Weigler, Fabian and Mellmann, Jochen},
  journal={Advanced Powder Technology},
  volume={27},
  number={2},
  pages={692--703},
  year={2016},
  publisher={Elsevier}
}

@book{goldsmith2001impact,
  title={Impact},
  author={Goldsmith, Werner},
  year={2001},
  publisher={Courier Corporation}
}

@article{coetzee2016calibration,
  title={Calibration of the discrete element method and the effect of particle shape},
  author={Coetzee, Corn{\'e} J},
  journal={Powder Technology},
  volume={297},
  pages={50--70},
  year={2016},
  publisher={Elsevier}
}

@article{fern2016role,
  title={The role of constitutive models in MPM simulations of granular column collapses},
  author={Fern, Elliot James and Soga, Kenichi},
  journal={Acta Geotechnica},
  volume={11},
  number={3},
  pages={659--678},
  year={2016},
  publisher={Springer}
}

@article{baumgarten2019generalb,
  title={A general constitutive model for dense, fine-particle suspensions validated in many geometries},
  author={Baumgarten, Aaron S and Kamrin, Ken},
  journal={Proceedings of the National Academy of Sciences},
  volume={116},
  number={42},
  pages={20828--20836},
  year={2019},
  publisher={National Academy of Sciences}
}

@article{wikeckowski2004material,
  title={The material point method in large strain engineering problems},
  author={Wi{\k{e}}ckowski, Zdzis{\l}aw},
  journal={Computer methods in applied mechanics and engineering},
  volume={193},
  number={39-41},
  pages={4417--4438},
  year={2004},
  publisher={Elsevier}
}

@inproceedings{andersen2009analysis,
  title={Analysis of stress updates in the material-point method},
  author={Andersen, S{\o}ren and Andersen, Lars},
  booktitle={The Nordic Seminar on Computational Mechanics},
  pages={129--134},
  year={2009},
  organization={Department of Civil Engineering, Aalborg University}
}

@article{mast2015simulating,
  title={Simulating granular column collapse using the material point method},
  author={Mast, Carter M and Arduino, Pedro and Mackenzie-Helnwein, Peter and Miller, Gregory R},
  journal={Acta Geotechnica},
  volume={10},
  number={1},
  pages={101--116},
  year={2015},
  publisher={Springer}
}

@article{xiao2025sensitivity,
  title={Sensitivity analysis on critical combinations of input parameters in DEM granular flow analysis},
  author={Xiao, Junsen and Tozato, Kenta and Nomura, Reika and Otake, Yu and Terada, Kenjiro and Moriguchi, Shuji},
  journal={Acta Geotechnica},
  volume={20},
  number={1},
  pages={387--412},
  year={2025},
  publisher={Springer}
}

@article{kamrin2010nonlinear,
  title={Nonlinear elasto-plastic model for dense granular flow},
  author={Kamrin, Ken},
  journal={International Journal of Plasticity},
  volume={26},
  number={2},
  pages={167--188},
  year={2010},
  publisher={Elsevier}
}

@article{staron2012granular,
  title={The granular silo as a continuum plastic flow: The hour-glass vs the clepsydra},
  author={Staron, Lydie and Lagr{\'e}e, P-Y and Popinet, St{\'e}phane},
  journal={Physics of Fluids},
  volume={24},
  number={10},
  year={2012},
  publisher={AIP Publishing}
}

@article{wei2019numerical,
  title={Numerical and experimental studies of the effect of iron ore particle shape on repose angle and porosity of a heap},
  author={Wei, Han and Tang, Xiaojiu and Ge, Yao and Li, Meng and Sax{\'e}n, Henrik and Yu, Yaowei},
  journal={Powder Technology},
  volume={353},
  pages={526--534},
  year={2019},
  publisher={Elsevier}
}

@article{cundall1979discrete,
  title={A discrete numerical model for granular assemblies},
  author={Cundall, Peter A and Strack, Otto DL},
  journal={geotechnique},
  volume={29},
  number={1},
  pages={47--65},
  year={1979},
  publisher={Thomas Telford Ltd}
}

@article{staron2014continuum,
  title={Continuum simulation of the discharge of the granular silo: a validation test for the $\mu$ (I) visco-plastic flow law},
  author={Staron, L and Lagr{\'e}e, P-Y and Popinet, S},
  journal={The European Physical Journal E},
  volume={37},
  number={1},
  pages={5},
  year={2014},
  publisher={Springer}
}

@article{staron2007spreading,
  title={The spreading of a granular mass: role of grain properties and initial conditions},
  author={Staron, Lydie and Hinch, EJ},
  journal={Granular Matter},
  volume={9},
  number={3},
  pages={205--217},
  year={2007},
  publisher={Springer}
}

@article{coombs2020lagrangian,
  title={On Lagrangian mechanics and the implicit material point method for large deformation elasto-plasticity},
  author={Coombs, William M and Augarde, Charles E and Brennan, Andrew J and Brown, Michael J and Charlton, Tim J and Knappett, Jonathan A and Motlagh, Yousef Ghaffari and Wang, Lei},
  journal={Computer Methods in Applied Mechanics and Engineering},
  volume={358},
  pages={112622},
  year={2020},
  publisher={Elsevier}
}

@article{yue2018hybrid,
  title={Hybrid grains: Adaptive coupling of discrete and continuum simulations of granular media},
  author={Yue, Yonghao and Smith, Breannan and Chen, Peter Yichen and Chantharayukhonthorn, Maytee and Kamrin, Ken and Grinspun, Eitan},
  journal={ACM Transactions on Graphics (TOG)},
  volume={37},
  number={6},
  pages={1--19},
  year={2018},
  publisher={ACM New York, NY, USA}
}

@article{dunatunga2015continuum,
  title={Continuum modelling and simulation of granular flows through their many phases},
  author={Dunatunga, Sachith and Kamrin, Ken},
  journal={Journal of Fluid Mechanics},
  volume={779},
  pages={483--513},
  year={2015},
  publisher={Cambridge University Press}
}

@article{baumgarten2019general,
  title={A general fluid--sediment mixture model and constitutive theory validated in many flow regimes},
  author={Baumgarten, Aaron S and Kamrin, Ken},
  journal={Journal of Fluid Mechanics},
  volume={861},
  pages={721--764},
  year={2019},
  publisher={Cambridge University Press}
}

@article{liang2023imposition,
  title={The imposition of nonconforming Neumann boundary condition in the material point method without boundary representation},
  author={Liang, Yong and Given, Joel and Soga, Kenichi},
  journal={Computer Methods in Applied Mechanics and Engineering},
  volume={404},
  pages={115785},
  year={2023},
  publisher={Elsevier}
}

@article{kularathna2021semi,
  title={A semi-implicit material point method based on fractional-step method for saturated soil},
  author={Kularathna, Shyamini and Liang, Weijian and Zhao, Tianchi and Chandra, Bodhinanda and Zhao, Jidong and Soga, Kenichi},
  journal={International Journal for Numerical and Analytical Methods in Geomechanics},
  volume={45},
  number={10},
  pages={1405--1436},
  year={2021},
  publisher={Wiley Online Library}
}

@article{venkataramani2001pattern,
  title={Pattern selection in extended periodically forced systems: A continuum coupled map approach},
  author={Venkataramani, Shankar C and Ott, Edward},
  journal={Physical Review E},
  volume={63},
  number={4},
  pages={046202},
  year={2001},
  publisher={APS}
}

@article{bougie2011continuum,
  title={Continuum simulations of shocks and patterns in vertically oscillated granular layers},
  author={Bougie, J and Duckert, K},
  journal={Physical Review E—Statistical, Nonlinear, and Soft Matter Physics},
  volume={83},
  number={1},
  pages={011303},
  year={2011},
  publisher={APS}
}

@article{bougie2005onset,
  title={Onset of patterns in an oscillated granular layer: continuum and molecular dynamics simulations},
  author={Bougie, Jonathan and Kreft, Jennifer and Swift, Jack B and Swinney, Harry L},
  journal={Physical Review E—Statistical, Nonlinear, and Soft Matter Physics},
  volume={71},
  number={2},
  pages={021301},
  year={2005},
  publisher={APS}
}

@article{eggers1999continuum,
  title={Continuum description of vibrated sand},
  author={Eggers, Jens and Riecke, Hermann},
  journal={Physical Review E},
  volume={59},
  number={4},
  pages={4476},
  year={1999},
  publisher={APS}
}

@article{moon2004role,
  title={Role of friction in pattern formation in oscillated granular layers},
  author={Moon, Sung Joon and Swift, JB and Swinney, Harry L},
  journal={Physical Review E},
  volume={69},
  number={3},
  pages={031301},
  year={2004},
  publisher={APS}
}

@article{watson20253d,
  title={3D waveforms and patterning behavior in thin monodisperse and multidisperse vertically-vibrated layers},
  author={Watson, Peter and Bonnieu, Sebastien Vincent and Anwar, Ali and Lappa, Marcello},
  journal={Granular Matter},
  volume={27},
  number={1},
  pages={19},
  year={2025},
  publisher={Springer}
}

@article{umbanhowar1996localized,
  title={Localized excitations in a vertically vibrated granular layer},
  author={Umbanhowar, Paul B and Melo, Francisco and Swinney, Harry L},
  journal={Nature},
  volume={382},
  number={6594},
  pages={793--796},
  year={1996},
  publisher={Nature Publishing Group UK London}
}

@article{melo1994transition,
  title={Transition to parametric wave patterns in a vertically oscillated granular layer},
  author={Melo, Francisco and Umbanhowar, Paul and Swinney, Harry L},
  journal={Physical review letters},
  volume={72},
  number={1},
  pages={172},
  year={1994},
  publisher={APS}
}

@article{smith2012reflections,
  title={Reflections on simultaneous impact},
  author={Smith, Breannan and Kaufman, Danny M and Vouga, Etienne and Tamstorf, Rasmus and Grinspun, Eitan},
  journal={ACM Transactions on Graphics (TOG)},
  volume={31},
  number={4},
  pages={1--12},
  year={2012},
  publisher={ACM New York, NY, USA}
}

@article{bizon1998patterns,
  title={Patterns in 3D vertically oscillated granular layers: simulation and experiment},
  author={Bizon, C and Shattuck, MD and Swift, JB and McCormick, WD and Swinney, Harry L},
  journal={Physical review letters},
  volume={80},
  number={1},
  pages={57},
  year={1998},
  publisher={APS}
}

@article{chandra2024stabilized,
  title={Stabilized mixed material point method for incompressible fluid flow analysis},
  author={Chandra, Bodhinanda and Hashimoto, Ryota and Matsumi, Shinnosuke and Kamrin, Ken and Soga, Kenichi},
  journal={Computer Methods in Applied Mechanics and Engineering},
  volume={419},
  pages={116644},
  year={2024},
  publisher={Elsevier}
}

@article{chandra2024mixed,
  title={Mixed material point method formulation, stabilization, and validation for a unified analysis of free-surface and seepage flow},
  author={Chandra, Bodhinanda and Hashimoto, Ryota and Kamrin, Ken and Soga, Kenichi},
  journal={Journal of Computational Physics},
  volume={519},
  pages={113457},
  year={2024},
  publisher={Elsevier}
}

\end{document}